\documentclass[twocolumn]{aastex701}

\usepackage{amsmath}
\usepackage{verbatim}
\usepackage{hyperref}

\usepackage[utf8]{inputenc}
\usepackage[normalem]{ulem}

\newcommand{\HL}[1]{\textcolor{black}{#1}}

\newcommand{\pivec}{\mbox{\boldmath $\pi_{\rm E}$}}
\newcommand{\pieE}{\mbox{$\pi_{{\rm E},E}$}}
\newcommand{\pieN}{\mbox{$\pi_{{\rm E},N}$}}

\newcommand{\xivec}{\mbox{\boldmath $\xi_{\rm E}$}}

\newcommand{\masyr}{\mbox{$\rm mas\, yr^{-1}$}}

\newcommand{\HJD}{\mbox{${\rm HJD}^{\prime}$}}

\newcommand{\muvechel}{\mbox{\boldmath $\mu_{\rm rel, \odot}$}}
\newcommand{\muvecgeo}{\mbox{\boldmath $\mu_{\rm rel, \oplus}$}}

\newcommand{\vEarth}{\mbox{\boldmath $v_{\oplus,\perp}$}}

\newcommand{\threethirtyeight}{KMT-2021-BLG-0338}       %  KMT-2021-BLG-0338(*)     MOA-2021-BLG-082
\newcommand{\fourtwentyfour}{KMT-2021-BLG-0424}         %  KMT-2021-BLG-0424(*)     MOA-2021-BLG-111
\newcommand{\fourfiftyseven}{KMT-2021-BLG-0457}         %  KMT-2021-BLG-0457(*)     ---
\newcommand{\sixninety}{KMT-2021-BLG-0690}              %  KMT-2021-BLG-0690(*)     MOA-2021-BLG-161
\newcommand{\tensixtythree}{KMT-2021-BLG-1063}          %  KMT-2021-BLG-1063(*)     ---
\newcommand{\sixteenninetyone}{KMT-2021-BLG-1691}       %  KMT-2021-BLG-1691(*)     ---
\newcommand{\twentytwothirteen}{KMT-2021-BLG-2213}      %  KMT-2021-BLG-2213(*)     ---
\newcommand{\thirtytwoninety}{KMT-2021-BLG-3290}        %  KMT-2021-BLG-3290(*)     ---
\newcommand{\thirteeneightyfive}{KMT-2021-BLG-1385}    %  KMT-2021-BLG-1385(*)     ---
\newcommand{\seventeenfiftyone}{KMT-2021-BLG-1751}      %  KMT-2021-BLG-1751(*)     ---
\newcommand{\nineteenseven}{KMT-2021-BLG-1907}          %  KMT-2021-BLG-1907(*)     ---
\newcommand{\onefourtyeight}{KMT-2021-BLG-0148}         %  KMT-2021-BLG-0148        MOA-2021-BLG-031(*)
\newcommand{\threefourtyone}{KMT-2021-BLG-0341}         %  KMT-2024-BLG-0341(*)     ---
\newcommand{\fivesixtyseven}{KMT-2021-BLG-0567}         %  KMT-2021-BLG-0567(*)     ---
\newcommand{\sixfiftyseven}{KMT-2021-BLG-0657}          %  KMT-2021-BLG-0657(*)     ---
\newcommand{\eightseventysix}{KMT-2021-BLG-0876}        %  KMT-2021-BLG-0876(*)     ---
\newcommand{\twentythreefourty}{KMT-2021-BLG-2340}      %  KMT-2021-BLG-2340(*)     MOA-2021-BLG-367
\newcommand{\twentythreefiftyeight}{KMT-2021-BLG-2358}  %  KMT-2021-BLG-2358(*)     MOA-2021-BLG-380
\newcommand{\twentyseventhirtyseven}{KMT-2021-BLG-2737} %  KMT-2021-BLG-2737(*)     ---
\newcommand{\twentysevenfourtyfive}{KMT-2021-BLG-2745}  %  KMT-2021-BLG-2745(*)     ---
\newcommand{\thirtyonetwelve}{KMT-2021-BLG-3112}        %  KMT-2021-BLG-3112(*)     ---
\newcommand{\thirtyonefourty}{KMT-2021-BLG-3140}        %  KMT-2021-BLG-3140(*)     ---
\begin{document}

\title{Systematic KMTNet Planetary Anomaly Search. XIII. Complete Sample of 2021 Prime Field Planets
}

% Author List ------------------------------------------------------------------------------------------------------------
% 1
\author[0000-0002-4355-9838]{In-Gu Shin}
\affiliation{Department of Astronomy, Westlake University, Hangzhou 310030, Zhejiang Province, China}
\email{ingushin@gmail.com}
% 2
\author[0000-0001-9481-7123]{Jennifer C. Yee}
\affiliation{Center for Astrophysics $|$ Harvard \& Smithsonian, 60 Garden St.,Cambridge, MA 02138, USA}
\email{jyee@cfa.harvard.edu}
% 3
\author[0000-0001-6000-3463]{Weicheng Zang}
\affiliation{Department of Astronomy, Westlake University, Hangzhou 310030, Zhejiang Province, China}
\email{zangweicheng@westlake.edu.cn}
% 4
\author[0000-0002-2641-9964]{Cheongho Han}
\affiliation{Department of Physics, Chungbuk National University, Cheongju 28644, Republic of Korea}
\email{cheongho@astroph.chungbuk.ac.kr}
% 5
\author{Andrew Gould} % No OrcID on purpose
\affiliation{Department of Astronomy, Ohio State University, 140 W. 18th Ave., Columbus, OH 43210, USA}
\email{gould.34@osu.edu}
% 6
\author[0000-0001-8317-2788]{Shude Mao}
\affiliation{Department of Astronomy, Westlake University, Hangzhou 310030, Zhejiang Province, China}
\email{shude.mao@westlake.edu.cn}
% 7
\author[0000-0003-0043-3925]{Chung-Uk Lee}
\affiliation{Korea Astronomy and Space Science Institute, Daejeon 34055, Republic of Korea}
\email{leecu@kasi.re.kr}
% 8
\author[0000-0001-9823-2907]{Yoon-Hyun Ryu} 
\affiliation{Korea Astronomy and Space Science Institute, Daejeon 34055, Republic of Korea}
\email{yhryu@kasi.re.kr}
% 9
\author{Ian A. Bond}
\affiliation{Institute of Natural and Mathematical Sciences, Massey University, Auckland 0745, New Zealand}
\email{i.a.bond@massey.ac.nz}
% 10
\author{Takahiro Sumi}
\affiliation{Department of Earth and Space Science, Graduate School of Science, Osaka University, Toyonaka, Osaka 560-0043, Japan}
\email{sumi@ess.sci.osaka-u.ac.jp}
\collaboration{11}{(Leading authors),}
%
% KMTNet -------------------------------------------------------------------------------------------------------------------
% KMT Science Team
% 1
\author[0000-0003-3316-4012]{Michael D. Albrow}
\affiliation{University of Canterbury, School of Physical and Chemical Sciences, Private Bag 4800, Christchurch 8020, New Zealand}
\email{michael.albrow@canterbury.ac.nz}
% 2
\author[0000-0001-6285-4528]{Sun-Ju Chung}
\affiliation{Korea Astronomy and Space Science Institute, Daejeon 34055, Republic of Korea}
\email{sjchung@kasi.re.kr}
% 3
\author[0000-0002-9241-4117]{Kyu-Ha Hwang}
\affiliation{Korea Astronomy and Space Science Institute, Daejeon 34055, Republic of Korea}
\email{kyuha@kasi.re.kr}
% 4
\author[0000-0002-0314-6000]{Youn Kil Jung}
\affiliation{Korea Astronomy and Space Science Institute, Daejeon 34055, Republic of Korea}
\affiliation{National University of Science and Technology (UST), Daejeon 34113, Republic of Korea}
\email{ykjung21@kasi.re.kr}
% 5
\author[0000-0003-1525-5041]{Yossi Shvartzvald}
\affiliation{Department of Particle Physics and Astrophysics, Weizmann Institute of Science, Rehovot 7610001, Israel}
\email{yossishv@gmail.com}
% 6
\author[0000-0003-0626-8465]{Hongjing Yang}
\affiliation{Westlake Institute for Advanced Study, Hangzhou 310030, Zhejiang Province, China}
\affiliation{Department of Astronomy, Westlake University, Hangzhou 310030, Zhejiang Province, China}
\email{hongjing.yang@qq.com}
% KMT Operations Team
% 7
\author[0000-0002-7511-2950]{Sang-Mok Cha} % ONLY for events with years <= 2023
\affiliation{Korea Astronomy and Space Science Institute, Daejeon 34055, Republic of Korea}
\affiliation{School of Space Research, Kyung Hee University, Yongin, Kyeonggi 17104, Republic of Korea} 
\email{chasm@kasi.re.kr}
% 8
\author{Dong-Jin Kim}
\affiliation{Korea Astronomy and Space Science Institute, Daejeon 34055, Republic of Korea}
\email{keaton03@kasi.re.kr}
% 9
\author[0000-0003-0562-5643]{Seung-Lee Kim} % ONLY for events with years <= 2023
\affiliation{Korea Astronomy and Space Science Institute, Daejeon 34055, Republic of Korea}
\email{slkim@kasi.re.kr}
% 10
\author[0009-0000-5737-0908]{Dong-Joo Lee} % ONLY for events with years <= 2023
\affiliation{Korea Astronomy and Space Science Institute, Daejeon 34055, Republic of Korea}
\email{marin678@kasi.re.kr}
% 11
\author[0000-0001-7594-8072]{Yongseok Lee} % ONLY for events with years <= 2023
\affiliation{Korea Astronomy and Space Science Institute, Daejeon 34055, Republic of Korea}
\affiliation{School of Space Research, Kyung Hee University, Yongin, Kyeonggi 17104, Republic of Korea}
\email{yslee@kasi.re.kr}
% 12
\author[0000-0002-6982-7722]{Byeong-Gon Park}
\affiliation{Korea Astronomy and Space Science Institute, Daejeon 34055, Republic of Korea}
\email{bgpark@kasi.re.kr}
% 13
\author[0000-0003-1435-3053]{Richard W. Pogge} % ONLY for events with years <= 2023
\affiliation{Department of Astronomy, Ohio State University, 140 West 18th Ave., Columbus, OH  43210, USA}
\affiliation{Center for Cosmology and AstroParticle Physics, Ohio State University, 191 West Woodruff Ave., Columbus, OH 43210, USA}
\email{pogge.1@osu.edu}
\collaboration{14}{(The KMTNet Collaboration),}
% MOA ---------------------------------------------------------------------------------------------------------------------------
%
% 1
\author{Fumio Abe}
\affiliation{Institute for Space-Earth Environmental Research, Nagoya University, Nagoya 464-8601, Japan}
\email{abe@isee.nagoya-u.ac.jp}
% 2
\author{David P.~Bennett}
\affiliation{Code 667, NASA Goddard Space Flight Center, Greenbelt, MD 20771, USA}
\affiliation{Department of Astronomy, University of Maryland, College Park, MD 20742, USA}
\email{bennett.moa@gmail.com}
% 3
\author{Aparna Bhattacharya}
\affiliation{Code 667, NASA Goddard Space Flight Center, Greenbelt, MD 20771, USA}
\affiliation{Department of Astronomy, University of Maryland, College Park, MD 20742, USA}
\email{aparna.bhattacharya@nasa.gov}
% 4
\author{Ryusei Hamada}
\affiliation{Department of Earth and Space Science, Graduate School of Science, Osaka University, Toyonaka, Osaka 560-0043, Japan}
\email{hryusei@iral.ess.sci.osaka-u.ac.jp}
% 5
\author{Yuki Hirao}
\affiliation{Institute of Astronomy, Graduate School of Science, The University of Tokyo, 2-21-1 Osawa, Mitaka, Tokyo 181-0015, Japan}
\email{hirao@ioa.s.u-tokyo.ac.jp}
% 6
\author{Stela Ishitani Silva}
\affiliation{Department of Physics, The Catholic University of America, Washington, DC 20064, USA}
\affiliation{Code 667, NASA Goddard Space Flight Center, Greenbelt, MD 20771, USA}
\email{ishitanisilva@cua.edu}
% 7
\author{Shota Miyazaki}
\affiliation{Institute of Space and Astronautical Science, Japan Aerospace Exploration Agency, 3-1-1 Yoshinodai, Chuo, Sagamihara, Kanagawa 252-5210, Japan}
\email{miyazaki@ir.isas.jaxa.jp}
% 8
\author{Yasushi Muraki}
\affiliation{Institute for Space-Earth Environmental Research, Nagoya University, Nagoya 464-8601, Japan}
\email{muraki@isee.nagoya-u.ac.jp}
% 9
\author{Kansuke NUNOTA}
\affiliation{Department of Earth and Space Science, Graduate School of Science, Osaka University, Toyonaka, Osaka 560-0043, Japan}
\email{unota@iral.ess.sci.osaka-u.ac.jp}
% 10
\author{Greg Olmschenk}
\affiliation{Code 667, NASA Goddard Space Flight Center, Greenbelt, MD 20771, USA}
\email{greg@olmschenk.com}
% 11
\author{Cl\'ement Ranc}
\affiliation{Sorbonne Universit\'e, CNRS, UMR 7095, Institut d'Astrophysique de Paris, 98 bis bd Arago, 75014 Paris, France}
\email{ranc@iap.fr}
% 12
\author{Nicholas J. Rattenbury}
\affiliation{Department of Physics, University of Auckland, Private Bag 92019, Auckland, New Zealand}
\email{n.rattenbury@auckland.ac.nz}
% 13
\author[0000-0002-1228-4122]{Yuki K. Satoh}
\affiliation{College of Science and Engineering, Kanto Gakuin University, Yokohama, Kanagawa 236-8501, Japan}
\email{yukisato@kanto-gakuin.ac.jp}
% 14
\author{Daisuke Suzuki}
\affiliation{Department of Earth and Space Science, Graduate School of Science, Osaka University, Toyonaka, Osaka 560-0043, Japan}
\email{dsuzuki@ir.isas.jaxa.jp}
% 15
\author{Takuto Tamaoki}
\affiliation{Department of Earth and Space Science, Graduate School of Science, Osaka University, Toyonaka, Osaka 560-0043, Japan}
\email{tamaoki@iral.ess.sci.osaka-u.ac.jp}
% 16
\author{Sean K. Terry}
\affiliation{Code 667, NASA Goddard Space Flight Center, Greenbelt, MD 20771, USA}
\affiliation{Department of Astronomy, University of Maryland, College Park, MD 20742, USA}
\email{skterry@umd.edu}
% 17
\author{Paul . J. Tristram}
\affiliation{University of Canterbury Mt.¥ John Observatory, P.O. Box 56, Lake Tekapo 8770, New Zealand}
\email{tristram.p@gmail.com}
% 18
\author{Aikaterini Vandorou}
\affiliation{Code 667, NASA Goddard Space Flight Center, Greenbelt, MD 20771, USA}
\affiliation{Department of Astronomy, University of Maryland, College Park, MD 20742, USA}
\email{aikaterini.vandorou@utas.edu.au}
% 19
\author{Hibiki Yama}
\affiliation{Department of Earth and Space Science, Graduate School of Science, Osaka University, Toyonaka, Osaka 560-0043, Japan}
\email{yama@iral.ess.sci.osaka-u.ac.jp}
\collaboration{20}{(the MOA Collaboration)}
% ------------------------------------------------------------------------------------------------------------------------
%\correspondingauthor{In-Gu Shin (\texttt{ingushin@gmail.com})}

\begin{abstract}
The Systematic KMTNet Planetary Anomaly Search series was conducted using the KMTNet data archived from $2016$ to $2019$. From this first phase of the series, we reported a total of $50$ planetary systems hidden in the data archive, which represent about $35\%$ of the total microlensing planets discovered from $2016$ to $2019$, demonstrating that this semi-machine-based search is a crucial channel for building a complete microlensing planet sample. We continue this series for $2021$ and beyond to expand the microlensing planet sample. In this work for the $2021$ KMTNet high-cadence fields (Prime fields), we find seven hidden planetary systems and three planet candidates. These new planets represent about $33\%$ of the total microlensing planets discovered within the Prime fields observed during the $2021$ bulge season. While the by-eye search is the primary channel for detecting microlensing planets (i.e., two-thirds of microlensing planet discoveries), this work clearly shows that a systematic search series is still necessary for constructing a complete microlensing planet sample. Such a sample is essential for conducting unbiased statistical studies of planet demographics in our Galaxy. \HL{Datasets for all the events used for analyses in this work are publicly available:\dataset[doi:10.5281/zenodo.21472225]{https://doi.org/10.5281/zenodo.21472225}.}
\end{abstract}

\section{Introduction}
The Systematic KMTNet Planetary Anomaly Search series was conducted to provide a complete microlensing planet sample by finding hidden microlensing planets in the data archive obtained by the Korea Microlensing Telescope Network \citep[KMTNet;][]{kim16}. The first phase of the search was conducted for the KMTNet data archive obtained from the $2016$ to $2019$ bulge seasons \citep{gould22, gui24, hwang22, jung22, jung23, ryu24, shin23b, shin24, zang21, wang22, zang22b, zang23}. Based on the findings of this phase, the series can support the construction of the largest microlensing planet sample to date. \citet{zang25} made a statistical study based on the sample covering the $-5.2 < \log_{10}(q) < -1.5$ range, where $q$ is the planet-host mass ratio. They provide updated mass-ratio distributions of microlensing planets, showing bimodal characteristics (i.e., double Gaussian) instead of the broken power laws found in the previous study \citep{suzuki16}. In addition, \citet{zang25} infer that the total planet frequency is $\sim 0.65$ per star, and the frequency for super-Earth-mass planets is $\sim 0.35$ per star.

While we built the largest microlensing planet sample in the first phase of the series, we are continuing this series to expand the sample for the $2021$ bulge season and beyond\footnote{For the 2020 season, most KMTNet observations were temporarily suspended due to the COVID-19 pandemic.}. In this work, we conduct the systematic planetary anomaly search for $2021$ high-cadence fields ($\Gamma = 2.0 - 4.0\, {\rm hr^{-1}}$), called Prime fields \citep[see Figure 12 of][]{kim18}, of the KMTNet data archive. \HL{The search using a semi-machine-based algorithm called AnomalyFinder \citep[AF;][]{zang21, zang22b} identified a total of $137$ events caused by binary-lens systems. This AF search recovered all the planetary events in the $2021$ KMTNet Prime fields that were already found by eye, which are either published or in preparation for publication. Among the $137$ events, the KMTNet team has already found that $70$ events were caused by normal binary star systems (i.e., the mass ratios ($q$) are much larger than $0.06$), and $15$ planetary events are already published or in preparation for publication (see Table \ref{table:prime_planets}). For the remaining $52$ events, we find that these events require systematic modeling to determine the nature of the lens systems, which are in the following categories: (1) there is no inital model, (2) the existing intial model is not robust, (3) the mass ratio is close to the borderline of our mass-ratio criterion (i.e., $q \simeq 0.06$), and/or (4) there is a possibility alternative models exist.} We find that $22$ of these events were caused by possible planetary-lens systems (i.e.,$q < 0.06$). For these cases, we proceed with re-reductions for more detailed modeling using the best-quality datasets. Consequently, we thereby find seven planetary events and three planet candidates. These events have $q < 0.03$, which is the mass-ratio definition of a planetary event in this work, unless there exist degenerate non-planetary solutions, in which case it is labeled a ``planetary-event candidate".  

The methodology for this series is well-developed in the previous works. Hence, we follow the methods described in \citet{shin23b, shin24}. To reduce unnecessary and repetitive descriptions of the methods, we will exclude overly detailed descriptions that have already been covered in previous works. In this paper, we present observational information in Section \ref{sec:obs}. Then, we describe the light-curve analysis of each planetary event and planet candidate in Section \ref{sec:LC_analysis}. We present the source information of each event obtained from the analysis of color-magnitude diagrams (CMDs) in Section \ref{sec:CMDs}. By combining the analysis results, we determine the lens properties and present them in Section \ref{sec:lens_properties}. Lastly, in Section \ref{sec:discussion}, we discuss the newly discovered planets and their contributions to the microlensing planet sample. Additionally, in the Appendix, we briefly present non-planetary events that we found during the detailed analyses in this work to prevent future redundant efforts.

\section{Observations} \label{sec:obs}

% Table 1 (Observation info: planetary events) ------------------------------
\begin{deluxetable}{cc|rrr|rc}
\tablecaption{Observations of $2021$ Planets and Planet Candidates \label{table:obs_planet}}
\tablewidth{0pt}
\tablehead{
% ---------------------------------------------------------------------------
\multicolumn{2}{c|}{Event} &
\multicolumn{3}{c|}{Location} &
\multicolumn{2}{c}{obs. info.} \\
% --------------------------------------------
\multicolumn{1}{c}{KMTNet} &
\multicolumn{1}{c|}{MOA} &
% --------------------------------------------
\multicolumn{1}{c}{R.A. (J2000)} &
\multicolumn{1}{c}{Dec (J2000)} &
\multicolumn{1}{c|}{$(\ell, b)$} &
% --------------------------------------------
\multicolumn{1}{c}{$A_{I}$} &
\multicolumn{1}{c}{$\Gamma$ (${\rm hr}^{-1}$)} 
% ------------------------------------
}
\startdata
% ---------------------------------------------------------------------------------------------------------------------------------------------
% Planets
\bf{0424} &     111  & $18^{h} 04^{m} 06^{s}.44$ & $-27^{\circ} 28{'} 12{''}50$ & $(+3^{\circ}.32, -2^{\circ}.78)$ & 1.33 & 4.0 \\
\bf{0457} & \nodata  & $17^{h} 53^{m} 38^{s}.88$ & $-28^{\circ} 37{'} 39{''}18$ & $(+1^{\circ}.17, -1^{\circ}.35)$ & 3.18 & 4.0 \\
\bf{0690} &     161  & $17^{h} 59^{m} 38^{s}.51$ & $-29^{\circ} 00{'} 39{''}31$ & $(+1^{\circ}.50, -2^{\circ}.68)$ & 1.10 & 3.0 \\
\bf{1063} & \nodata  & $17^{h} 49^{m} 38^{s}.16$ & $-29^{\circ} 32{'} 47{''}94$ & $(-0^{\circ}.07, -1^{\circ}.07)$ & 4.87 & 2.0 \\
\bf{1691} & \nodata  & $17^{h} 52^{m} 45^{s}.67$ & $-28^{\circ} 07{'} 04{''}94$ & $(+1^{\circ}.51, -0^{\circ}.92)$ & 3.90 & 4.0 \\
\bf{2213} & \nodata  & $17^{h} 57^{m} 31^{s}.50$ & $-28^{\circ} 50{'} 54{''}82$ & $(+1^{\circ}.41, -2^{\circ}.20)$ & 1.48 & 8.0 \\
\bf{3290} & \nodata  & $17^{h} 54^{m} 21^{s}.65$ & $-31^{\circ} 46{'} 18{''}01$ & $(-1^{\circ}.47, -3^{\circ}.07)$ & 1.99 & 2.0 \\
% ---------------------------------------------------------------------------------------------------------------------------------------------
\hline
% Planet Candidates
\bf{1385} & \nodata  & $17^{h} 56^{m} 11^{s}.26$ & $-30^{\circ} 17{'} 30{''}98$ & $(+0^{\circ}.01, -2^{\circ}.67)$ & 1.71 & 4.0 \\
\bf{1751} & \nodata  & $17^{h} 57^{m} 25^{s}.54$ & $-29^{\circ} 32{'} 28{''}61$ & $(+0^{\circ}.79, -2^{\circ}.52)$ & 1.64 & 4.0 \\
\bf{1907} & \nodata  & $17^{h} 51^{m} 44^{s}.66$ & $-29^{\circ} 53{'} 18{''}49$ & $(-0^{\circ}.13, -1^{\circ}.63)$ & 2.51 & 4.0 \\
% ---------------------------------------------------------------------------------------------------------------------------------------------
\enddata
\tablecomments{
The boldface indicates the ``discovery" name of each event. 
The horizontal line separates planetary events from planet candidates.
The cadence ($\Gamma$) presented in this table is the KMTNet cadence of each event.
}
\end{deluxetable}
% ---------------------------------------------------------------------------

The AnomalyFinder is applied to the KMTNet datasets alone to find anomalous events. Once we find the anomalous events, we analyze the light curves for each event by combining all available datasets, which were independently obtained by other microlensing surveys, such as the Optical Gravitational Lensing Experiment \citep[OGLE-IV;][]{udalski15} or the Microlensing Observations in Astrophysics \citep[MOA;][]{bond01, sumi03}. However, in the 2021 bulge season, OGLE-IV observations are unavailable because the observatory was shut down from 2020 to 2021 due to the COVID-19 pandemic. Thus, we have only checked the MOA datasets for the anomalous events. To the best of our knowledge, there were no follow-up observations of any of the $10$ events that are analyzed here. In Table \ref{table:obs_planet}, we present a list of planetary events, including planet candidates with their location, cadence, and extinction in the $I$ band\footnote{{\scriptsize\HL{For planetary events, the extinction in the $I$ band ($A_{I}$) is a byproduct of the CMD analysis (see Section \ref{sec:CMDs}), where it plays the role of an intermediate result. That is, the fact that the photometry is not calibrated has no effect on the final measurement of the source radii, $\theta_{\ast}$. In contrast, for planet candidates, we used the relation, $A_{I} \simeq 7A_{K}$, which was derived by Y.S. by regression of $A_{I}$ \citep{nataf13} on $A_{K}$ \citep{gonzalez12} over the intersection of these two catalogs, which was restricted to the relatively low extinction regions that were accessible to good $V$-band photometry in the underlying OGLE-III optical catalog \citep{OIIIcat}.  See Figure $5$ of \citet{nataf13}. The original purpose was to serve as a rough guide to optical and infrared extinctions, which were needed for rapid decision making when choosing targets for {\it Spitzer} microlensing observations (e.g., \citealt{yee15}). They are useful when a precise measurement is not required. For example, for the four planets analyzed in this paper that have calibrated (i.e., OGLE-III) photometry available, the offset between this estimate and a proper measurement is $\langle(A_{I,{\rm YS}} - A_{I,{\rm true}})\rangle = -0.11 \pm 0.03$.}}}. Following convention, we refer to these events by the names designated by the surveys that first alerted them, highlighted in boldface.

The Prime fields of the KMTNet are BLG01, BLG02, BLG03, BLG41, BLG42, and BLG43, which have a nominal cadence of $\Gamma = 2\, {\rm hr^{-1}}$. If an event is located in an overlapping region of $2$--$4$ fields, the cadence would be increased to $\Gamma = 4$--$8\, {\rm hr^{-1}}$. The KMTNet is designed for near-continuous observation using three identical $1.6$-meter telescopes, each equipped with a wide field-of-view science camera that covers $4$ square degrees. These telescopes are located at distinct observatories in well-separated time zones: the Cerro Tololo Inter-American Observatory in Chile (KMTC), the South African Astronomical Observatory in South Africa (KMTS), and the Siding Spring Observatory in Australia (KMTA). Thus, we refer to the KMTNet datasets using a combination of the observatory and field number, e.g., KMTC02. The KMTNet images are processed using a pipeline implemented by pySIS \citep{albrow09}, which employs the difference image analysis (DIA) method \citep{tomaney96, alard98}. The final KMTNet datasets are re-reduced using an optimized version of the pySIS package \citep{yang23}. We note that the KMTNet observations are primarily conducted in the $I$ band. Additionally, one $V$-band observation is taken for every 10th $I$-band observation. These $V$-band observations are used to measure the source color.

As shown in Table \ref{table:obs_planet}, the MOA survey independently observed three of these planetary events. These observations were conducted using the MOA-Red band ($R$-band), which is comparable to a sum of the Cousins $R$ and $I$ bands. The MOA observations were taken with a $1.8$-meter telescope at the Mt. John University Observatory in New Zealand. The MOA images were processed using their DIA pipeline \citep{bond01}. The final MOA datasets were subsequently re-reduced to achieve the best quality, utilizing an optimized DIA technique.

\section{Light curve Analysis} \label{sec:LC_analysis}
\HL{The methodology of this systematic search series is already well defined in the previous works. Hence, we omit these redundant descriptions in this article. Rather than presenting them directly, we offer specific references that detail them thoroughly. We follow the methodologies described in these references. \citet{shin23a} describe the general 2L1S\footnote{The abbreviation ``$n$L$m$S" means ``$n$" lenses and ``$m$" sources.} modeling approach, strategies for incorporating higher-order effects, and techniques for addressing model degeneracies in Section $3$. Especially for the 1L2S modeling approaches, \citet{shin19} describe them in detail in the appendix. For the 3L1S modeling, \citet{gould14} describe them in the supplemental material. In addition, \citet{kuang22} provide a detailed description of the 3L1S modeling approaches in the appendix. In this work, we follow the identical procedures for light-curve analysis described in \citet{shin23b}. The acronyms and definitions of the models and their parameters we used are provided in Table $2$ of \citet{shin24}.} 

\HL{Here, we briefly present the light-curve analysis procedures. We conduct the full grid search over the ranges of $\log_{10}(s) \in [-1.0, 1.0]$ and $\log_{10}(q) \in [-5.0, 1.0]$ with $200\times200$ grids.} For each grid point, the $\chi^{2}$ minimization method searches for local minima, starting from the $21$ $\alpha$ seeds in $\alpha \in [0.0, 2\pi]$ radians. For each local minimum, a denser grid search is performed around that minimum. All locals are further refined with free-varying parameters. In addition, we provide the result of the heuristic analysis described in \citet{hwang22, ryu22, shin23a}. Lastly, we investigate the annual microlensing parallax \citep[APRX;][]{gould92} effect exclusively for events whose timescales are longer than $15$ days\footnote{Indeed, the microlensing parallax can be measured using three methods: annual microlensing parallax \citep[APRX;][]{gould92}, space-based microlensing parallax \citep[SPRX;][]{refsdal66}, and terrestrial microlensing parallax \citep[TPRX;][]{gould97}. However, in this work, we cannot adopt the methods for SPRX or TPRX to measure the microlensing parallax, because we do not have any space-based observations, nor are any of the events extremely magnified.}, which is an empirical criterion. 

We find a total of seven planetary events that were hidden in the 2021 prime fields. We claim planet detection when the event meets three specific criteria: (a) the best-fit model(s) must have $q < 0.03$; (b) the planet/binary degeneracy should be resolved with $\Delta \chi^{2} > 10$; and (c) the 2L1S/1L2S degeneracy \citep{gaudi98} must be resolved with $\Delta \chi^{2} > 15$, which were used for the previous anomaly search series. If there exists degeneracy in fiducial models, all fiducial solutions must comply with these conditions to claim a planet detection. We will provide details of the light-curve analysis on each planetary event in the following sections \ref{sec:KB210424} -- \ref{sec:KB213290}.

In addition, we find three additional events for which at least one model has a value of $q < 0.03$. In these events, we find that unresolvable degeneracies exist, such as planet/binary or 2L1S/1L2S degeneracies. Thus, these events are treated as planet candidates. We describe each planet candidate in the following sections \ref{sec:KB211385} -- \ref{sec:KB211907}.

% Table 2 (KB-21-0424: Model Parameters) ------------------------------------
%\begin{longrotatetable}
\begin{deluxetable}{lrrrr|lr}
\tablecaption{The parameters of 2L1S and 1L1S models for \fourtwentyfour\ \HL{(planetary event)} \label{table:model_0424}}
\tablewidth{0pt}
\tablehead{
% ---------------------------------------------------------------------------
\multicolumn{1}{c}{} &
\multicolumn{2}{c}{2L1S (STD)} &
\multicolumn{2}{c}{2L1S (APRX)} &
\multicolumn{2}{|c}{1L1S} \\
\multicolumn{1}{c}{Parameter} &
\multicolumn{1}{c}{Inner} & 
\multicolumn{1}{c}{Outer} & 
\multicolumn{1}{c}{Inner ($u_{0}<0$)} & 
\multicolumn{1}{c}{Inner ($u_{0}>0$)} & 
\multicolumn{1}{|c}{Parameter} &
\multicolumn{1}{c}{}
% ---------------------------------------------------------------------------
}
\startdata
% -----------------------------------------------------------------------------------
$\chi^{2} / {\rm N}_{\rm data}$ & $ 18100.721 / 17679  $ & $ 18455.026 / 17679  $ & $\mathbf{ 17681.803 / 17679  }$ & $\mathbf{ 17679.880 / 17679  }$ & $\chi^{2} / {\rm N}_{\rm data}$ & $ 22825.902 / 17679  $ \\
$\Delta\chi^{2}$                & $ 420.841            $ & $ 775.146            $ & $\mathbf{ 1.923              }$ &      \bf{\nodata (best-fit)  }  & $\Delta\chi^{2}$                & $ 5146.022           $ \\
$t_0$ [${\rm HJD'}$]            & $ 9370.817 \pm 0.004 $ & $ 9370.802 \pm 0.003 $ & $\mathbf{ 9370.844 \pm 0.004 }$ & $\mathbf{ 9370.844 \pm 0.004 }$ & $t_0$ [${\rm HJD'}$]            & $ 9370.716 \pm 0.003 $ \\
$u_0$                           & $    0.080 \pm 0.001 $ & $    0.079 \pm 0.001 $ & $\mathbf{   -0.082 \pm 0.001 }$ & $\mathbf{    0.081 \pm 0.001 }$ & $u_0$                           & $    0.074 \pm 0.001 $ \\
$t_{\rm E}$ [days]              & $   54.747 \pm 0.260 $ & $   55.696 \pm 0.246 $ & $\mathbf{   53.551 \pm 0.280 }$ & $\mathbf{   54.368 \pm 0.259 }$ & $t_{\rm E}$ [days]              & $   58.699 \pm 0.272 $ \\
$s$                             & $    1.095 \pm 0.001 $ & $    1.021 \pm 0.001 $ & $\mathbf{    1.097 \pm 0.001 }$ & $\mathbf{    1.096 \pm 0.001 }$ & \nodata                         & \nodata                \\
$q$ ($\times10^{-4}$)           & $    1.580 \pm 0.063 $ & $    1.278 \pm 0.050 $ & $\mathbf{    1.829 \pm 0.081 }$ & $\mathbf{    1.730 \pm 0.079 }$ & \nodata                         & \nodata                \\
$\langle\log_{10} q\rangle$     & $   -3.835 \pm 0.019 $ & $   -3.882 \pm 0.016 $ & $\mathbf{   -3.732 \pm 0.019 }$ & $\mathbf{   -3.746 \pm 0.019 }$ & \nodata                         & \nodata                \\
$\alpha$ [rad]                  & $    0.808 \pm 0.001 $ & $    0.805 \pm 0.001 $ & $\mathbf{   -0.808 \pm 0.001 }$ & $\mathbf{    0.815 \pm 0.001 }$ & \nodata                         & \nodata                \\
$\rho_{\ast}$ ($\times10^{-4}$) & $   10.749 \pm 0.777 $ & $    8.904 \pm 0.975 $ & $\mathbf{   13.102 \pm 0.836 }$ & $\mathbf{   12.561 \pm 0.830 }$ & \nodata                         & \nodata                \\
$\pieN$                         & \nodata                & \nodata                & $\mathbf{   -0.474 \pm 0.082 }$ & $\mathbf{   -0.288 \pm 0.070 }$ & \nodata                         & \nodata                \\
$\pieE$                         & \nodata                & \nodata                & $\mathbf{   -0.236 \pm 0.020 }$ & $\mathbf{   -0.235 \pm 0.027 }$ & \nodata                         & \nodata                \\
$|\pivec|$                      & \nodata                & \nodata                & $\mathbf{    0.530 \pm 0.081 }$ & $\mathbf{    0.372 \pm 0.070 }$ & \nodata                         & \nodata                \\
$f_{\rm S,KMTC}$                & $    0.600 \pm 0.003 $ & $    0.587 \pm 0.003 $ & $\mathbf{    0.616 \pm 0.004 }$ & $\mathbf{    0.611 \pm 0.004 }$ & $f_{\rm S,KMTC}$                & $    0.547 \pm 0.003 $ \\
$f_{\rm B,KMTC}$                & $    0.050 \pm 0.003 $ & $    0.061 \pm 0.002 $ & $\mathbf{    0.027 \pm 0.003 }$ & $\mathbf{    0.030 \pm 0.003 }$ & $f_{\rm B,KMTC}$                & $    0.095 \pm 0.002 $ \\
% -----------------------------------------------------------------------------------
\enddata
\tablecomments{
${\rm HJD' \equiv HJD - 2450000.0}$.
The boldface indicates our fiducial solutions for this event. 
}
\end{deluxetable}
%\end{longrotatetable}
% -----------------------------------------------------------------------------------

\subsection{\fourtwentyfour} % Planet : KMT-2021-BLG-0424(*) == MOA-2021-BLG-111
\label{sec:KB210424}

% Figure 1 (KB-21-0424) : planet -------------------------------------------------------------------
\begin{figure}[t]
\epsscale{1.00}
\plotone{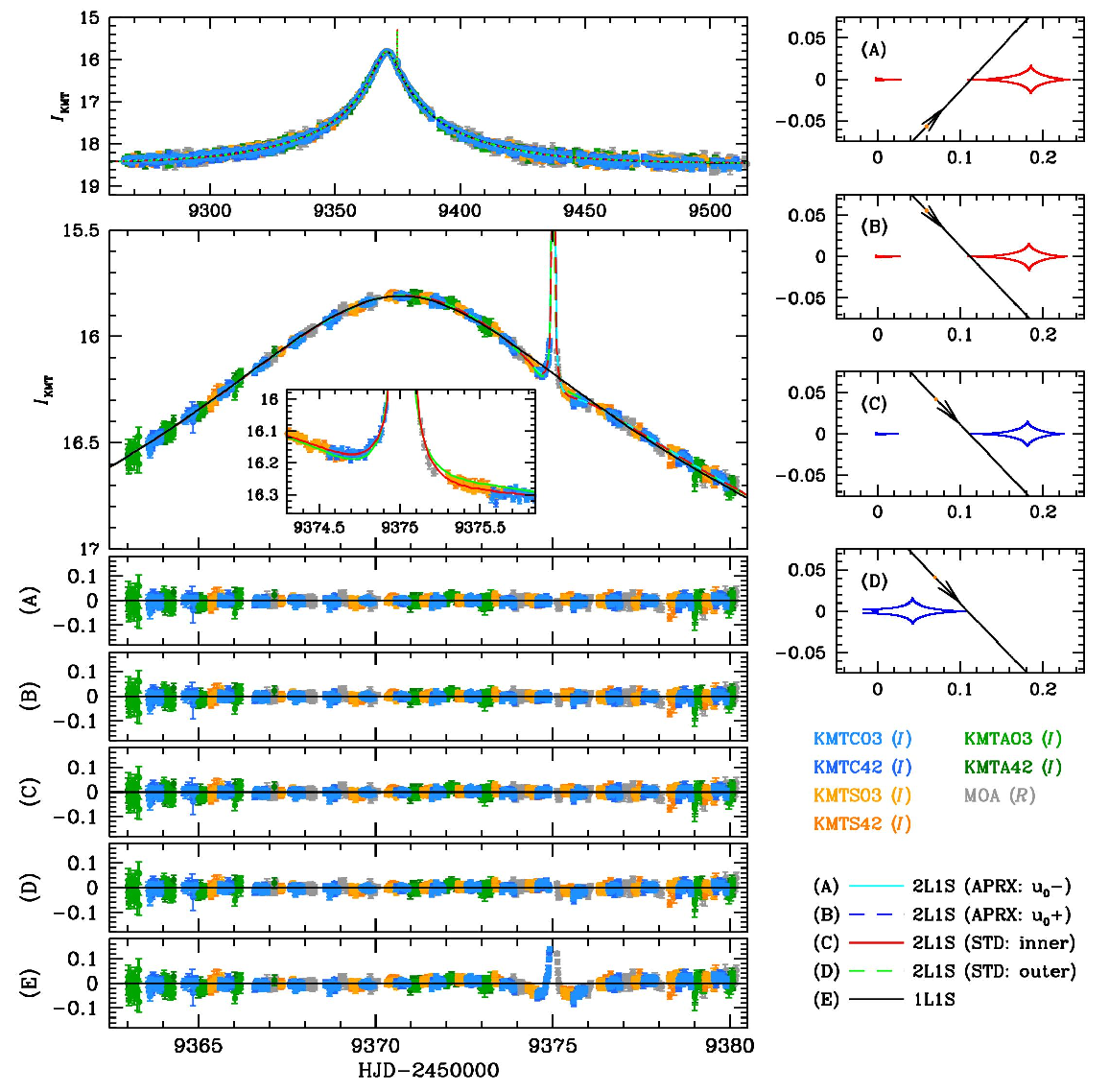}
\caption{Light curve of \fourtwentyfour\ \HL{(planetary event)} with 2L1S and 1L1S models.  
\label{fig:lc_0424}}
\end{figure}
% --------------------------------------------------------------------------------------------------

In Figure \ref{fig:lc_0424}, we present the light curve of \fourtwentyfour. The light curve exhibits a clear bump-shaped anomaly at $\HJD \sim 9735.0$. The rising and declining parts of the anomaly are densely covered by KMTC and MOA observations, respectively. Quantitatively, the anomaly yields $\Delta\chi^{2} = 5146$ from the single-lens/single-source (1L1S) model. We find that the anomaly can be explained by a binary-lens/single-source (2L1S) model indicating a planetary system (i.e., $q \sim 2\times10^{-4} < 0.03$). We find plausible models caused by the inner/outer degeneracy from the grid search (see geometry panels (C) and (D) in Figure \ref{fig:lc_0424}). However, we can resolve the degeneracy due to two reasons. First, the trajectory angle is not vertical in this case. As pointed out by \citet{zhang22}, the intrinsic symmetry of the lens equation is close to perfect only for vertical trajectories. This symmetry deteriorates as the trajectory angle moves farther away from vertical. In this case, the shallow bump from the off-axis cusp comes after the main bump anomaly in the inner geometry and before it in the outer geometry. Hence, the inner/outer degeneracy can be resolved. Second, the anomaly was densely covered by KMTNet and MOA observations. Thus, the models can be resolved quantitatively by their $\Delta\chi^{2} \rm{(outer - inner)}= 354$ difference (see the zoom-in of Figure \ref{fig:lc_0424}). Indeed, based on the heuristic analysis, we find that $s_{+}^{\dagger} = 1.057$ for the major image perturbation, which is derived from $\tau_{\rm anom} \equiv (t_{\rm anom}  - t_{0})/t_{\rm E}  = 0.076$ and $u_{\rm anom} \equiv (\tau_{\rm anom}  + u_{0})^{1/2} = 0.111$ where $(t_{\rm anom}, t_{0}, t_{\rm E}, u_{0}) = (9375.0, 9370.817, 54.747, 0.080)$. This $s_{+}^{\dagger}$ prediction is consistent with $s^{\dagger} = \sqrt{s_{\rm inner}s_{\rm outer}} = 1.057$.

\begin{figure}[t]
\epsscale{1.00}
\plotone{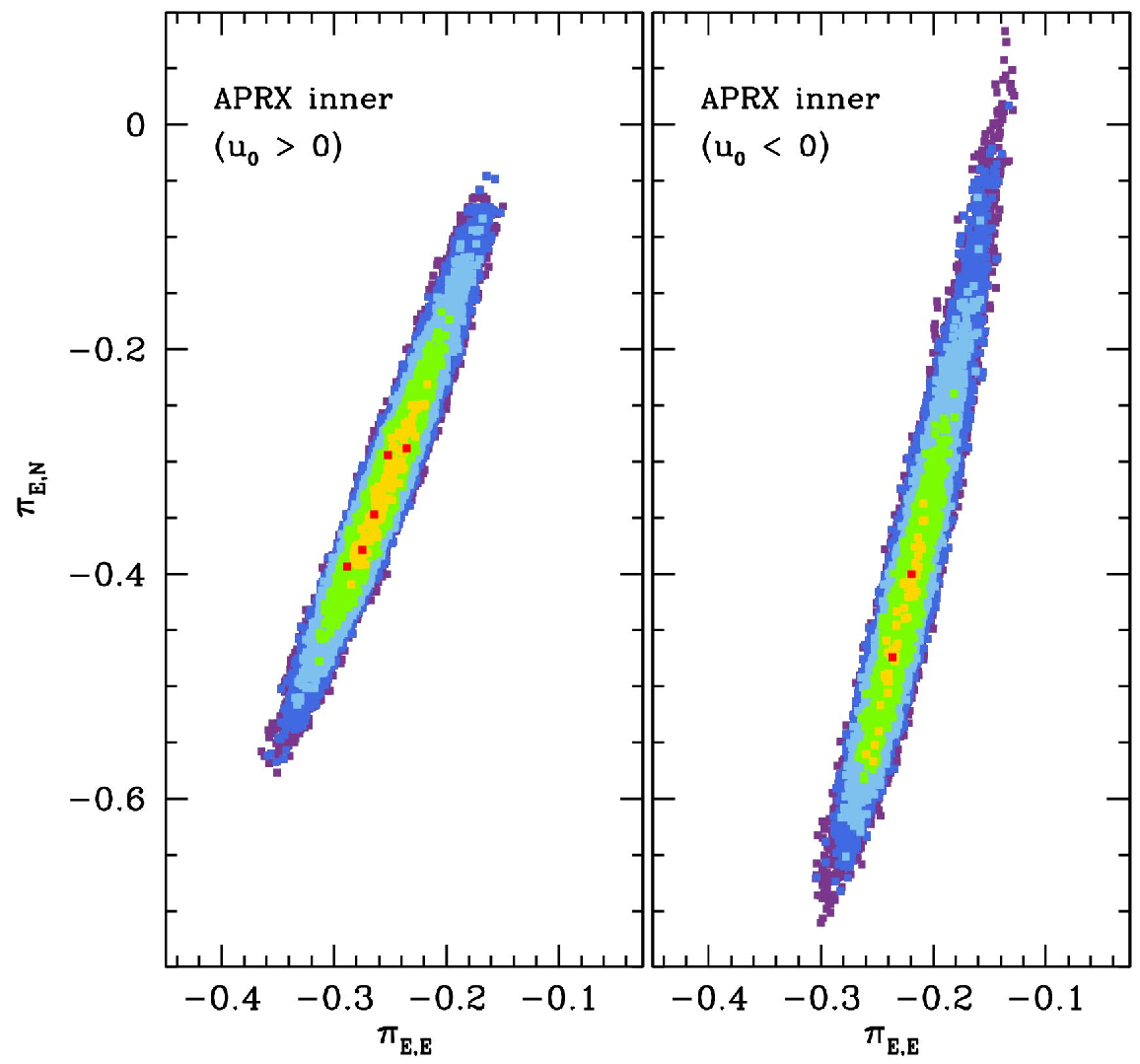}
\caption{$\pivec$ distributions of \fourtwentyfour\ \HL{(planetary event)} for APRX inner ($u_{0} > 0$; left panel) and 
($u_{0} < 0$; right panel) cases. The colored dots represent $\Delta\chi^{2} \leq n^{2}$ with 
respect to the best-fit $\chi^{2}$ value of each case, where $n = 1$ (red), $2$ (yellow), 
$3$ (green), $4$ (light blue), $5$ (blue), and $6$ (purple).
\label{fig:APRX_0424}}
\end{figure}

We test the annual microlensing parallax (APRX) effect because the standard/static-lens (STD) models show that $t_{\rm E} \sim 54$ days, which is longer than our criterion for attempting APRX modeling (i.e., $t_{\rm E} > 15$ days). We find that there exists a strong APRX effect, which improves the fits by $\Delta\chi^{2} = 421$ compared to the STD inner case. In Figure \ref{fig:APRX_0424}, we present the $\pivec$ distributions of APRX ($u_{0} > 0$) and ($u_{0} < 0$) cases, which are well constrained. In addition, we present the model parameters of 2L1S (STD and APRX cases) and 1L1S interpretations in Table \ref{table:model_0424}.

% Figure 2 (KB-21-0424: APRX) : planet -------------------------------------------------------------------
% --------------------------------------------------------------------------------------------------

% Table 3 (KB-21-0457: Model Parameters) ------------------------------------
\begin{deluxetable}{lrr|lr|lr}
\tablecaption{The parameters of 2L1S and 1L2S models for \fourfiftyseven\ \HL{(planetary event)} \label{table:model_0457}}
\tablewidth{0pt}
\tablehead{
% ---------------------------------------------------------------------------
\multicolumn{3}{c}{2L1S} &
\multicolumn{2}{|c}{1L2S} & 
\multicolumn{2}{|c}{1L1S} \\
\multicolumn{1}{c}{Parameter} &
\multicolumn{1}{c}{Outer} & 
\multicolumn{1}{c}{Inner} &
\multicolumn{1}{|c}{Parameters} &
\multicolumn{1}{c}{} &
\multicolumn{1}{|c}{Parameters} &
\multicolumn{1}{c}{}
% ---------------------------------------------------------------------------
}
\startdata
% -----------------------------------------------------------------------------------
$\chi^{2} / {\rm N}_{\rm data}$  & $\mathbf{ 9958.847 / 9962    }$ & $ 9983.682 / 9962    $ & $\chi^{2} / {\rm N}_{\rm data}$ & $ 10096.809 / 9962   $ & $\chi^{2} / {\rm N}_{\rm data}$ & $ 10426.240 / 9962   $ \\
$\Delta\chi^{2}$                 &      \bf{\nodata             }  & $ 25.151             $ & $\Delta\chi^{2}$                & $ 137.962            $ & $\Delta\chi^{2}$                & $ 467.393            $ \\
$t_0$ [${\rm HJD'}$]             & $\mathbf{ 9329.427 \pm 0.034 }$ & $ 9329.470 \pm 0.032 $ & $t_{0,S1}$ [${\rm HJD'}$]       & $ 9329.307 \pm 0.037 $ & $t_0$ [${\rm HJD'}$]            & $ 9329.574 \pm 0.028 $ \\
$u_0$                            & $\mathbf{    0.796 \pm 0.059 }$ & $    0.820 \pm 0.062 $ & $u_{0,S1}$                      & $    1.074 \pm 0.029 $ & $u_0$                           & $    0.806 \pm 0.055 $ \\
$t_{\rm E}$ [days]               & $\mathbf{    9.507 \pm 0.456 }$ & $    9.421 \pm 0.453 $ & $t_{\rm E}$ [days]              & $    8.111 \pm 0.123 $ & $t_{\rm E}$ [days]              & $    9.425 \pm 0.399 $ \\
$s$                              & $\mathbf{    1.289 \pm 0.048 }$ & $    1.792 \pm 0.051 $ & $t_{0,S2}$ [${\rm HJD'}$]       & $ 9331.604 \pm 0.035 $ & \nodata                         & \nodata                \\
$q$ ($\times 10^{-4}$)           & $\mathbf{   47.094 \pm 7.240 }$ & $   57.572 \pm 7.746 $ & $u_{0,S2}$                      & $    0.006 \pm 0.022 $ & \nodata                         & \nodata                \\
$\langle\log_{10} q\rangle$      & $\mathbf{   -2.345 \pm 0.071 }$ & $   -2.247 \pm 0.060 $ & $q_{\rm flux}$                  & $    0.005 \pm 0.001 $ & \nodata                         & \nodata                \\
$\alpha$ [rad]                   & $\mathbf{    1.289 \pm 0.011 }$ & $    1.299 \pm 0.011 $ & $\rho_{\ast,S1}$                & $    0.723 \pm 0.092 $ & \nodata                         & \nodata                \\
$\rho_{\ast,{\rm limit}}$        & $\mathbf{  < 0.112           }$ & $  < 0.106           $ & $\rho_{\ast,S2}$                & $    0.084 \pm 0.006 $ & \nodata                         & \nodata                \\
$f_{\rm S, KMTC}$                & $\mathbf{    0.854 \pm 0.114 }$ & $    0.907 \pm 0.129 $ & $f_{\rm S, KMTC}$               & $    1.185 \pm 0.001 $ & $f_{\rm S, KMTC}$               & $    0.885 \pm 0.116 $ \\
$f_{\rm B, KMTC}$                & $\mathbf{    0.332 \pm 0.114 }$ & $    0.279 \pm 0.129 $ & fixed $f_{\rm B} = 0.0$         &      \nodata           & $f_{\rm B, KMTC}$               & $    0.301 \pm 0.116 $ \\
% -----------------------------------------------------------------------------------
\enddata
\tablecomments{
${\rm HJD' \equiv HJD - 2450000.0}$.
The boldface indicates our fiducial solution for this event.
We note that the inequality sign of the $\rho_{\ast}$ parameters indicates 
the upper limit on $\rho_{\ast}$ (i.e., $3\sigma$). 
}
\end{deluxetable}
% -----------------------------------------------------------------------------------

Note that we also test the orbital-lens (OBT) effect because it can affect the APRX measurement \citep{shin12}. We find that the OBT effect is minor. Although the OBT effect improves the fits (by $\Delta\chi^{2} = 12$), the $|\pivec|$ value is identical to well within $1\sigma$ (i.e., $|\pivec| = 0.37 \pm 0.07$ and $0.38 \pm 0.07$ for APRX only and APRX+OBT cases, respectively). In addition, the unusually large OBT parameter (i.e., $d\alpha/dt \sim -5.91\, {\rm rad\, yr^{-1}}$) would imply that the system is not bound \citep[i.e., the dynamical constraint of the lens system (${\rm KE/PE} \gg 1$);][]{dong09}. Hence, we conclude that the 2L1S solutions with the APRX effect are the fiducial model for this planetary event.

\subsection{\fourfiftyseven} % Planet : KMT-2021-BLG-0457(*)
\label{sec:KB210457}
% inner/outer degen. | 2L1S/1L2S degen. is resolved. (fixed Fb case)

\begin{figure}[t]
\epsscale{1.00}
\plotone{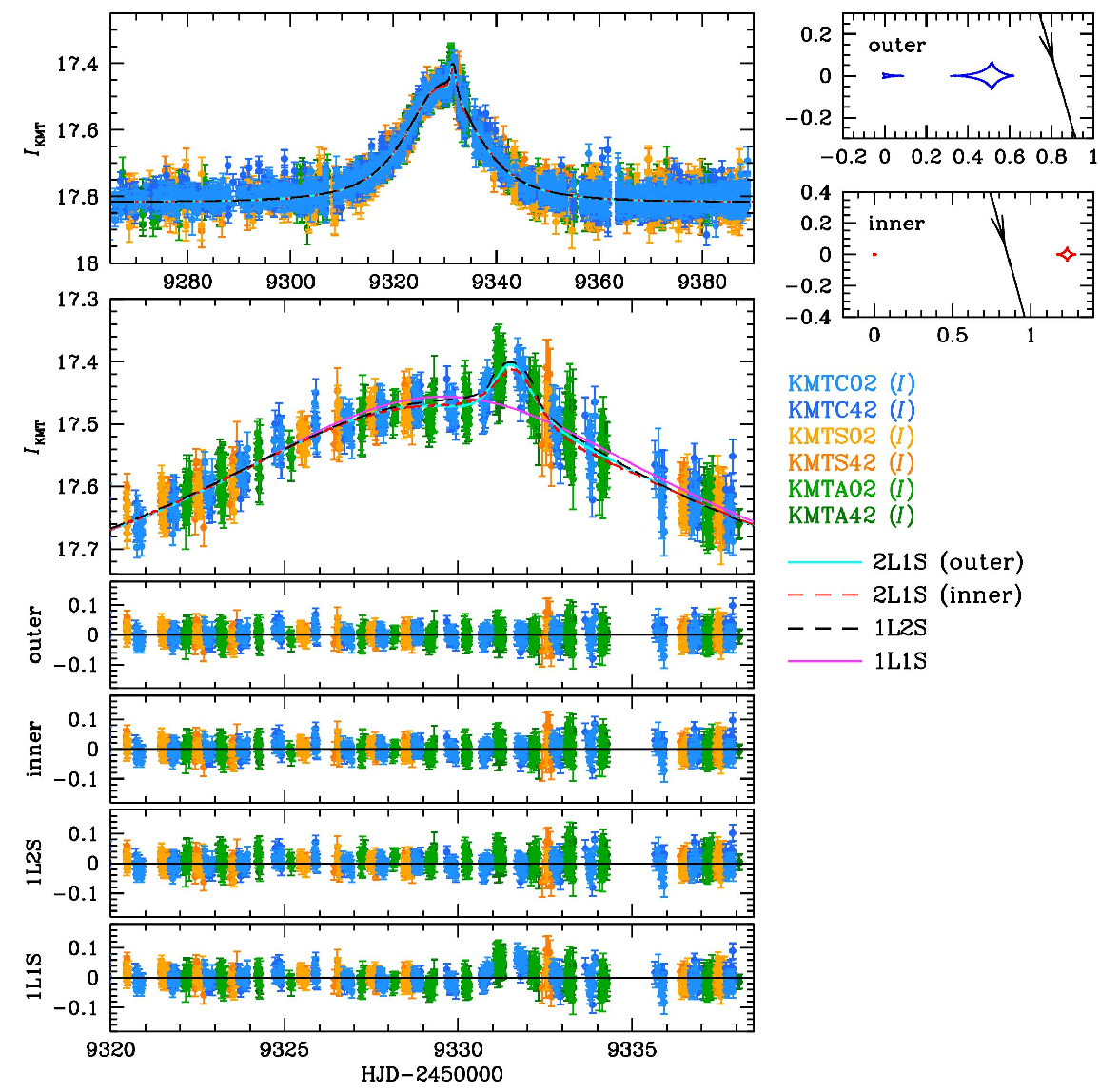}
\caption{Light curve of \fourfiftyseven\ \HL{(planetary event)} with 2L1S and 1L2S models.  
\label{fig:lc_0457}}
\end{figure}

We present the observed light curve of \fourfiftyseven\ with 2L1S and 1L2S (single-lens/binary-source) models in Figure \ref{fig:lc_0457}. The light curve exhibits a bump-shaped anomaly at $\HJD \sim 9331.5$, which yields $\Delta\chi^{2} = 467$ relative to the 1L1S model. The anomaly can be explained by the 2L1S or 1L2S models presented in Table \ref{table:model_0457}. However, the 1L2S model is disfavored by $\Delta\chi^{2} = 138$, which is sufficient to resolve the 2L1S/1L2S degeneracy (i.e., $\Delta\chi^{2} \gg 15$). The 2L1S models indicate that this event was caused by a planetary system (i.e., $q = 47\times10^{-4} < 0.03$). Note that we find a possible inner/outer degeneracy. Indeed, the heuristic analysis indicates that $s_{+}^{\dagger} = 1.505$ from $(\tau_{\rm anom}, u_{\rm anom}, t_{\rm anom}, t_{0}, t_{\rm E}, u_{0}) = (0.271, 0.841, 9332.000, 9329.427, 9.507, 0.796)$. The $s_{+}^{\dagger}$ prediction is similar to $s^{\dagger} = \sqrt{s_{\rm inner}s_{\rm outer}} = 1.520$. However, the inner solution is disfavored by $\Delta\chi^{2} = 25$. Similar to the case of \fourtwentyfour, the trajectory is not vertical (see geometries in Figure \ref{fig:lc_0457}). However, in this case, the angle is closer to vertical compared to those of \fourtwentyfour. Thus, the degeneracy is much stronger (i.e., $\Delta\chi^{2} = 25$). Although this event shows much greater inner/outer similarity, the $\Delta\chi^{2}$ value is large enough to resolve the degeneracy considering our $\chi^{2}$ thresholds. Hence, we conclude that the 2L1S-outer model is the fiducial solution for this planetary event. Note that we do not consider the APRX effect for this event because $t_{\rm E}$ is only $9.5 \pm 0.5$ days, which is much shorter than our criterion ($t_{\rm E} > 15$ days) to test the APRX effect.

% Figure 3 (KB-21-0457) : planet -------------------------------------------------------------------
% --------------------------------------------------------------------------------------------------

% Figure 4 (KB-21-0690) : planet -------------------------------------------------------------------
\begin{figure}[t]
\epsscale{1.00}
\plotone{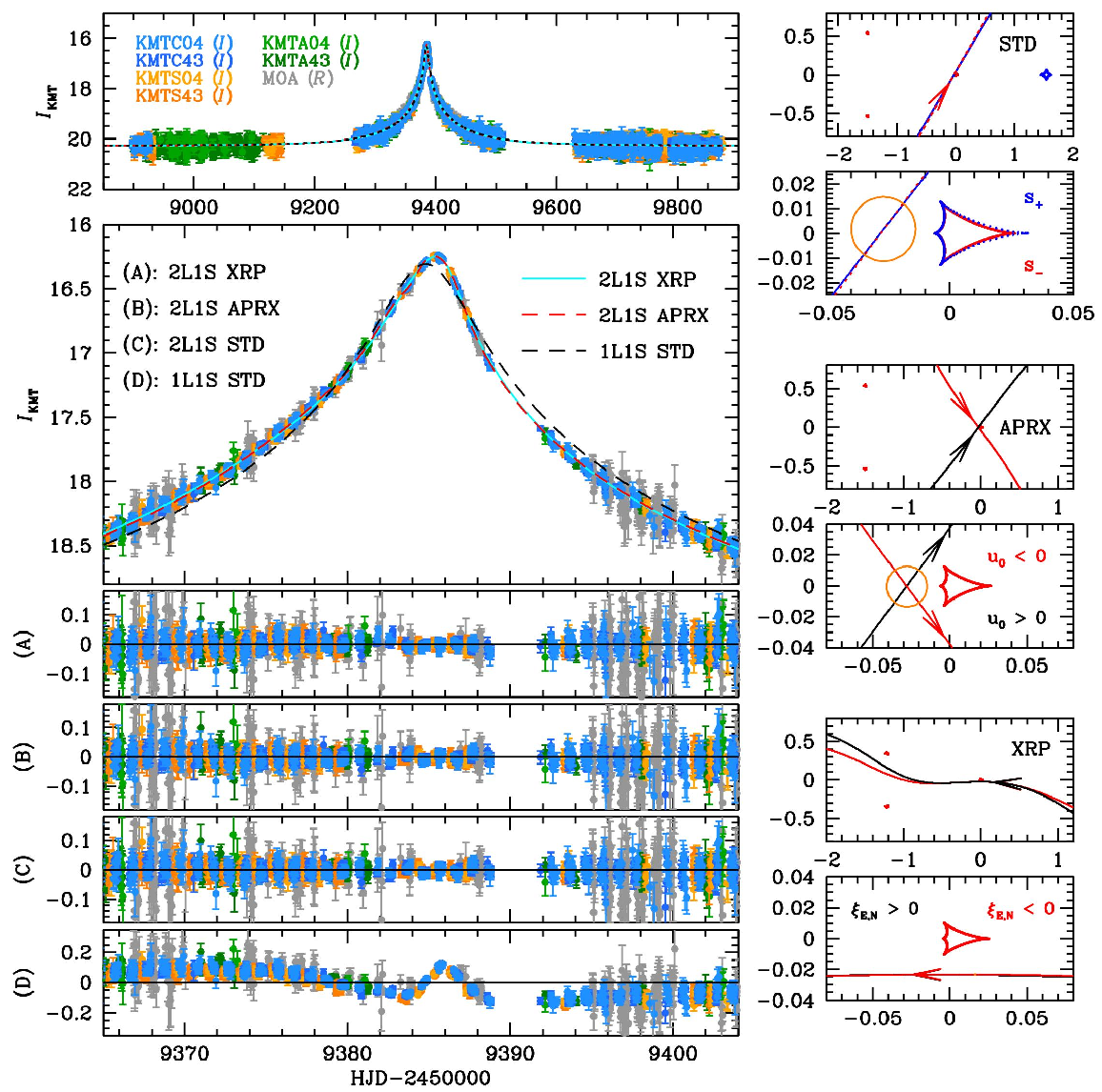}
\caption{Light curve of \sixninety\ \HL{(planetary event)} with 2L1S and 1L1S models.  
\label{fig:lc_0690}}
\end{figure}
% --------------------------------------------------------------------------------------------------

\subsection{\sixninety} % Planet : KMT-2021-BLG-0690(*) == MOA-2021-BLG-161
\label{sec:KB210690}
The observed light curve of \sixninety\ exhibits asymmetric deviations around the peak as shown in Figure \ref{fig:lc_0690}, which induces clear residuals from the 1L1S model. Quantitatively, the residuals yield $\Delta\chi^{2} = 23070$. This kind of asymmetric anomaly can, in principle, be explained by a 1L1S model with the APRX effect. However, for this event, the 1L1S APRX model cannot explain the anomaly. There are clear residuals similar to those of the 1L1S STD model, which yield $\Delta\chi^{2} = 14261$.

We find that the binary lens (2L) is essential for explaining the anomaly of this event, and both 2L1S models, $s_{\pm}$\footnote{The notations of $s_{+}$ and $s_{-}$ represent the conventional close ($s < 1$) and wide ($s > 1$) solutions, respectively. In this work, by following the notation in \citet{shin23b, shin24}, we use the $s_{\pm}$ notation instead of the close/wide notation.}, can well describe the asymmetric anomaly. Indeed, the heuristic analysis indicates that $s_{-}^{\dagger} = 0.989$ and $s_{+}^{\dagger} = 1.011$ from $(\tau_{\rm anom}, u_{\rm anom}, t_{\rm anom}, t_{0}, t_{\rm E}, u_{0}) = (0.000, 0.022, 9384.3, 9384.314, 123.362, 0.022)$. This $s_{-}^{\dagger}$ prediction is similar to $s^{\dagger} = \sqrt{s_{-}s_{+}} = 0.996$. In Table \ref{table:model_0690}, we present the parameters of the 2L1S models with those of the 1L1S model for comparison. Even though both STD $s_{\pm}$ cases seem to describe the light curve well, we find that the $s_{-}$ case shows much better fits compared with the $s_{+}$ case by $\Delta\chi^{2} = 83$. We find that this $\chi^{2}$ difference comes from ${\rm HJD}^{\prime} \sim 9368$ to $9388$ continuously. Thus, we choose the $s_{-}$ case as our fiducial case because of the sufficient $\Delta\chi^{2}$. 

Based on the $s_{-}$ STD case, we begin to investigate higher-order effects by following the normal course. First, we investigate the APRX effect because this event shows a long timescale (i.e., $t_{\rm E} \sim 120$ days). We find that the $\chi^{2}$ improves by only $\sim 2$ (see Table \ref{table:model_0690}) when the APRX effect is included. Although the $\chi^{2}$ improvements are negligible, we find that the $\pivec$ distributions are well constrained, which can provide meaningful information to estimate the lens properties. In Figure \ref{fig:XRP_APRX_0690}, we present the $\pivec$ distributions of the APRX ($u_{0} > 0$) and ($u_{0} < 0$) cases. We note that the best-fit value of $|\pivec|$ cannot be used for conclusive determinations of the lens properties because the improvement in $\Delta\chi^{2}$ is too small to consider the best-fit $\pivec$ values as reliable. Second, although the $\Delta\chi^{2}$ in the APRX models is minor, we check the OBT effect because the effect can affect the $\pivec$ distributions without $\chi^{2}$ improvements. 

\subsubsection{STD, APRX, and APRX$+$OBT Models Have Extremely Low Prior Probabilities}
\label{sec:KB210690_low_prob}
In our usual procedure, we find various light-curve models in the ``Light Curve Analysis" section (Section \ref{sec:LC_analysis}) and then evaluate the prior probabilities of various such models using a Galactic model in the ``Planet Properties" section (Section \ref{sec:lens_properties}). Then, it can happen that one solution is so disfavored compared to another that it strongly affects our assessment of the most likely interpretation, even competing as a criterion with the light-curve-based $\chi^2$ values of the different solutions.

However, as we now show, the STD solution and its APRX and APRX$+$OBT variants have such extremely low prior probabilities that we are forced to search for competing solutions by somewhat unconventional means.

There are two independent characteristics of this event that have low prior probability, namely, low proper motion and low orbital motion. While only the first of these applies the STD and APRX models, the other one arises in the APRX$+$OBT not because the two additional degrees of model freedom (dof) lead to unreasonable fits that could be dominated by systematics, but because these dof are extremely constrained. Thus, we assess both together.

In all three models, the light-curve measurement yields $\rho_{\ast} = 0.013 \pm 0.001$, and $ t_{\ast}\equiv \rho_{\ast} t_{\rm E} = 1.61 \pm 0.013$ days, while from Section \ref{sec:CMDs}, we have $\theta_{\ast}=0.47\pm0.03\,\mu{\rm as}$. These values yield $\mu_{\rm rel}=\theta_{\ast}/t_{\ast}=0.11\pm0.01\, {\rm mas\, yr^{-1}}$ and $\theta_{\rm E}= 36\pm3\,\mu{\rm as}$.

There are two separate issues related to this measurement. First, \citet{MASADA} showed that the occurrence rate (in nature) of such low-proper motion events is $p(\mu_{\rm rel})=(\mu_{\rm rel}/\sigma_{\mu})^{3}/6\sqrt{\pi}\rightarrow 5\times 10^{-6}$ where $\sigma_{\mu}=3\,{\rm mas\, yr^{-1}}$. If this probability were fairly weighed against the light-curve likelihood of the model, then it would correspond, by itself, to $\Delta\chi^{2}=-2\ln p = 25$, and would show that models that would otherwise be very improbable need to be considered.

Before continuing, we note that \citet{MASADA} also found that for $69$ archival planetary events with measured proper motions, the $\mu_{\rm rel}$ distribution was skewed toward lower values, presumably because it is easier to detect the finite source effect in slower-moving events. According to his formula for planetary events $p(\mu_{\rm rel})= (\mu_{\rm rel}/\sigma_{\mu})^{2}/4\rightarrow 3\times 10^{-4}$. However, this boosted probability for planetary detections does not change the fact that the underlying microlensing event is extremely rare, occurring only one time per $100$ microlensing seasons. Moreover, the lowest such archival proper motion that \citet{MASADA} actually found was $\mu_{\rm rel}=0.6\pm 0.1\,{\rm mas\, yr^{-1}}$ (for OGLE-2018-BLG-1647), i.e., with $162$ times larger prior probability than this event.

In addition, the same large $\rho_{\ast}$ measurement implies the unusually small $\theta_{\rm E}=36\,\mu{\rm as}$, which is just above the so-called ``Einstein Desert" \citep{gould22}, implying that it is likely to be a very low-mass brown dwarf. Such objects clearly exist, so this is not, by itself, a reason to doubt the solution. However, it would be unusual for reasons that are completely independent of the kinematics.

The second extreme characteristic is that the best fit for the ratio of projected kinetic to potential energy $\beta\equiv{\rm KE/PE}=0.001$, while at the $3\,\sigma$ level $\beta<0.006$. It is straightforward to show that if a planet in a circular orbit is viewed at random angles, then $p(\beta<\beta_{0})\simeq \beta_{0}\rightarrow 0.006$. By itself, this would be notable, but it would not disqualify the model. But added to the extremely low prior probability of $\mu_{\rm rel} = 0.11\,{\rm mas\, yr^{-1}}$, it concatenates the concern.

\subsubsection{Searching for Alternative Models}
\label{sec:KB210690_XRP}

Therefore, although the APRX models describe the observed light curve quite well, we nevertheless consider an alternative interpretation to explain the light curve, which is the xallarap effect \citep[XRP;][]{griest92,han97,paczynski97,dominik98,poindexter05}. Initially, we began our search for XRP models by seeding them at the 1L1S model. However, we found that the 1L1S XRP model cannot explain the deviations at the peak, similar to the 1L1S APRX model, which shows $\Delta\chi^{2} \simeq 10713$. Thus, for the XRP model, the binary lenses are still essential to describe the observations. For the 2L1S XRP interpretation, we find an alternative model that produces a nearly identical $\chi^{2}$ value to those of the APRX models. In addition, in contrast to the 2L1S STD and APRX models, the finite source effect is not detected for the XRP model. That is, we can only obtain an upper limit on the angular source radius (see Table \ref{table:model_0690}). \HL{We note that this 2L1S XRP model only considers the effect caused by the orbital motion of the binary source system without the brightness contribution of the second source, by the definition of the XRP effect, because we require a counterpart to the 2L1S APRX models that already describe the light curve well.} 

\begin{figure}[t]
\epsscale{1.00}
\plotone{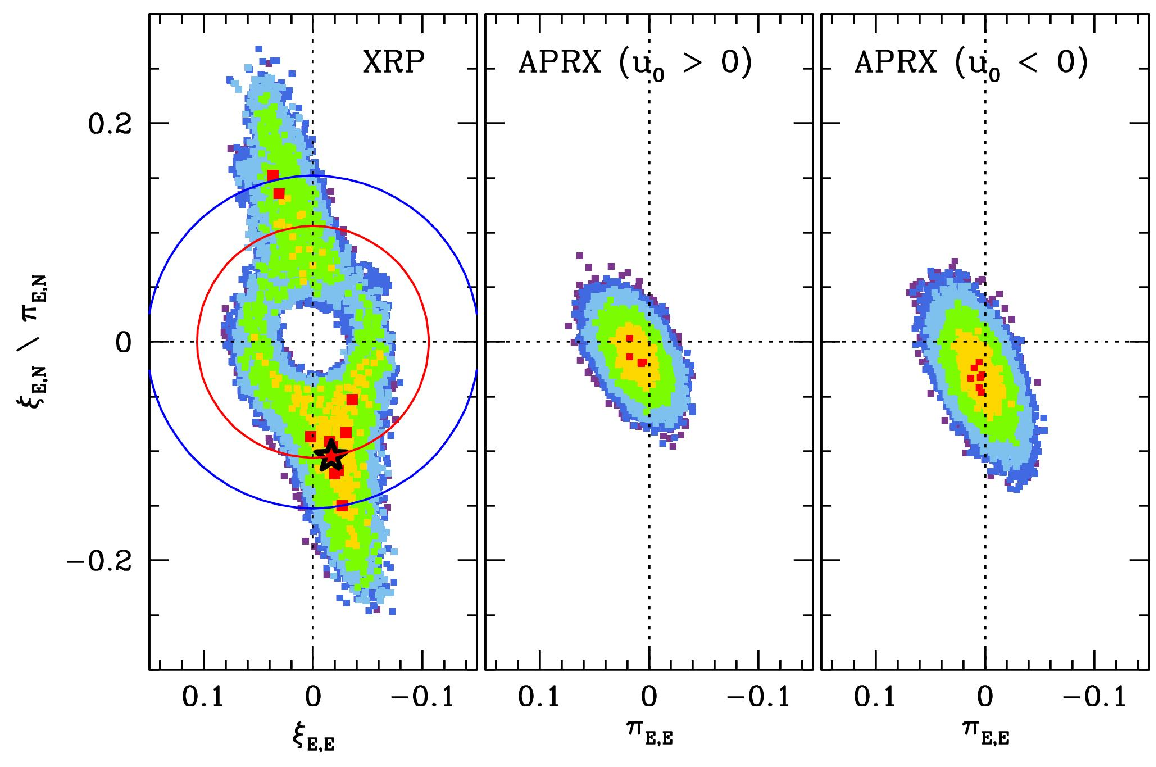}
\caption{$\xivec$ and $\pivec$ distributions of \sixninety\ \HL{(planetary event)} for XRP and APRX 
($u_{0} > 0$ and $u_{0} < 0$) cases. The color scheme is identical to Figure \ref{fig:APRX_0424}.
\HL{In the XRP panel, the red and blue circles represent $|\xi_{\rm E}|$ values derived from 
Equation \ref{eqn:xivec_test} for $q_{\rm c} = 0.4$ and $0.65$, respectively. The star symbol 
indicates the best-fit $\xivec$ value, and red squares indicate the $1\sigma$ cases 
($\Delta\chi^{2} \le 1$) relative to the best-fit XRP model.}
\label{fig:XRP_APRX_0690}}
\end{figure}

Regardless of the lens properties from the XRP model (we will discuss these in Section \ref{sec:lens_KB210690}), the XRP model parameters must obey basic physics to be reasonable. Thus, we check whether or not the parameter values are viable. First, we define the mass ratio of the binary sources ($q_{c}$), and use the definitions of the $\rho_{\ast}$ parameter and the amplitude of the xallarap vector ($\xivec = (\xi_{{\rm E}, N}, \xi_{{\rm E}, E})$) as described in follow equations:
\begin{equation}
\begin{split}
q_{c} \equiv \frac{M_{\rm S2}}{M_{\rm S1}}~;~ 
\rho_{\ast} \equiv \frac{\theta_{\ast}}{\theta_{\rm E}}~;~ \\
|\xivec| \equiv \frac{a_{\rm S1}}{\theta_{\rm E} D_{\rm S}} = \sqrt{\xi_{{\rm E}, N}^{2} + \xi_{{\rm E}, E}^2},
\end{split}
\end{equation}
where $M_{\rm S1}$ is the mass of the primary source involved in the lensing event, $M_{\rm S2}$ is the mass of the secondary source, $\theta_{\ast}$ is the angular source radius, $\theta_{\rm E}$ is the angular Einstein ring radius, $D_{\rm S}$ is the distance to the sources, and $a_{\rm S1}$ is the distance from the center of mass to the primary source in the system. Then, we rewrite Kepler's third law and Newton's third law using our definitions. That is,  
\begin{equation}
\begin{split}
\left(\frac{a}{\rm au}\right)^3 = \left( \frac{M_{\rm S1}+M_{\rm S2}}{M_{\odot}}\right) \left( \frac{P}{\rm yr}\right)^2 = \\ M_{\rm S1}(1+q_{c})P^{2} ~;~ 
a_{\rm S1} = a \left(\frac{q_{c}}{1+q_{c}}\right),
\end{split}
\end{equation}
where $a$ is the semi-major axis of the system and $P$ is the orbital period. The rewritten equation with respect to the XRP parameters is
\begin{equation}
|\xivec| = \frac{\rho_{\ast}}{\theta_{\ast} D_{\rm S}} \frac{q_{c}}{(1+q_{c})^{\frac{2}{3}}} M_{\rm S1}^{\frac{1}{3}}P^{\frac{2}{3}}.
\label{eqn:xie}
\end{equation}
We set $M_{\rm S1} = 1\, M_{\odot}$ and $D_{\rm S} \simeq 8$ kpc, $P \simeq 0.95$ years, and $\theta_{\ast} \simeq 0.47\times10^{-3}\, {\rm mas}$. In addition, we have the $3\sigma$ upper limit: $\rho_{\ast} \leq 0.0013$. Thus, Equation \ref{eqn:xie} beomces that 
\begin{equation}
|\xivec| = \sqrt{\xi_{{\rm E}, N}^{2} + \xi_{{\rm E}, E}^2} \leq 0.3272 \times \left[ \frac{q_{c}}{(1+q_{c})^{\frac{2}{3}}} \right].
\label{eqn:xivec_test}
\end{equation}
For the 2L1S XRP model to be valid, the $\xivec$ distributions must satisfy Equation \ref{eqn:xivec_test} for a given $q_{\rm c}$. We find that $|\xivec| = 0.105$ is equivalent to $q_{\rm c} = 0.4$, which is identical to the best-fit value $|\xivec| = 0.106 \pm 0.039$ as shown in Figure \ref{fig:XRP_APRX_0690} (see the red circle).

% Table 4 (KB-21-0690: Model Parameters) ------------------------------------
\begin{longrotatetable}
\begin{deluxetable}{lrrrr|lr|lr}
\tablecaption{The parameters of 2L1S and 1L1S models for \sixninety\ \HL{(planetary event)} \label{table:model_0690}}
\tablewidth{0pt}
\tabletypesize{\scriptsize}
\tablehead{
% ---------------------------------------------------------------------------
\multicolumn{1}{c}{} &
\multicolumn{2}{c}{2L1S (STD)} &
\multicolumn{2}{c}{2L1S (APRX)} &
\multicolumn{2}{|c}{2L1S (XRP)} &
\multicolumn{2}{|c}{1L1S (STD)} \\
% ---------------------------------------------------------------------------
\multicolumn{1}{c}{Parameter} &
\multicolumn{1}{c}{$s_{-}$} & 
\multicolumn{1}{c}{$s_{+}$} & 
\multicolumn{1}{c}{$\mathbf{s_{-}\, (u_{0} > 0)}$} & 
\multicolumn{1}{c}{$\mathbf{s_{-}\, (u_{0} < 0)}$} & 
\multicolumn{1}{|c}{Parameter} &
\multicolumn{1}{c}{} & 
\multicolumn{1}{|c}{Parameter} &
\multicolumn{1}{c}{}
% ---------------------------------------------------------------------------
}
\startdata
% -----------------------------------------------------------------------------------
$\chi^{2} / {\rm N}_{\rm data}$  & $ 9106.584 / 9109    $ & $ 9189.603 / 9109    $ & $\mathbf{ 9104.638 / 9109    }$ & $\mathbf{ 9104.225 / 9109    }$ & $\chi^{2} / {\rm N}_{\rm data}$  & $\mathbf{ 9103.102 / 9109     }$ & $\chi^{2} / {\rm N}_{\rm data}$ & $ 35173.533 / 9109   $ \\
$\Delta\chi^{2}$                 & $    3.482           $ & $   86.501           $ & $\mathbf{   1.536            }$ & $\mathbf{    1.123           }$ & $\Delta\chi^{2}$                 & \bf{\nodata (best-fit)        }  & $\Delta\chi^{2}$                & $ 26070.431          $ \\
$t_0$ [${\rm HJD'}$]             & $ 9384.314 \pm 0.005 $ & $ 9384.335 \pm 0.005 $ & $\mathbf{ 9384.310 \pm 0.007 }$ & $\mathbf{ 9384.306 \pm 0.006 }$ & $t_0$ [${\rm HJD'}$]             & $\mathbf{ 9384.196 \pm  0.013 }$ & $t_0$ [${\rm HJD'}$]            & $ 9384.806 \pm 0.003 $ \\
$u_0$                            & $    0.022 \pm 0.001 $ & $    0.022 \pm 0.001 $ & $\mathbf{    0.022 \pm 0.001 }$ & $\mathbf{   -0.022 \pm 0.001 }$ & $u_0$                            & $\mathbf{    0.023 \pm  0.001 }$ & $u_0$                           & $    0.019 \pm 0.001 $ \\
$t_{\rm E}$ [days]               & $  123.362 \pm 1.184 $ & $  127.876 \pm 1.241 $ & $\mathbf{  123.539 \pm 2.295 }$ & $\mathbf{  123.375 \pm 2.251 }$ & $t_{\rm E}$ [days]               & $\mathbf{  122.240 \pm  2.104 }$ & $t_{\rm E}$ [days]              & $  126.852 \pm 1.000 $ \\
$s$                              & $    0.487 \pm 0.002 $ & $    2.037 \pm 0.009 $ & $\mathbf{    0.488 \pm 0.002 }$ & $\mathbf{    0.487 \pm 0.002 }$ & $s$                              & $\mathbf{    0.555 \pm  0.003 }$ & \nodata                         & \nodata                \\
$q$ ($\times 10^{-4}$)           & $  230.767 \pm 3.860 $ & $  231.008 \pm 4.043 $ & $\mathbf{  230.068 \pm 5.289 }$ & $\mathbf{  233.154 \pm 5.432 }$ & $q$ ($\times 10^{-4}$)           & $\mathbf{  131.179 \pm  2.913 }$ & \nodata                         & \nodata                \\
$\langle\log_{10} q\rangle$      & $   -1.636 \pm 0.007 $ & $   -1.630 \pm 0.007 $ & $\mathbf{   -1.641 \pm 0.010 }$ & $\mathbf{   -1.637 \pm 0.010 }$ & $\langle\log_{10} q\rangle$      & $\mathbf{   -1.886 \pm  0.010 }$ & \nodata                         & \nodata                \\
$\alpha$ [rad]                   & $    5.353 \pm 0.003 $ & $   -0.909 \pm 0.003 $ & $\mathbf{    5.351 \pm 0.003 }$ & $\mathbf{   -5.353 \pm 0.003 }$ & $\alpha$ [rad]                   & $\mathbf{    3.147 \pm  0.007 }$ & \nodata                         & \nodata                \\
$\rho_{\ast}$                    & $    0.013 \pm 0.001 $ & $    0.013 \pm 0.001 $ & $\mathbf{    0.013 \pm 0.001 }$ & $\mathbf{    0.013 \pm 0.001 }$ & $\rho_{\ast}$                    & $\mathbf{  < 0.0013           }$ & \nodata                         & \nodata                \\
$\pi_{{\rm E},N}$                & \nodata                & \nodata                & $\mathbf{   -0.018 \pm 0.019 }$ & $\mathbf{   -0.042 \pm 0.025 }$ & $\xi_{{\rm E},N}$                & $\mathbf{   -0.105 \pm  0.043 }$ & \nodata                         & \nodata                \\
$\pi_{{\rm E},E}$                & \nodata                & \nodata                & $\mathbf{    0.005 \pm 0.014 }$ & $\mathbf{    0.005 \pm 0.015 }$ & $\xi_{{\rm E},E}$                & $\mathbf{   -0.017 \pm  0.026 }$ & \nodata                         & \nodata                \\
$|\pivec|$                       & \nodata                & \nodata                & $\mathbf{    0.019 \pm 0.012 }$ & $\mathbf{    0.042 \pm 0.019 }$ & $|\xivec|$                       & $\mathbf{    0.106 \pm  0.039 }$ & \nodata                         & \nodata                \\
\nodata                          & \nodata                & \nodata                & \nodata                         & \nodata                         & $\psi$  [deg]                    & $\mathbf{  307.132 \pm 47.293 }$ & \nodata                         & \nodata                \\
\nodata                          & \nodata                & \nodata                & \nodata                         & \nodata                         & {\it i} [deg]                    & $\mathbf{  -24.667 \pm 19.648 }$ & \nodata                         & \nodata                \\
\nodata                          & \nodata                & \nodata                & \nodata                         & \nodata                         & $P$ [year]                       & $\mathbf{    0.948 \pm  0.019 }$ & \nodata                         & \nodata                \\
$f_{\rm S, KMTC}$                & $    0.097 \pm 0.001 $ & $    0.096 \pm 0.001 $ & $\mathbf{    0.097 \pm 0.002 }$ & $\mathbf{    0.097 \pm 0.002 }$ & $f_{\rm S, KMTC}$                & $\mathbf{    0.102 \pm  0.002 }$ & $f_{\rm S, KMTC}$               & $    0.094 \pm 0.001 $ \\
$f_{\rm B, KMTC}$                & $    0.027 \pm 0.001 $ & $    0.028 \pm 0.001 $ & $\mathbf{    0.028 \pm 0.002 }$ & $\mathbf{    0.027 \pm 0.002 }$ & $f_{\rm B, KMTC}$                & $\mathbf{    0.022 \pm  0.002 }$ & $f_{\rm B, KMTC}$               & $    0.031 \pm 0.001 $ \\
% -----------------------------------------------------------------------------------
\enddata
\tablecomments{
${\rm HJD' \equiv HJD - 2450000.0}$. 
The boldface indicates our fiducial solution(s) for this event.
}
\tabletypesize{\small}
\end{deluxetable}
\end{longrotatetable}
% -----------------------------------------------------------------------------------

% Figure 5 (KB-21-0690) : XRP & APRX contours -------------------------------------------------------------------
% --------------------------------------------------------------------------------------------------

Indeed, $q_{\rm c} = 0.4$ implies that the secondary source's mass ($M_{\rm S2}$) is about $0.35\, M_{\odot}$, derived from the primary source's visible color (see Section \ref{sec:CMDs}) and the dwarf stellar properties presented in \citet{dwarftable}.
Hence, the absolute $I$-band magnitude of the secondary source ($M_{I, {\rm S2}}$) is approximately $8.9$, while for the primary source, it is $M_{I, {\rm S1}} \simeq 5.0$. Consequently, the flux ratio between the two sources ($q_{\rm flux} \equiv f_{\rm S2}/f_{\rm S1}$), is about $0.028$. Therefore, the flux of the secondary source ($f_{\rm S2}$) is calculated to be $0.0028$, based on the total flux ($f_{\rm S} \equiv (f_{\rm S1} + f_{\rm S2}) = 0.102$). This corresponds to an $I$-band magnitude for the secondary of $I_{\rm S2} \simeq 24.4$ on the KMTNet magnitude scale. This value is significantly fainter than the baseline magnitude, which is $I_{\rm base} = 20.3$. Even at the maximum magnification ($A_{\rm max} \simeq 49$), the secondary source's magnitude ($I_{\rm S2} \simeq 20.2$) remains nearly unchanged relative to its baseline magnitude. Therefore, the brightness of the secondary source cannot produce a signal that would affect the light curve for this event.

% Figure 6 (KB-21-0690) : Cum Dchi2 APRX - XRP -------------------------------------------------------------------
\begin{figure}[t]
\epsscale{1.00}
\plotone{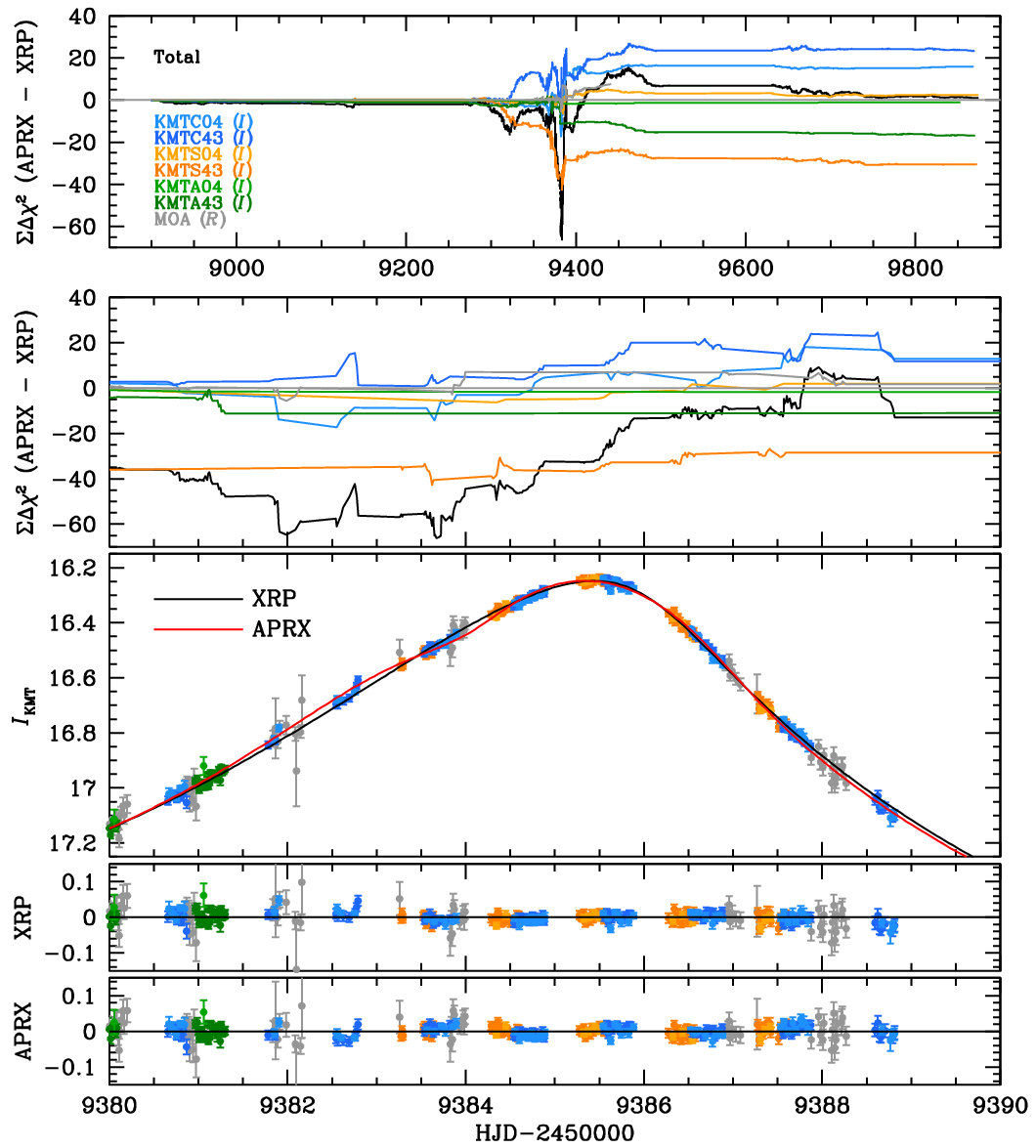}
\caption{$\Sigma\Delta\chi^{2}$ plots of \sixninety\ \HL{(planetary event)} comparing the APRX and XRP models 
with their light curves and residuals. 
The top panel shows the $\Sigma\Delta\chi^{2} {\rm (APRX-XPR)}$ plot of the whole range of the light curve. 
The middle panel shows a zoom-in plot around the peak (${\rm HJD}^{\prime} = 9380$ to $9390$), 
where the biggest difference between the APRX and XRP models is exhibited. The bottom three panels show 
the light curve and residuals of the APRX and XRP models. 
\label{fig:cumDchi2_0690}}
\end{figure}
% --------------------------------------------------------------------------------------------------

Furthermore, for nearly half of the $1\sigma$ cases ($\Delta\chi^{2} \le 1$) with $\xivec \le 0.106$ (i.e., red squares are located inside the red circle in Figure \ref{fig:XRP_APRX_0690}), the secondary source is much fainter than the best-fit case. In addition, we also find that the $1\sigma$ cases at $q_{c} \simeq 0.65$ (see the blue circle in Figure \ref{fig:XRP_APRX_0690}) imply that the secondary could be a white dwarf, which is several magnitudes fainter.

Although the XRP model is more likely than the APRX models, it is difficult to conclude that the XRP model is a fiducial solution of this event. Because we cannot conclusively rule out the APRX models without clear evidence, even if their lens properties are suspicious. Thus, we compare the XRP and APRX models. In Figure \ref{fig:cumDchi2_0690}, we present cumulative $\Delta\chi^{2}$ plots with the light curve and residuals of both models. We find that there exists a big difference around the peak (see zoom-ins of Figure \ref{fig:cumDchi2_0690}). This difference in both fits arises from the finite source effect, although the source does not cross the caustic. For the APRX cases, the finite source effect is detected. In contrast, for the XRP case, the effect is not detected (see geometries of Figure \ref{fig:lc_0690}). Indeed, it is not clear whether or not the $\rho_{\ast}$ measurements of the APRX models are caused by the fits of systematics around the peak. However, the $\Delta\chi^{2}$ of the peak is canceled out by the better fits of the other parts for the XRP model. As a result, both models show almost the same $\chi^{2}$. From this investigation, we can understand the reason for the different $\rho_{\ast}$ measurements in the APRX and XRP models. However, there is no clear evidence that could decisively resolve them at this moment. The only fact we know is that the lens system has a planet, considering the mass ratios of both models. Therefore, we treat this event as a planetary event with APRX and XRP solutions.

% Table 5 (KB-21-1063: Model Parameters) ------------------------------------
\begin{deluxetable}{lrr|lr|lr}
\tablecaption{The parameters of 2L1S, 1L2S, and 1L1S models for \tensixtythree\ \HL{(planetary event)} \label{table:model_1063}}
\tablewidth{0pt}
\tablehead{
% ---------------------------------------------------------------------------
\multicolumn{3}{c}{2L1S} &
\multicolumn{2}{|c}{1L2S} & 
\multicolumn{2}{|c}{1L1S} \\
\multicolumn{1}{c}{Parameter} &
\multicolumn{1}{c}{$\mathbf{s_{-}}$} & 
\multicolumn{1}{c}{$s_{+}$} & 
\multicolumn{1}{|c}{Parameter} &
\multicolumn{1}{c}{} &
\multicolumn{1}{|c}{Parameter} &
\multicolumn{1}{c}{}
% ---------------------------------------------------------------------------
}
\startdata
% -----------------------------------------------------------------------------------
$\chi^{2} / {\rm N}_{\rm data}$  & $\mathbf{ 5079.212 / 5080    }$ & $\mathbf{ 5079.266 / 5080    }$ & $\chi^{2} / {\rm N}_{\rm data}$ & $ 5094.851 / 5080    $ & $\chi^{2} / {\rm N}_{\rm data}$ & $ 5189.318 / 5080    $ \\  
$\Delta\chi^{2}$                 &      \bf{\nodata (best-fit)  }  & $\mathbf{ 0.054              }$ & $\Delta\chi^{2}$                & $ 15.639             $ & $\Delta\chi^{2}$                & $ 110.106            $ \\ 
$t_0$ [${\rm HJD'}$]             & $\mathbf{ 9366.938 \pm 0.054 }$ & $\mathbf{ 9366.892 \pm 0.045 }$ & $t_{0,S1}$ [${\rm HJD'}$]       & $ 9367.000 \pm 0.048 $ & $t_0$ [${\rm HJD'}$]            & $ 9366.822 \pm 0.045 $ \\  
$u_0$                            & $\mathbf{    0.182 \pm 0.016 }$ & $\mathbf{    0.177 \pm 0.015 }$ & $u_{0,S1}$                      & $    0.213 \pm 0.028 $ & $u_0$                           & $    0.213 \pm 0.019 $ \\  
$t_{\rm E}$ [days]               & $\mathbf{   19.741 \pm 1.206 }$ & $\mathbf{   20.176 \pm 1.219 }$ & $t_{\rm E}$ [days]              & $   19.211 \pm 1.328 $ & $t_{\rm E}$ [days]              & $   17.945 \pm 1.087 $ \\  
$s$                              & $\mathbf{    0.816 \pm 0.012 }$ & $\mathbf{    1.332 \pm 0.024 }$ & $t_{0,S2}$ [${\rm HJD'}$]       & $ 9359.437 \pm 0.053 $ & \nodata                         & \nodata                \\  
$q$ ($\times 10^{-4}$)           & $\mathbf{    6.703 \pm 3.515 }$ & $\mathbf{   15.386 \pm 4.876 }$ & $u_{0,S2}$                      & $    0.003 \pm 0.007 $ & \nodata                         & \nodata                \\  
$\langle\log_{10} q\rangle$      & $\mathbf{   -3.103 \pm 0.146 }$ & $\mathbf{   -2.782 \pm 0.122 }$ & $q_{\rm flux}$                  & $    0.006 \pm 0.002 $ & \nodata                         & \nodata                \\  
$\alpha$ [rad]                   & $\mathbf{    5.952 \pm 0.025 }$ & $\mathbf{    2.692 \pm 0.013 }$ & $\rho_{\ast,S1}$                & $  < 0.341           $ & \nodata                         & \nodata                \\  
$\rho_{\ast}$                    & $\mathbf{  < 0.032           }$ & $\mathbf{  < 0.029           }$ & $\rho_{\ast,S2}$                & $  < 0.023           $ & \nodata                         & \nodata                \\  
$f_{\rm S, KMTC}$                & $\mathbf{    0.063 \pm 0.007 }$ & $\mathbf{    0.061 \pm 0.006 }$ & $f_{\rm S, KMTC}$               & $    0.065 \pm 0.008 $ & $f_{\rm S, KMTC}$               & $    0.076 \pm 0.008 $ \\  
$f_{\rm B, KMTC}$                & $\mathbf{    0.135 \pm 0.006 }$ & $\mathbf{    0.137 \pm 0.006 }$ & $f_{\rm B, KMTC}$               & $    0.134 \pm 0.007 $ & $f_{\rm B, KMTC}$               & $    0.123 \pm 0.008 $ \\  
% -----------------------------------------------------------------------------------
\enddata
\tablecomments{
${\rm HJD' \equiv HJD - 2450000.0}$. 
The boldface indicates our fiducial solutions for this event.
We note that the inequality sign of the $\rho_{\ast}$ parameters indicates 
the upper limit on $\rho_{\ast}$ (i.e., $3\sigma$).
}
\end{deluxetable}
% -----------------------------------------------------------------------------------

\subsection{\tensixtythree} % Planet : KMT-2021-BLG-1063(*)
\label{sec:KB211063}

% Figure 7 (KB-21-1063) : planet -------------------------------------------------------------------
\begin{figure}[t]
\epsscale{1.00}
\plotone{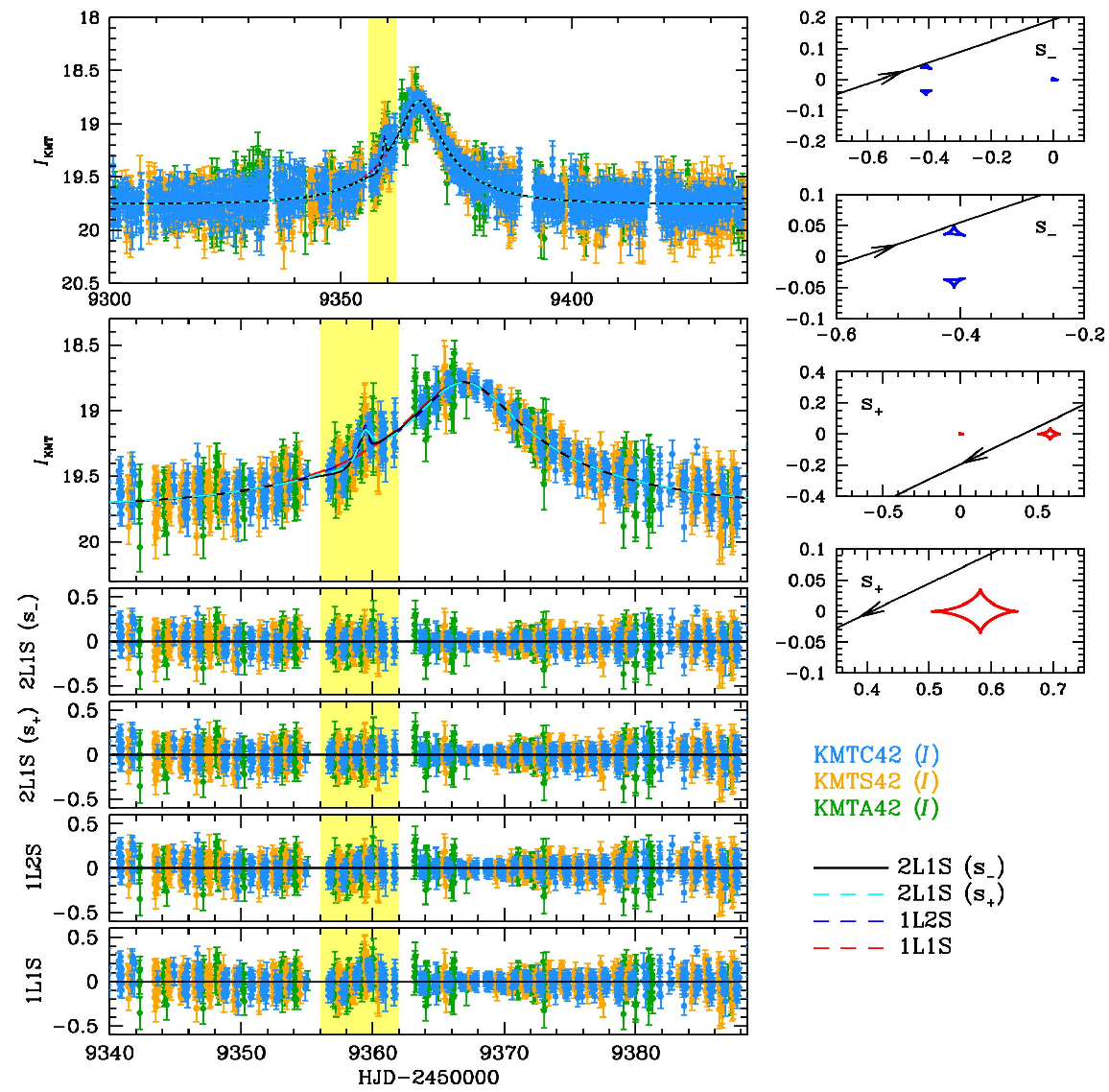}
\caption{Light curve of \tensixtythree\ \HL{(planetary event)} with 2L1S, 1L2S, and 1L1S models.  
\HL{The residual parts highlighted in yellow show the noticeable differences in different models. 
Compared to the best-fit model, 2L1S $(s_{-})$, both 1L2S and 1L1S models show worse fits at this
 anomaly part.}
\label{fig:lc_1063}}
\end{figure}
% --------------------------------------------------------------------------------------------------

The light curve of \tensixtythree\ has a small bump-shaped anomaly at ${\rm HJD}^{\prime} \sim 9359.5$ on the 1L1S-like light curve as shown in Figure \ref{fig:lc_1063}. This anomaly is buried in the data scatter, so it is difficult to notice by eye. However, the AnomalyFinder can identify this anomaly (see the residuals to the 1L1S model in Figure \ref{fig:lc_1063}). Quantitatively, the anomaly yields $\Delta\chi^{2} = 110$ between 2L1S (i.e., the best fit) and 1L1S models. The best-fit 2L1S model (i.e., the $s_{-}$ case) indicates that the event was caused by a planetary lens system (i.e., $q \sim 7\times10^{-4})$). Indeed, the heuristic analysis indicate that $s_{-}^{\dagger} = 0.813$ and $s_{+}^{\dagger} = 1.229$ from $(\tau_{\rm anom}, u_{\rm anom}, t_{\rm anom}, t_{0}, t_{\rm E}, u_{0}) = (-0.375, 0.416, 9359.5, 9397.0, 20.0, 0.180)$. This value of $s_{-}^{\dagger}$ is consistent with $s_{-}  \simeq 0.816$, which is the planetary-caustic-crossing model \citep[see][]{hwang22}. We also find that there is a competing 2L1S model (i.e., the $s_{+}$ case) with $\Delta\chi^{2} = 0.05$. The weak amorphous bump can be produced by either a major-image perturbation ($s_{+}$ case; far from the caustic) or a minor-image perturbation ($s_{-}$ case; hitting the planetary caustic). In principle, we expect this degeneracy to be resolved because the trajectory angle is far from vertical.  However, in this case, the anomaly is faint ($I \simeq 19.3$) and the error bars are too large to resolve the two cases. Although this $s_{-}/s_{+}$ degeneracy cannot be resolved, the $s_{+}$ case also indicates that the lens is a planetary system (i.e., $q \sim 15\times10^{-4}$). 

Because the anomaly is a bump-shaped, we check the 2L1S/1L2S degeneracy. We find that the 1L2S model can describe the anomaly. However, the fits are worse than 2L1S fits by $\Delta\chi^{2} = 16$, which satisfies our $\chi^{2}$ criterion for resolving the 2L1S/1L2S degeneracy. That is, the 1L2S model is rejected. Indeed, the worse fit of the 1L2S model is noticeable by eye over the anomaly ($9358.0 \lesssim {\rm HJD}^{\prime} \lesssim 9361.0$) as shown in the residuals of Figure \ref{fig:lc_1063}. In Table \ref{table:model_1063}, we present the two planetary models together with the 1L2S and 1L1S models, which are shown for comparison.

% Table 6 (KB-21-1691: Model Parameters) ------------------------------------
%\begin{longrotatetable}
\begin{deluxetable}{lrr|lr}
\tablecaption{The parameters of 2L1S and 1L1S models for \sixteenninetyone\ \HL{(planetary event)} \label{table:model_1691_01}}
\tablewidth{0pt}
%\tabletypesize{\scriptsize}
\tablehead{
% ---------------------------------------------------------------------------
\multicolumn{3}{c}{2L1S} &
\multicolumn{2}{|c}{1L1S} \\
% ---------------------------------------------------------------------------
\multicolumn{1}{c}{Parameter} &
\multicolumn{1}{c}{$s_{-}$} & 
\multicolumn{1}{c}{$s_{+}$} & 
\multicolumn{1}{|c}{Parameter} &
\multicolumn{1}{c}{} 
% ---------------------------------------------------------------------------
}
\startdata
% -----------------------------------------------------------------------------------
$\chi^{2} / {\rm N}_{\rm data}$ & $ 9653.730 / 9569    $ & $ 9654.200 / 9569    $ & $\chi^{2} / {\rm N}_{\rm data}$ & $ 9882.585 / 9569     $ \\
$\Delta\chi^{2}$                & $ 88.255             $ & $ 88.725             $ & $\Delta\chi^{2}$                & $ 317.11              $ \\
$t_0$ [${\rm HJD'}$]            & $ 9406.070 \pm 0.019 $ & $ 9406.066 \pm 0.019 $ & $t_{0}$ [${\rm HJD'}$]          & $ 9406.055 \pm  0.019 $ \\
$u_0$                           & $   -0.029 \pm 0.005 $ & $   -0.028 \pm 0.006 $ & $u_{0}$                         & $    0.024 \pm  0.005 $ \\
$t_{\rm E}$ [days]              & $   31.865 \pm 6.339 $ & $   32.979 \pm 5.841 $ & $t_{\rm E}$ [days]              & $   40.627 \pm 10.483 $ \\
$s$                             & $    0.894 \pm 0.021 $ & $    1.155 \pm 0.027 $ & \nodata                         & \nodata                 \\
$q$ ($\times10^{-4}$)           & $   21.973 \pm 6.494 $ & $   21.450 \pm 6.979 $ & \nodata                         & \nodata                 \\
$\langle\log_{10} q\rangle$     & $   -2.638 \pm 0.119 $ & $   -2.610 \pm 0.118 $ & \nodata                         & \nodata                 \\
$\alpha$ [rad]                  & $    4.597 \pm 0.021 $ & $    4.602 \pm 0.021 $ & \nodata                         & \nodata                 \\
$\rho_{\ast}$ ($\times10^{-3}$) & $    2.523 \pm 0.655 $ & $    2.376 \pm 0.697 $ & \nodata                         & \nodata                 \\
$f_{\rm S, KMTC}$               & $    0.007 \pm 0.001 $ & $    0.006 \pm 0.001 $ & $f_{\rm S, KMTC}$               & $    0.005 \pm  0.001 $ \\
$f_{\rm B, KMTC}$               & $    0.099 \pm 0.001 $ & $    0.099 \pm 0.001 $ & $f_{\rm B, KMTC}$               & $    0.100 \pm  0.001 $ \\
% -----------------------------------------------------------------------------------
\enddata
\tablecomments{
${\rm HJD' \equiv HJD - 2450000.0}$.
The presented $\Delta\chi^{2}$ values are calculated comparing with the best-fit model (i.e., the 2L2S $s_{+}$ case) presented in Table \ref{table:model_1691_02}.
}
%\tabletypesize{\small}
\end{deluxetable}
%\end{longrotatetable}
% -----------------------------------------------------------------------------------

% Figure 8 (KB-21-1691) : planet -------------------------------------------------------------------
\begin{figure}[t]
\epsscale{0.85}
\plotone{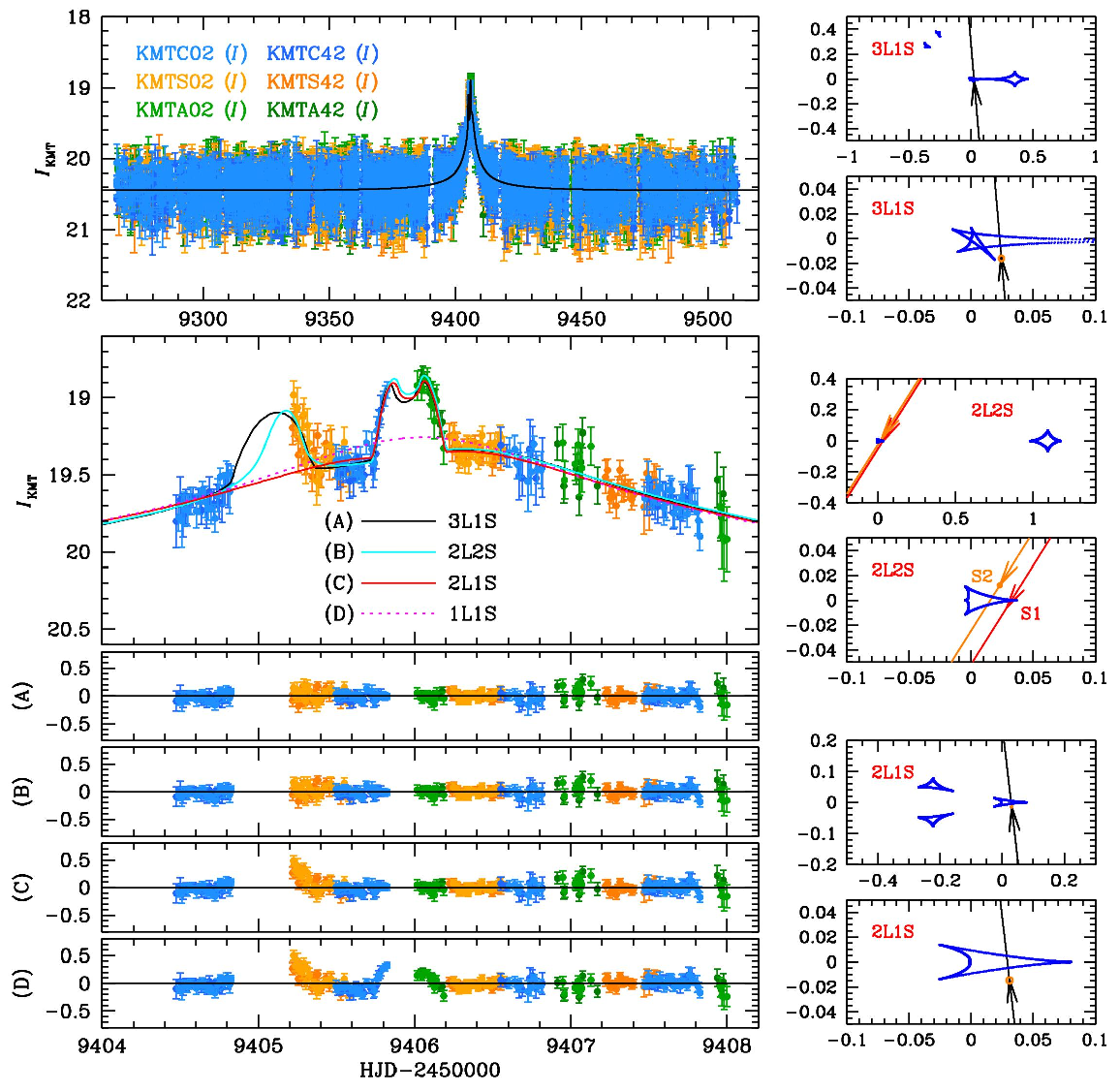}
\caption{Light curve of \sixteenninetyone\ \HL{(planetary event)} with 2L1S, 2L2S, 3L1S, and 1L1S models.  
\label{fig:lc_1691_01}}
\end{figure}
% --------------------------------------------------------------------------------------------------

Lastly, we test the APRX effect because the timescale is about $20$ days, which is longer than our criterion. We find that the APRX model shows improvements in $\chi^{2}$ by $13$. However, we find that the APRX effect is not reliable for three reasons. First, the amount of $\Delta\chi^{2}$ improvements is marginal to robustly claim the APRX effect. Second, we investigate the $\Delta\chi^{2}$ origin. We find inconsistencies in the $\chi^{2}$ improvements for different datasets. Indeed, for the KMTS data, $\chi^{2}$ improves on the rising wing of the light curve. By contrast, the fits to the falling wing are worse than those of the STD model. Overall, the net $\Delta\chi^{2}$ becomes negligible for this dataset. For the KMTA data, there is no improvement. These inconsistencies imply that the $\Delta\chi^{2}$ derives from fits to systematics in the KMTC data. Therefore, the APRX signal is highly doubtful. Third, the $\pivec$ distributions are not constrained, but rather exhibit a broad dispersion toward the $\pi_{E,E}$ direction, reaching the boundary of the values given in the code. These clues strongly support that the APRX effect is unreliable. Hence, we conclude that the 2L1S STD models are fiducial solutions for this event.

\subsection{\sixteenninetyone} % Planet : KMT-2021-BLG-1691(*)
\label{sec:KB211691}
The light curve of \sixteenninetyone\ exhibits clear and complex features at the peak from ${\rm HJD}^{\prime} = 9404.8$ to $9406.2$. As shown in Figure \ref{fig:lc_1691_01}, the anomalies consist of a half-covered bump and a well-covered caustic-crossing feature, which were captured by different KMTNet observations. Quantitatively, these anomalies yield $\Delta\chi^{2} = 317$ relative to a 1L1S model. We find that no 2L1S model can describe both features simultaneously. As shown in Figure \ref{fig:lc_1691_01}, the 2L1S model cannot explain the first bump-shaped anomaly. It can only describe the second bump, which it does with a central caustic-crossing geometry. In Table \ref{table:model_1691_01}, we present the 2L1S $s_{\pm}$ and 1L1S models for comparison. Although the 2L1S models cannot explain anomalies, we conduct the heuristic analysis for them. The result indicates that  $s_{-}^{\dagger} = 0.985$ and $s_{+}^{\dagger} = 1.016$ from $(\tau_{\rm anom}, u_{\rm anom}, t_{\rm anom}, t_{0}, t_{\rm E}, u_{0}) = (-0.012, 0.031, 9405.7, 9406.07, 32.0, 0.029)$. The $s_{+}^{\dagger}$ prediction is consistent with $s^{\dagger} = \sqrt{s_{-}s_{+}} = 1.016$.

% Table 7 (KB-21-1691: Model Parameters) ------------------------------------
\begin{longrotatetable}
\begin{deluxetable}{lrr|lrrrr}
\tablecaption{The parameters of 2L2S and 3L1S models for \sixteenninetyone\ \HL{(planetary event)} \label{table:model_1691_02}}
\tablewidth{0pt}
%\tabletypesize{\scriptsize}
\tablehead{
% ---------------------------------------------------------------------------
\multicolumn{3}{c}{2L2S} &
\multicolumn{5}{|c}{3L1S} \\
% ---------------------------------------------------------------------------
\multicolumn{1}{c}{Parameter} &
\multicolumn{1}{c}{$s_{-}$} & 
\multicolumn{1}{c}{$s_{+}$} & 
\multicolumn{1}{|c}{Parameter} &
\multicolumn{1}{c}{$s_{-,-}$} & 
\multicolumn{1}{c}{$s_{-,+}$} & 
\multicolumn{1}{c}{$s_{+,-}$} & 
\multicolumn{1}{c}{$s_{+,+}$} 
% ---------------------------------------------------------------------------
}
\startdata
% -----------------------------------------------------------------------------------
$\chi^{2} / {\rm N}_{\rm data}$           & $ 9565.543 / 9569     $     & $ 9565.475 / 9569     $     & $\chi^{2} / {\rm N}_{\rm data}$ & $ 9567.321 / 9569    $      & $ 9567.491 / 9569    $      & $ 9567.054 / 9569    $      & $ 9568.019 / 9569    $      \\
$\Delta\chi^{2}$                          & $ 0.068               $     & \nodata (best-fit)          & $\Delta\chi^{2}$                & $ 1.846              $      & $ 2.016              $      & $ 1.579              $      & $ 2.544              $      \\
$t_{0,{\rm S1}}$ [${\rm HJD'}$]           & $ 9405.758 \pm  0.094 $     & $ 9405.763 \pm  0.103 $     & $t_{0}$ [${\rm HJD'}$]          & $ 9406.085 \pm 0.030 $      & $ 9406.081 \pm 0.027 $      & $ 9406.116 \pm 0.025 $      & $ 9406.101 \pm 0.026 $      \\
$u_{0,{\rm S1}}$                          & $    0.026 \pm  0.006 $     & $    0.028 \pm  0.006 $     & $u_{0}$                         & $   -0.024 \pm 0.006 $      & $   -0.025 \pm 0.005 $      & $   -0.023 \pm 0.005 $      & $   -0.023 \pm 0.004 $      \\
$t_{\rm E}$ [days]                        & $   40.642 \pm  5.660 $     & $   39.051 \pm  6.366 $     & $t_{\rm E}$ [days]              & $   39.391 \pm 8.441 $      & $   38.057 \pm 6.235 $      & $   37.880 \pm 5.060 $      & $   38.093 \pm 5.678 $      \\
$s$                                       & $    0.635 \pm  0.058 $     & $    1.688 \pm  0.161 $     & $s_{1}$                         & $    0.810 \pm 0.041 $      & $    0.815 \pm 0.041 $      & $    1.192 \pm 0.037 $      & $    1.190 \pm 0.036 $      \\
$q$ ($\times10^{-4}$)                     & $  100.130 \pm 41.957 $     & $  122.221 \pm 33.121 $     & $q_{1}$ ($\times10^{-4}$)       & $   14.108 \pm 7.736 $      & $   14.725 \pm 7.490 $      & $   26.384 \pm 9.071 $      & $   25.803 \pm 8.004 $      \\
$\langle\log_{10} q\rangle$               & $   -1.926 \pm  0.148 $     & $   -2.107 \pm  0.169 $     & $\langle\log_{10} q_{1}\rangle$ & $   -2.794 \pm 0.196 $      & $   -2.816 \pm 0.203 $      & $   -2.508 \pm 0.120 $      & $   -2.555 \pm 0.119 $      \\
$\alpha$ [rad]                            & $    2.125 \pm  0.051 $     & $    2.133 \pm  0.072 $     & $\alpha$ [rad]                  & $    3.860 \pm 0.030 $      & $    3.857 \pm 0.031 $      & $    4.627 \pm 0.029 $      & $    4.640 \pm 0.036 $      \\
$t_{0,{\rm S2}}$ [${\rm HJD'}$]           & $ 9406.267 \pm  0.038 $     & $ 9406.254 \pm  0.043 $     & $s_{2}$                         & $    0.877 \pm 0.026 $      & $    1.185 \pm 0.037 $      & $    0.800 \pm 0.040 $      & $    1.285 \pm 0.062 $      \\
$u_{0,{\rm S2}}$                          & $    0.013 \pm  0.003 $     & $    0.013 \pm  0.004 $     & $q_{2}$ ($\times10^{-4}$)       & $   20.989 \pm 8.814 $      & $   24.962 \pm 8.228 $      & $   18.273 \pm 7.414 $      & $   17.841 \pm 8.777 $      \\
$\rho_{\ast, {\rm S1}}$                   & $  < 0.009            $     & $  < 0.007            $     & $\langle\log_{10} q_{2}\rangle$ & $   -2.569 \pm 0.134 $      & $   -2.550 \pm 0.122 $      & $   -2.812 \pm 0.201 $      & $   -2.807 \pm 0.218 $      \\
$\rho_{\ast, {\rm S2}}$ ($\times10^{-3}$) & $ 1.513_{-0.151}^{+1.262} $ & $ 1.600_{-0.041}^{+1.004} $ & $\psi$ [rad]                    & $    0.754 \pm 0.045 $      & $    0.768 \pm 0.043 $      & $    5.496 \pm 0.042 $      & $    5.492 \pm 0.049 $      \\
$q_{\rm flux}$                            & $ 0.576_{-0.036}^{+0.426} $ & $ 0.518_{-0.131}^{+1.209} $ & $\rho_{\ast}$ ($\times10^{-3}$) & $ 2.091_{-0.234}^{+1.414} $ & $ 2.212_{-0.148}^{+1.094} $ & $ 2.222_{-0.111}^{+1.510} $ & $ 2.222_{-0.130}^{+1.096} $ \\
$f_{\rm S, KMTC}$                         & $    0.005 \pm  0.001 $     & $    0.005 \pm  0.001 $     & $f_{\rm S, KMTC}$               & $    0.005 \pm 0.001 $      & $    0.006 \pm 0.001 $      & $    0.006 \pm 0.001 $      & $    0.006 \pm 0.001 $      \\
$f_{\rm B, KMTC}$                         & $    0.100 \pm  0.001 $     & $    0.100 \pm  0.001 $     & $f_{\rm B, KMTC}$               & $    0.100 \pm 0.001 $      & $    0.100 \pm 0.001 $      & $    0.100 \pm 0.001 $      & $    0.100 \pm 0.001 $      \\                        
% -----------------------------------------------------------------------------------
\enddata
\tablecomments{
${\rm HJD' \equiv HJD - 2450000.0}$. 
We note that the inequality sign of the $\rho_{\ast,{\rm S1}}$ parameters indicates 
the upper limit on $\rho_{\ast,{\rm S1}}$ (i.e., $3\sigma$).
}
%\tabletypesize{\small}
\end{deluxetable}
\end{longrotatetable}
% -----------------------------------------------------------------------------------

\begin{figure}[t]
\epsscale{1.00}
\plotone{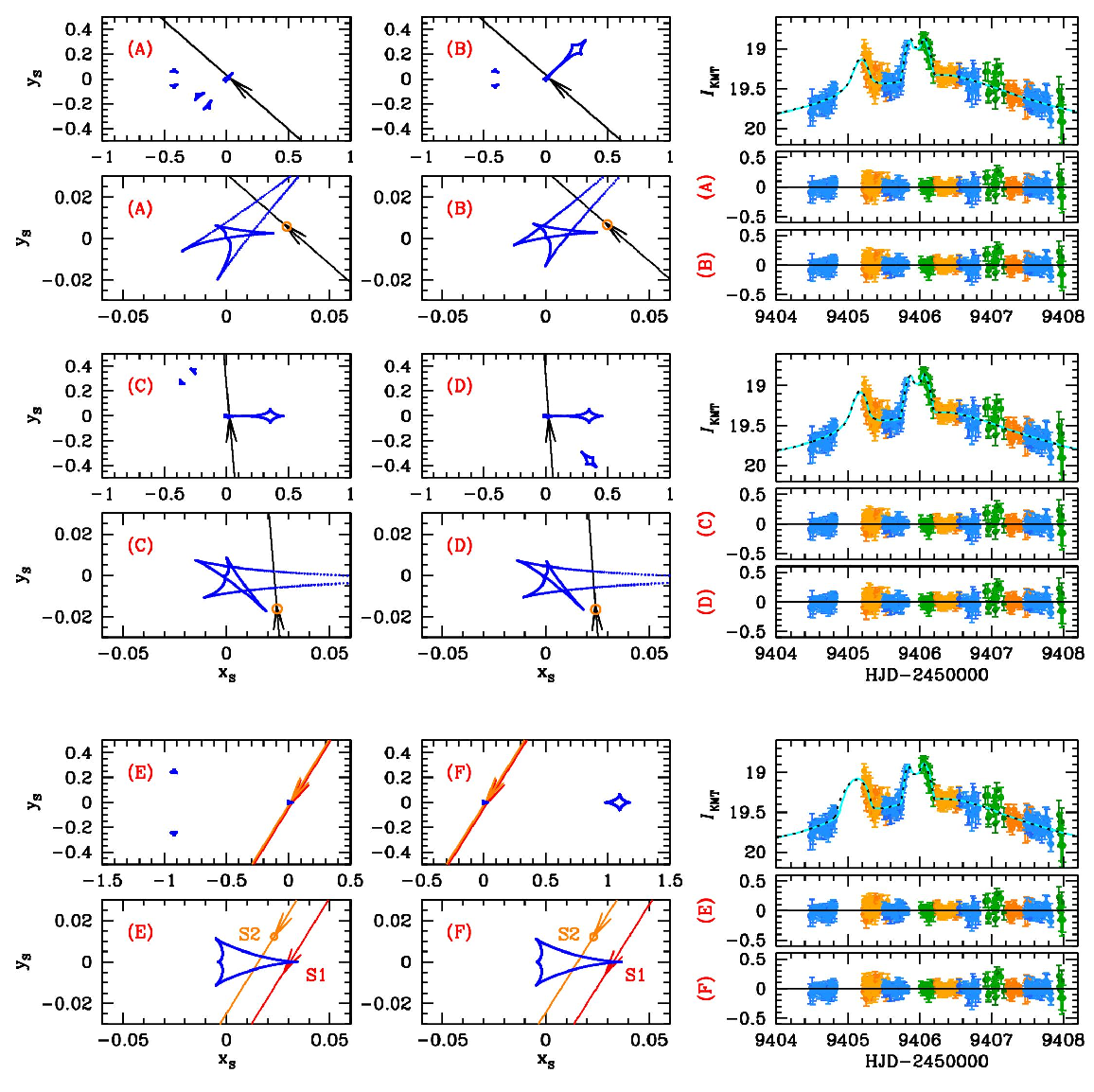}
\caption{Comparison for 2L2S and 3L1S models of \sixteenninetyone\ \HL{(planetary event)}.  
\label{fig:lc_1691_02}}
\end{figure}

Because the 2L1S model cannot explain the anomalies, we attempt 3L1S and 2L2S modeling. We find that both interpretations can describe all features of the anomalies as shown in Figure \ref{fig:lc_1691_01}. We also find that there are several competing 3L1S and 2L2S models, which are caused by the close/wide degeneracy. In Figure \ref{fig:lc_1691_02}, we present the geometries of these degenerate models with the light curve and their residuals over the anomaly, which demonstrate that these degeneracies cannot be resolved. Among these models, the 2L2S $s_{+}$ model is nominally the best-fit model. However, the $\Delta\chi^{2}$ values of the other models are less than $3$ compared with the best-fit case, which is \HL{insufficient} to resolve them. Although these models cannot be resolved, the mass ratios of all models are less than our criterion of planet detection, i.e., $q < 0.03$. Thus, all the 2L2S or 3L1S models indicate that at least one of the companions in their lens system is a planet. Hence, we treat this event as a planetary event, regardless of the number of planets in the system. In Table \ref{table:model_1691_02}, we present their model parameters. We note that the $\rho_{\ast}$ values related to the second anomaly can be measured (i.e., $\rho_{\ast,{\rm S2}}$ in the 2L2S models and $\rho_{\ast}$ in the 3L1S models). For the $\rho_{\ast, {\rm S1}}$ values in the 2L2S models, which are related to the first anomaly, we only obtain upper limits because this anomaly was only partially covered.

% Figure 9 (KB-21-1691) : planet -------------------------------------------------------------------
% --------------------------------------------------------------------------------------------------

Lastly, these STD models show longer timescales ($\sim 40$ days) than our criterion for testing the APRX effect. We find that there are negligible $\chi^{2}$ improvements ($\Delta\chi^{2} < 2$) in all cases. However, in contrast to the case of \sixninety, the $\pivec$ distributions are not converged for all cases. Hence, we conclude that 2L2S and 3L1S STD models are fiducial solutions for this planetary event.

\subsection{\twentytwothirteen} % Planet : KMT-2021-BLG-2213(*)
\label{sec:KB212213}

% Figure 10 (KB-21-2213) : planet -------------------------------------------------------------------
\begin{figure}[t]
\epsscale{1.00}
\plotone{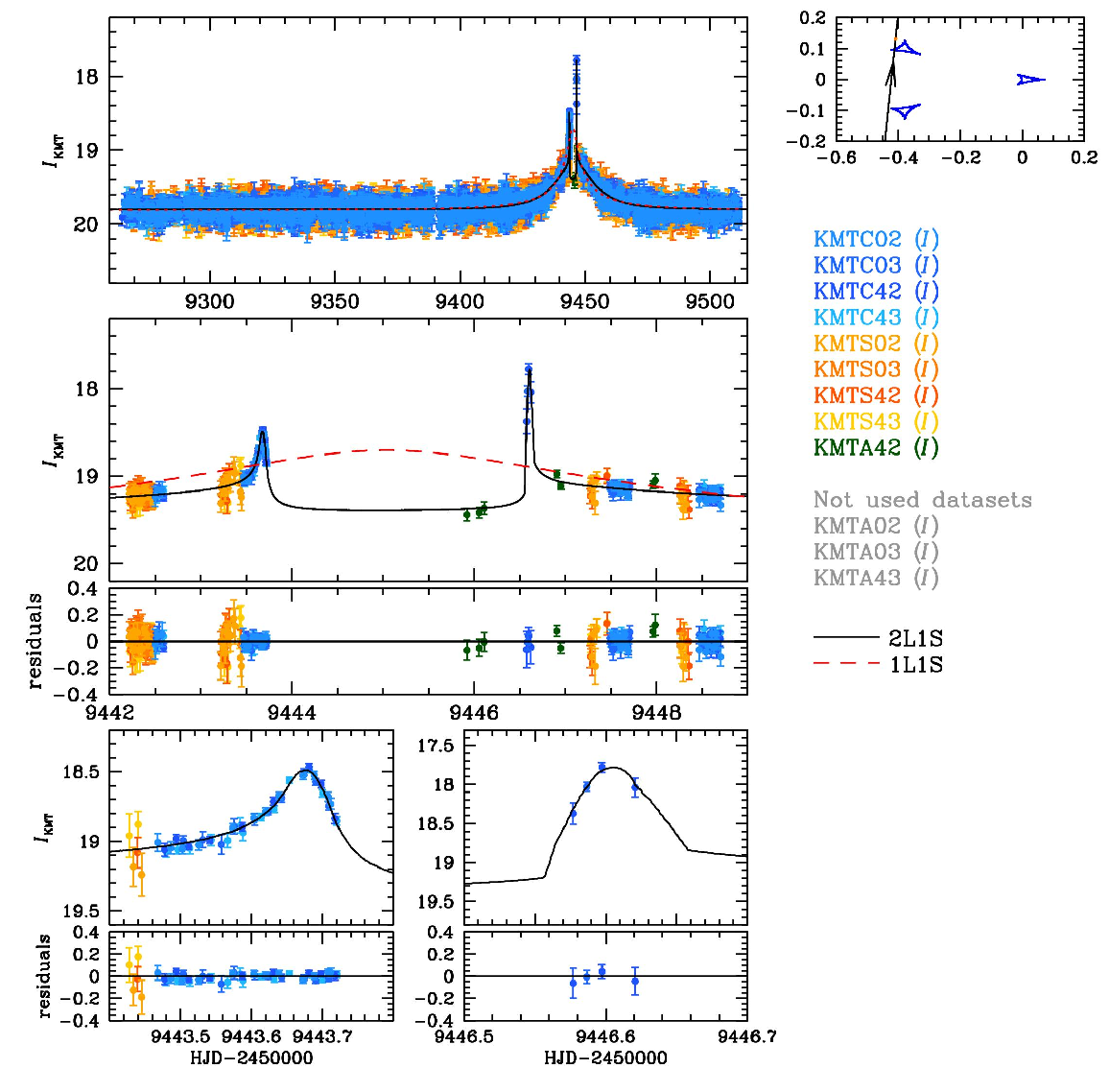}
\caption{Light curve of \twentytwothirteen\ \HL{(planetary event)} with 2L1S and 1L1S models.  
\label{fig:lc_2213}}
\end{figure}
% --------------------------------------------------------------------------------------------------

In Figure \ref{fig:lc_2213}, we present the observed light curve of \twentytwothirteen\ with the 2L1S model curve and caustic geometry. In addition, we present the 1L1S model curve for comparison, but it is disfavored by $\Delta\chi^{2} = 3205$. The light curve of \twentytwothirteen\ shows two bump anomalies. The first bump appeared at ${\rm HJD}^{\prime} \sim 9443.7$. After about 3 days, the second bump appeared at ${\rm HJD}^{\prime} \sim 9446.6$. Between these anomalies, the light curve exhibits a flat feature (so-called ``flat-trough" anomaly), which is caused the near annihilation of the minor image when the planet lies very close to the minor image of the unperturbed event. Although this flat-trough part was not covered by reliable KMTNet observations (we excluded KMTA02, KMTA03, and KMTA43 observations due to their extremely poor qualities), the bump-shaped anomalies were optimally covered by KMTC observations. These observed bump anomalies require a specific trajectory and angle that approaches the outer part of the planetary caustics to explain them (see the geometry in Figure \ref{fig:lc_2213}). 

As a result, the inner trajectory between the central and planetary caustics cannot explain the observed light curve of this event. Therefore, we can uniquely determine the solution for this event, which is presented in Table \ref{table:model_2213}. Indeed, the inner model cannot properly describe the second anomaly at ${\rm HJD^{\prime}}\sim9446.6$, which yields $\Delta\chi^{2} > 100$. This minor-image anomaly can be anticipated through the heuristic analysis. The analysis predicts that $s_{-}^{\dagger} = 0.811$ from $(\tau_{\rm anom}, u_{\rm anom}, t_{\rm anom}, t_{0}, t_{\rm E}, u_{0}) = (0.052, 0.422, 9446.6, 9445.8, 15.4, 0.419)$, which is similar to $s = 0.826 \pm 0.012$. Additionally, $\Delta t_{\rm dip} \simeq 3.5$ day, thus, the heuristic analysis predicts $q \simeq 47\times 10^{-4}$, which is consistent with the empirical result $q = (45.713 \pm 2.151)\times10^{-4}$.

% Table 8 (KB-21-2213: Model Parameters) ------------------------------------
\begin{deluxetable}{lr|lr}
\tablecaption{The parameters of the best-fit model for \twentytwothirteen\ \HL{(planetary event)} \label{table:model_2213}}
\tablewidth{0pt}
\tabletypesize{\scriptsize}
\tablehead{
% ---------------------------------------------------------------------------
\multicolumn{1}{c}{Parameter} &
\multicolumn{1}{c}{2L1S (outer)} &  
\multicolumn{1}{|c}{Parameter} &
\multicolumn{1}{c}{1L1S}  
% ---------------------------------------------------------------------------
}
\startdata
% -----------------------------------------------------------------------------------
$\chi^{2} / {\rm N}_{\rm data}$  & $ 11723.541 / 11725  $ & $\chi^{2} / {\rm N}_{\rm data}$ & $ 14928.785 / 11725  $ \\  
$\Delta\chi^{2}$                 & \nodata (best-fit)     & $\Delta\chi^{2}$                & $ 3205.244           $ \\ 
$t_0$ [${\rm HJD'}$]             & $ 9445.807 \pm 0.037 $ & $t_0$ [${\rm HJD'}$]            & $ 9445.041 \pm 0.018 $ \\  
$u_0$                            & $    0.419 \pm 0.025 $ & $u_0$                           & $    0.040 \pm 0.003 $ \\  
$t_{\rm E}$ [days]               & $   15.403 \pm 0.586 $ & $t_{\rm E}$ [days]              & $   44.638 \pm 2.392 $ \\  
$s$                              & $    0.826 \pm 0.012 $ & \nodata                         & \nodata                \\  
$q$ ($\times 10^{-4}$)           & $   45.603 \pm 0.889 $ & \nodata                         & \nodata                \\  
$\langle\log_{10} q\rangle$      & $   -2.343 \pm 0.009 $ & \nodata                         & \nodata                \\  
$\alpha$ [rad]                   & $    4.817 \pm 0.006 $ & \nodata                         & \nodata                \\  
$\rho_{\ast}$ ($\times 10^{-4}$) & $   20.713 \pm 2.151 $ & \nodata                         & \nodata                \\  
$f_{\rm S, KMTC}$                & $    0.096 \pm 0.008 $ & $f_{\rm S, KMTC}$               & $    0.014 \pm 0.001 $ \\  
$f_{\rm B, KMTC}$                & $    0.095 \pm 0.008 $ & $f_{\rm B, KMTC}$               & $    0.175 \pm 0.001 $ \\  
% -----------------------------------------------------------------------------------
\enddata
\tablecomments{
${\rm HJD' \equiv HJD - 2450000.0}$. 
}
\tabletypesize{\small}
\end{deluxetable}
% -----------------------------------------------------------------------------------

We note that the $\rho_{\ast}$ value can be robustly measured because the bump anomalies produced by cusp-crossing planetary caustics were optimally covered by KMTC. Additionally, we examine the APRX effect, although the timescale ($t_{\rm E} \sim 15.4$ days) is on the border of our criterion for this test. As expected, the $\pivec$ distributions show broad dispersions without $\chi^{2}$ improvement, which cannot provide meaningful constraints. Hence, we conclude that the 2L1S STD outer model is a fiducial solution of this planetary event without any degeneracy.

\subsection{\thirtytwoninety} % Planet : KMT-2021-BLG-3290(*)
\label{sec:KB213290}

% Figure 11 (KB-21-3290) : planet -------------------------------------------------------------------
\begin{figure}[t]
\epsscale{1.00}
\plotone{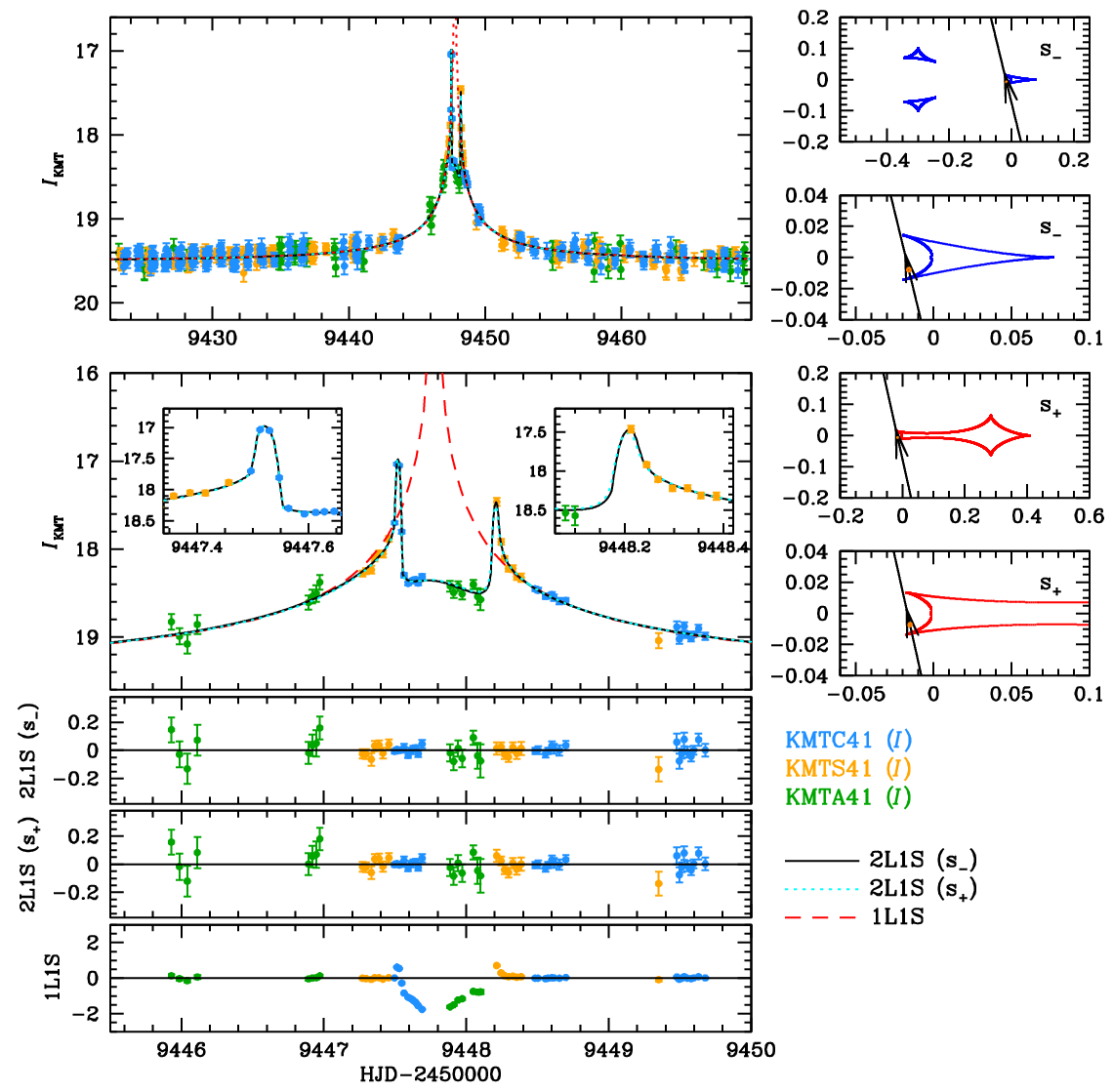}
\caption{Light curve of \thirtytwoninety\ \HL{(planetary event)} with 2L1S and 1L1S models.  
\label{fig:lc_3290}}
\end{figure}
% --------------------------------------------------------------------------------------------------

% Table 9 (KB-21-3290: Model Parameters) ------------------------------------
\begin{deluxetable}{lrr|lr}
\tablecaption{The parameters of 2L1S and 1L1S models for \thirtytwoninety\ \HL{(planetary event)} \label{table:model_3290}}
\tablewidth{0pt}
\tablehead{
% ---------------------------------------------------------------------------
\multicolumn{3}{c}{2L1S} &
\multicolumn{2}{|c}{1L1S} \\
\multicolumn{1}{c}{Parameter} &
\multicolumn{1}{c}{$s_{-}$} & 
\multicolumn{1}{c}{$s_{+}$} & 
\multicolumn{1}{|c}{Parameter} &
\multicolumn{1}{c}{}
% ---------------------------------------------------------------------------
}
\startdata
% -----------------------------------------------------------------------------------
$\chi^{2} / {\rm N}_{\rm data}$  & $\mathbf{ 2387.566 / 2394    }$ & $\mathbf{ 2391.915 / 2394    }$ & $\chi^{2} / {\rm N}_{\rm data}$ & $ 112927.460 / 2394   $ \\
$\Delta\chi^{2}$                 &      \bf{\nodata (best-fit)  }  & $\mathbf{ 4.349              }$ & $\Delta\chi^{2}$                & $ 110539.894          $ \\
$t_0$ [${\rm HJD'}$]             & $\mathbf{ 9447.741 \pm 0.010 }$ & $\mathbf{ 9447.753 \pm 0.009 }$ & $t_0$ [${\rm HJD'}$]            & $ 9447.776 \pm  0.010 $ \\
$u_0$                            & $\mathbf{    0.017 \pm 0.002 }$ & $\mathbf{    0.016 \pm 0.002 }$ & $u_0$ ($\times 10^{-4}$)        & $    1.148 \pm 16.585 $ \\
$t_{\rm E}$ [days]               & $\mathbf{   23.595 \pm 1.850 }$ & $\mathbf{   24.691 \pm 2.317 }$ & $t_{\rm E}$ [days]              & $   29.498 \pm  3.766 $ \\
$s$                              & $\mathbf{    0.860 \pm 0.004 }$ & $\mathbf{    1.152 \pm 0.006 }$ & \nodata                         & \nodata                 \\
$q$ ($\times 10^{-4}$)           & $\mathbf{   32.770 \pm 2.458 }$ & $\mathbf{   32.693 \pm 2.874 }$ & \nodata                         & \nodata                 \\
$\langle\log_{10} q\rangle$      & $\mathbf{   -2.484 \pm 0.033 }$ & $\mathbf{   -2.483 \pm 0.038 }$ & \nodata                         & \nodata                 \\
$\alpha$ [rad]                   & $\mathbf{    4.475 \pm 0.019 }$ & $\mathbf{    4.494 \pm 0.018 }$ & \nodata                         & \nodata                 \\
$\rho_{\ast}$ ($\times 10^{-4}$) & $\mathbf{   10.401 \pm 0.826 }$ & $\mathbf{    9.907 \pm 0.916 }$ & \nodata                         & \nodata                 \\
$f_{\rm S, KMTC}$                & $\mathbf{    0.013 \pm 0.001 }$ & $\mathbf{    0.013 \pm 0.001 }$ & $f_{\rm S, KMTC}$               & $    0.010 \pm  0.001 $ \\
$f_{\rm B, KMTC}$                & $\mathbf{    0.238 \pm 0.001 }$ & $\mathbf{    0.239 \pm 0.001 }$ & $f_{\rm B, KMTC}$               & $    0.241 \pm  0.001 $ \\
% -----------------------------------------------------------------------------------
\enddata
\tablecomments{
${\rm HJD' \equiv HJD - 2450000.0}$. 
The boldface indicates our fiducial solutions for this event.
}
\end{deluxetable}
% -----------------------------------------------------------------------------------

As shown in Figure \ref{fig:lc_3290}, the light curve of \thirtytwoninety\ exhibits the flat-trough anomaly similar to \twentytwothirteen, which yields $\Delta\chi^{2} > 10^{5}$ relative to the 1L1S model. We find that the 2L1S models can describe the anomaly on the light curve. This anomaly was produced by a pair cusp crossings that are separated by the broad demagnified region behind the central caustic, as shown in the geometries of Figure \ref{fig:lc_3290}. Although the coverage of the cusp crossings is sparse, we can measure $\rho_{\ast}$ robustly because both crossings (see the zoom-ins of Figure \ref{fig:lc_3290}) are adequately covered. These 2L1S models indicate that the event was caused by a planetary lens system, i.e., $q \simeq 3.3\times10^{-3}$. In Table \ref{table:model_3290}, we present the parameters of 2L1S models with the 1L1S model parameters for comparison. We find that the best-fit model is the $s_{-}$ case, which cannot be reliably distinguished from the $s_{+}$ case (i.e., $\Delta\chi^{2} = 4.35$). Indeed, the heuristic analysis indicates that $s_{-}^{\dagger} = 0.991$ from $(\tau_{\rm anom}, u_{\rm anom}, t_{\rm anom}, t_{0}, t_{\rm E}, u_{0}) = (-0.008, 0.019, 9447.5, 9447.7, 24.0, 0.17)$. The $s_{-}^{\dagger}$ prediction is similar to $s^{\dagger} = \sqrt{s_{-}s_{+}} = 0.995$. Additionally, $\Delta t_{\rm dip} \simeq 0.7$ days, thus, the heuristic analysis predicts $q \simeq 1.3\times 10^{-3}$, which is qualitatively similar to the empirical $q \simeq 3.3\times10^{-3}$. The reason for the relatively large difference between the prediction and empirical result is easily explained by inspection of Figure \ref{fig:lc_3290}. The formula for $q$ \citep{hwang22, ryu22, shin23a} approximates that $\Delta t_{\rm dip}$ is the time required to travel between the two planetary caustics, whereas the interval between the two cusps at the back end of the central caustic is much shorter. We check the APRX effect because of the relatively long timescale (i.e., $t_{\rm E} \sim 24$ days). We find that the $\pivec$ distributions show random scatters in the ranges set by the parameter limits (i.e., $-10 < \pivec < 10$) without $\chi^{2}$ improvement, which is meaningless and cannot be used for constraints. Hence, we conclude that the 2L1S STD models ($s_{\pm}$) are fiducial solutions for this event.

\newpage

\subsection{\thirteeneightyfive} % Planet Candidate : KMT-2021-BLG-1385(*)
\label{sec:KB211385}

% Figure 12 (KB-21-1385) : planet candidate ---------------------------------------------------------
\begin{figure}[t]
\epsscale{1.00}
\plotone{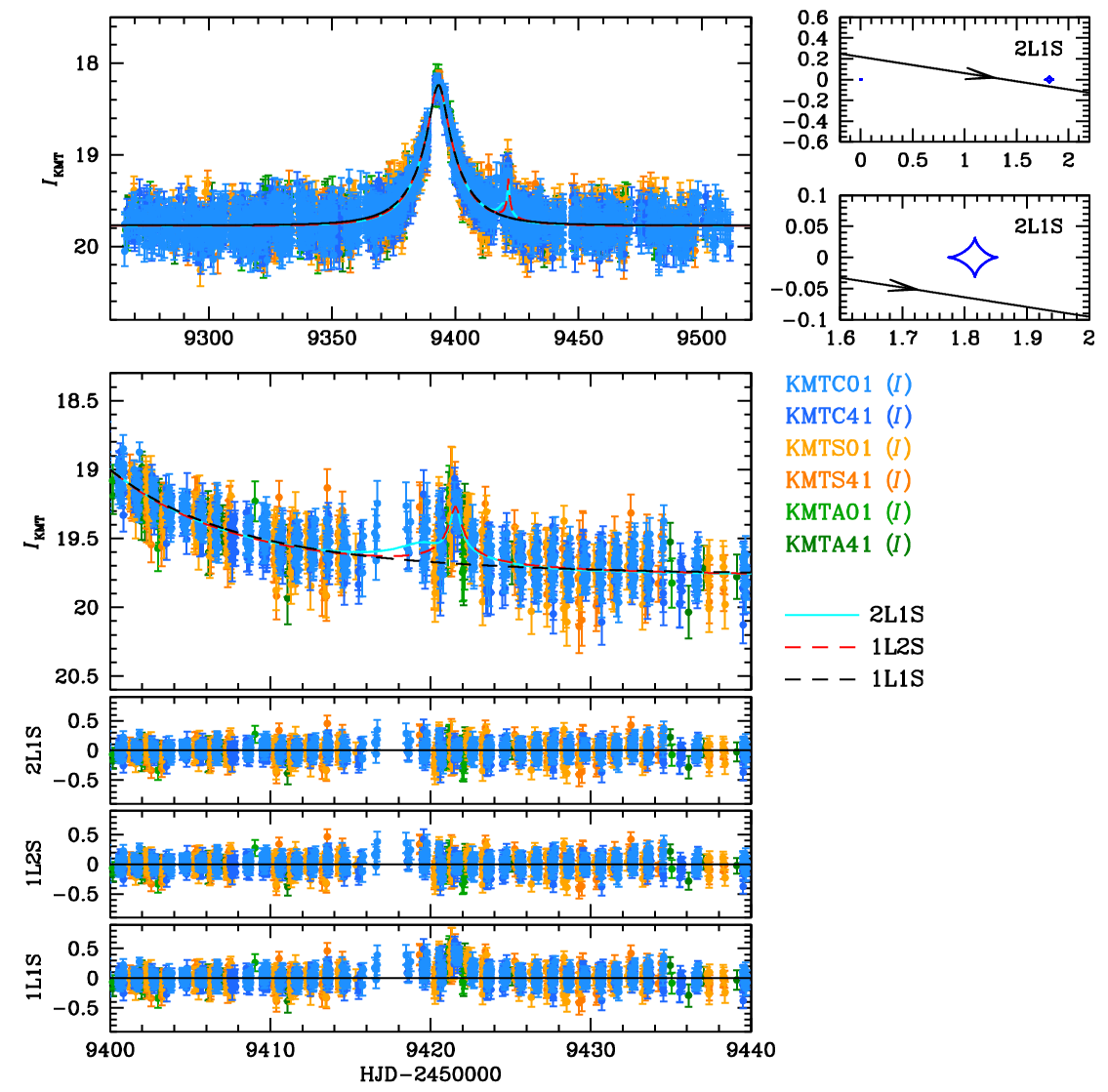}
\caption{Light curve of \thirteeneightyfive\ \HL{(planet candidate)} with 2L1S, 1L2S, and 1L1S models.  
\label{fig:lc_1385}}
\end{figure}
% --------------------------------------------------------------------------------------------------

% Figure 13 (KB-21-1751) : planet candidate ---------------------------------------------------------
\begin{figure}[t]
\epsscale{1.00}
\plotone{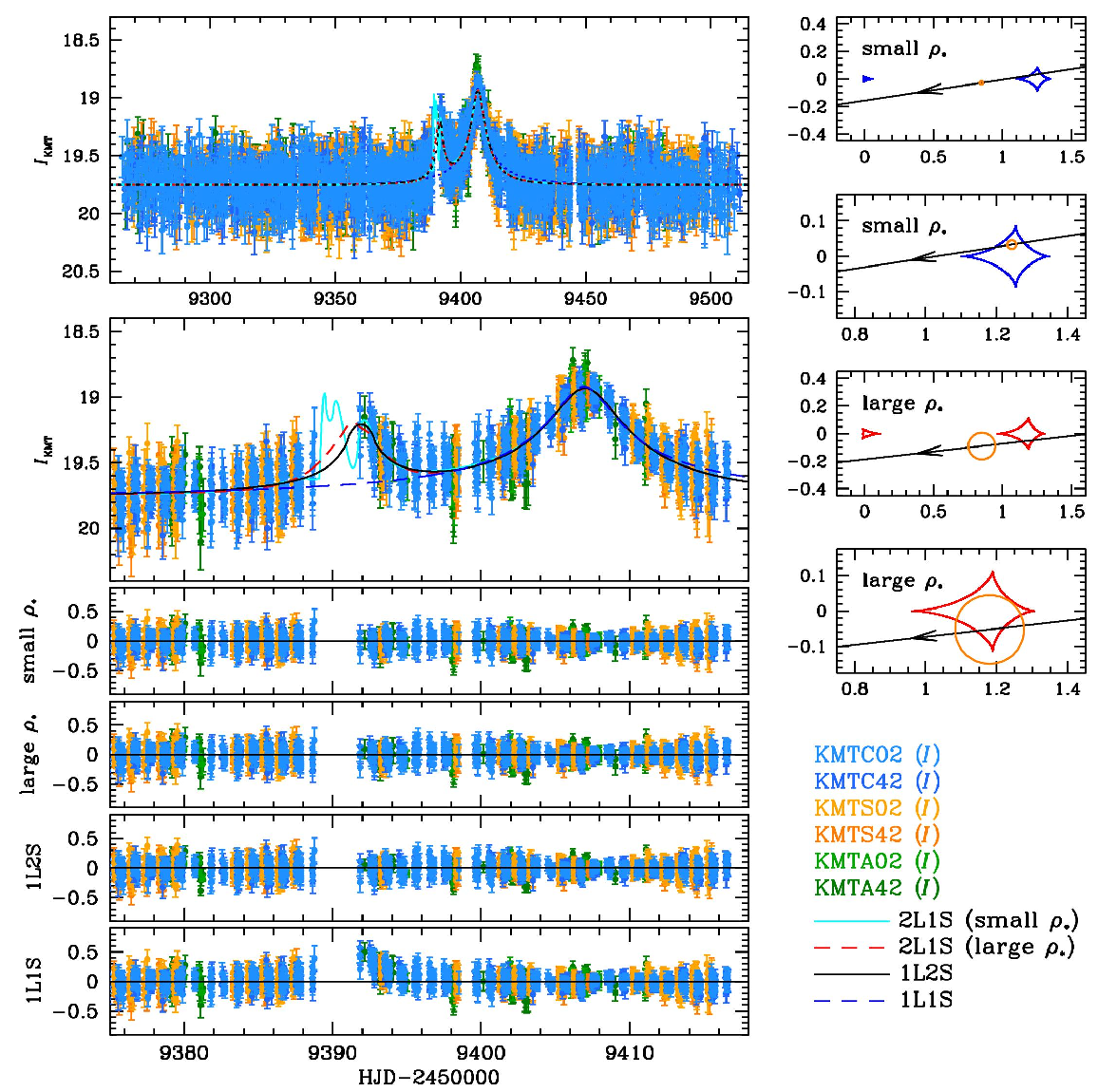}
\caption{Light curve of \seventeenfiftyone\ \HL{(planet candidate)} with 2L1S, 1L2S, and 1L1S models.  
\label{fig:lc_1751}}
\end{figure}
% --------------------------------------------------------------------------------------------------

Now, we move on planet candidate cases. The light curve of \thirteeneightyfive\ shows a 1L1S-like light curve with a bump-shaped anomaly on the falling wing at ${\rm HJD}^{\prime} \sim 9421.5$. This anomaly is noticeable by eye and quantitatively yields $\Delta\chi^{2} = 812$ relative to a 1L1S model. As shown in Figure \ref{fig:lc_1385}, we find that the anomaly can be described by both 2L1S and 1L2S interpretations. In Table \ref{table:model_1385}, we present the best-fit model parameters of both interpretations. Indeed, the heuristic analysis indicates that $s_{+}^{\dagger} = 2.258$ from $(\tau_{\rm anom}, u_{\rm anom}, t_{\rm anom}, t_{0}, t_{\rm E}, u_{0}) = (1.803, 1.815, 9421.5, 9393.2, 15.7, 0.216)$, which is consistent with $s_{+} = 2.259$. We find that the 2L1S model indicates that the lens system could have a planet companion (i.e., $q \sim 0.008 \pm 0.001$). However, the 2L1S model competes with the 1L2S model, which cannot be reliably distinguished from it ($\Delta\chi^{2} = 7$). Because the 2L1S/1L2S degeneracy cannot be resolved, we conclude that this event should be treated as a planet candidate.

% Table 10 (KB-21-1385: Model Parameters) ------------------------------------
\begin{deluxetable}{lr|lr|lr}
\tablecaption{The parameters of 2L1S, 1L2S, and 1L1S models for \thirteeneightyfive\ \HL{(planet candidate)} \label{table:model_1385}}
\tablewidth{0pt}
\tablehead{
% ---------------------------------------------------------------------------
\multicolumn{1}{c}{Parameter} &
\multicolumn{1}{c}{2L1S} & 
\multicolumn{1}{|c}{Parameter} &
\multicolumn{1}{c}{1L2S} & 
\multicolumn{1}{|c}{Parameter} &
\multicolumn{1}{c}{1L1S}
% ---------------------------------------------------------------------------
}
\startdata
% -----------------------------------------------------------------------------------
$\chi^{2} / {\rm N}_{\rm data}$  & $ 5406.439 / 5408     $ & $\chi^{2} / {\rm N}_{\rm data}$ & $ 5413.177 / 5408    $ & $\chi^{2} / {\rm N}_{\rm data}$ & $ 6218.363 / 5408    $ \\
$\Delta\chi^{2}$                 & \nodata (best-fit)      & $\Delta\chi^{2}$                & $ 6.738              $ & $\Delta\chi^{2}$                & $ 811.924            $ \\
$t_0$ [${\rm HJD'}$]             & $ 9393.169 \pm  0.025 $ & $t_{0,S1}$ [${\rm HJD'}$]       & $ 9393.157 \pm 0.025 $ & $t_0$ [${\rm HJD'}$]            & $ 9393.198 \pm 0.025 $ \\
$u_0$                            & $    0.216 \pm  0.008 $ & $u_{0,S1}$                      & $    0.204 \pm 0.016 $ & $u_0$                           & $    0.165 \pm 0.008 $ \\
$t_{\rm E}$ [days]               & $   15.736 \pm  0.404 $ & $t_{\rm E}$ [days]              & $   16.788 \pm 0.560 $ & $t_{\rm E}$ [days]              & $   19.205 \pm 0.667 $ \\
$s$                              & $    2.259 \pm  0.038 $ & $t_{0,S2}$ [${\rm HJD'}$]       & $ 9421.543 \pm 0.046 $ & \nodata                         & \nodata                \\
$q$ ($\times 10^{-4}$)           & $   80.864 \pm 12.915 $ & $u_{0,S2}$                      & $    0.027 \pm 0.004 $ & \nodata                         & \nodata                \\
$\langle\log_{10} q\rangle$      & $   -2.084 \pm  0.069 $ & $q_{\rm flux}$                  & $    0.015 \pm 0.001 $ & \nodata                         & \nodata                \\
$\alpha$ [rad]                   & $    0.155 \pm  0.003 $ & $\rho_{\ast,S1}$                & $  < 0.273           $ & \nodata                         & \nodata                \\
$\rho_{\ast}$                    & $  < 0.046            $ & $\rho_{\ast,S2}$                & $  < 0.049           $ & \nodata                         & \nodata                \\
$f_{\rm S, KMTC}$                & $    0.164 \pm  0.007 $ & $f_{\rm S, KMTC}$               & $    0.150 \pm 0.009 $ & $f_{\rm S, KMTC}$               & $    0.119 \pm 0.006 $ \\
$f_{\rm B, KMTC}$                & $    0.031 \pm  0.007 $ & $f_{\rm B, KMTC}$               & $    0.045 \pm 0.008 $ & $f_{\rm B, KMTC}$               & $    0.077 \pm 0.006 $ \\
% -----------------------------------------------------------------------------------
\enddata
\tablecomments{
${\rm HJD' \equiv HJD - 2450000.0}$. 
We note that the inequality sign of the $\rho_{\ast}$ parameters indicates 
the upper limit on $\rho_{\ast}$ (i.e., $3\sigma$).
}
\end{deluxetable}
% -----------------------------------------------------------------------------------

% Table 11 (KB-21-1751: Model Parameters) ------------------------------------
\begin{deluxetable}{lrr|lr|lr}
\tablecaption{The parameters of 2L1S, 1L2S, and 1L1S models for \seventeenfiftyone\ \HL{(planet candidate)} \label{table:model_1751}}
\tablewidth{0pt}
\tablehead{
% ---------------------------------------------------------------------------
\multicolumn{3}{c}{2L1S} &
\multicolumn{2}{|c}{1L2S} &
\multicolumn{2}{|c}{1L1S} \\
\multicolumn{1}{c}{Parameter} &
\multicolumn{1}{c}{small $\rho_{\ast}$} & 
\multicolumn{1}{c}{large $\rho_{\ast}$} & 
\multicolumn{1}{|c}{Parameter} &
\multicolumn{1}{c}{} & 
\multicolumn{1}{|c}{Parameter} &
\multicolumn{1}{c}{}
% ---------------------------------------------------------------------------
}
\startdata
% -----------------------------------------------------------------------------------
$\chi^{2} / {\rm N}_{\rm data}$  & $ 8893.731 / 8896    $ & $ 8896.634 / 8896    $ & $\chi^{2} / {\rm N}_{\rm data}$ & $ 8886.077 / 8896    $ & $\chi^{2} / {\rm N}_{\rm data}$ & $ 9759.601 / 8896    $ \\
$\Delta\chi^{2}$                 & $ 7.654              $ & $ 10.557             $ & $\Delta\chi^{2}$                & \nodata (best-fit)     & $\Delta\chi^{2}$                & $ 873.524            $ \\
$t_0$ [${\rm HJD'}$]             & $ 9406.782 \pm 0.031 $ & $ 9406.671 \pm 0.041 $ & $t_{0,S1}$ [${\rm HJD'}$]       & $ 9391.900 \pm 0.166 $ & $t_0$ [${\rm HJD'}$]            & $ 9406.922 \pm 0.024 $ \\
$u_0$                            & $    0.156 \pm 0.014 $ & $    0.186 \pm 0.013 $ & $u_{0,S1}$                      & $   -0.087 \pm 0.019 $ & $u_0$                           & $    0.062 \pm 0.006 $ \\
$t_{\rm E}$ [days]               & $   13.750 \pm 1.050 $ & $   12.743 \pm 0.607 $ & $t_{\rm E}$ [days]              & $   12.553 \pm 0.980 $ & $t_{\rm E}$ [days]              & $   31.020 \pm 2.851 $ \\
$s$                              & $    1.805 \pm 0.068 $ & $    1.754 \pm 0.051 $ & $t_{0,S2}$ [${\rm HJD'}$]       & $ 9407.015 \pm 0.025 $ & \nodata                         & \nodata                \\
$q$                              & $    0.027 \pm 0.005 $ & $    0.044 \pm 0.008 $ & $u_{0,S2}$                      & $    0.177 \pm 0.020 $ & \nodata                         & \nodata                \\
$\langle\log_{10} q\rangle$      & $   -1.616 \pm 0.083 $ & $   -1.340 \pm 0.072 $ & $q_{\rm flux}$                  & $    5.208 \pm 0.444 $ & \nodata                         & \nodata                \\
$\alpha$ [rad]                   & $    2.989 \pm 0.007 $ & $    3.028 \pm 0.007 $ & $\rho_{\ast,S1}$                & $  < 0.178           $ & \nodata                         & \nodata                \\
$\rho_{\ast}$                    & $    0.012 \pm 0.011^{\star} $ & $    0.097 \pm 0.017 $ & $\rho_{\ast,S2}$                & $  < 0.230           $ & \nodata                         & \nodata                \\
$f_{\rm S, KMTC}$                & $    0.042 \pm 0.004 $ & $    0.050 \pm 0.004 $ & $f_{\rm S, KMTC}$               & $    0.054 \pm 0.007 $ & $f_{\rm S, KMTC}$               & $    0.015 \pm 0.002 $ \\
$f_{\rm B, KMTC}$                & $    0.157 \pm 0.004 $ & $    0.150 \pm 0.004 $ & $f_{\rm B, KMTC}$               & $    0.145 \pm 0.007 $ & $f_{\rm B, KMTC}$               & $    0.183 \pm 0.001 $ \\
% -----------------------------------------------------------------------------------
\enddata
\tablecomments{
${\rm HJD' \equiv HJD - 2450000.0}$. 
We note that the inequality sign of the $\rho_{\ast}$ parameters indicates 
the upper limit on $\rho_{\ast}$ (i.e., $3\sigma$).
$^\star$For the small $\rho_{\ast}$ case, the $\rho_{\ast}$ distributions concatenates 
to those of the large $\rho_{\ast}$ case within $3\sigma$ level (i.e., $\Delta\chi^{2} < 3^{2}$). 
In addition, the small $\rho_{\ast}$ distributions reach zero value within $3\sigma$ level. 
Hence, only the $1\sigma$ error for the small $\rho_{\ast}$ case is meaningful. 
}
\end{deluxetable}
% -----------------------------------------------------------------------------------

% Table 12 (KB-21-1907: Model Parameters) ------------------------------------
%\begin{rotate}
\begin{longrotatetable}
\begin{deluxetable}{lrrrr|lr|lr}
\tablecaption{The parameters of 2L1S, 1L2S, and 1L1S models for \nineteenseven\ \HL{(planet candidate)} \label{table:model_1907}}
\tabletypesize{\scriptsize}
\tablewidth{0pt}
\tablehead{
% ---------------------------------------------------------------------------
\multicolumn{5}{c}{2L1S} &
\multicolumn{2}{|c}{1L2S} &
\multicolumn{2}{|c}{1L1S} \\
\multicolumn{1}{c}{Parameter} &
\multicolumn{1}{c}{Binary $s_{-}$} & 
\multicolumn{1}{c}{Binary $s_{+}$} & 
\multicolumn{1}{c}{Planet $s_{-}$} & 
\multicolumn{1}{c}{Planet $s_{+}$} & 
\multicolumn{1}{|c}{Parameter} &
\multicolumn{1}{c}{} & 
\multicolumn{1}{|c}{Parameter} &
\multicolumn{1}{c}{}
% ---------------------------------------------------------------------------
}
\startdata
% -----------------------------------------------------------------------------------
$\chi^{2} / {\rm N}_{\rm data}$ & $ 8732.139 / 8756    $ & $ 8735.275 / 8756    $ & $ 8733.909 / 8756    $ & $ 8734.275 / 8756    $ & $\chi^{2} / {\rm N}_{\rm data}$ & $ 8750.539 / 8756    $ & $\chi^{2} / {\rm N}_{\rm data}$ & $ 9017.892 / 8756    $ \\
$\Delta\chi^{2}$                & \nodata (best-fit)     & $ 3.136              $ & $ 1.770              $ & $ 2.136              $ & $\Delta\chi^{2}$                & $ 18.400             $ & $\Delta\chi^{2}$                & $ 285.751            $ \\
$t_0$ [${\rm HJD'}$]            & $ 9422.933 \pm 0.062 $ & $ 9422.480 \pm 0.042 $ & $ 9422.555 \pm 0.025 $ & $ 9422.539 \pm 0.024 $ & $t_{0,S1}$ [${\rm HJD'}$]       & $ 9421.492 \pm 0.052 $ & $t_0$ [${\rm HJD'}$]            & $ 9422.556 \pm 0.032 $ \\
$u_0$                           & $    0.199 \pm 0.019 $ & $    0.104 \pm 0.019 $ & $    0.180 \pm 0.026 $ & $    0.183 \pm 0.024 $ & $u_{0,S1}$                      & $   -0.084 \pm 0.014 $ & $u_0$                           & $    0.669 \pm 0.116 $ \\
$t_{\rm E}$ [days]              & $    6.856 \pm 0.577 $ & $   13.624 \pm 2.896 $ & $    7.175 \pm 0.600 $ & $    7.052 \pm 0.605 $ & $t_{\rm E}$ [days]              & $    9.268 \pm 0.838 $ & $t_{\rm E}$ [days]              & $    3.463 \pm 0.378 $ \\
$s$                             & $    0.534 \pm 0.036 $ & $    3.319 \pm 0.332 $ & $    0.690 \pm 0.022 $ & $    1.231 \pm 0.042 $ & $t_{0,S2}$ [${\rm HJD'}$]       & $ 9423.488 \pm 0.050 $ & \nodata                         & \nodata                \\
$q$                             & $    0.417 \pm 0.109 $ & $    0.457 \pm 0.194 $ & $    0.025 \pm 0.003 $ & $    0.030 \pm 0.004 $ & $u_{0,S2}$                      & $    0.116 \pm 0.019 $ & \nodata                         & \nodata                \\
$\langle\log_{10} q\rangle$     & $   -0.316 \pm 0.086 $ & $   -0.372 \pm 0.195 $ & $   -1.616 \pm 0.065 $ & $   -1.565 \pm 0.062 $ & $q_{\rm flux}$                  & $    1.542 \pm 0.410 $ & \nodata                         & \nodata                \\
$\alpha$ [rad]                  & $    4.233 \pm 0.063 $ & $    3.986 \pm 0.035 $ & $    4.749 \pm 0.023 $ & $    4.740 \pm 0.020 $ & $\rho_{\ast,S1}$                & $  < 0.135           $ & \nodata                         & \nodata                \\
$\rho_{\ast}$                   & $  < 0.122           $ & $  < 0.081           $ & $  < 0.090           $ & $  < 0.094           $ & $\rho_{\ast,S2}$                & $  < 0.176           $ & \nodata                         & \nodata                \\
$f_{\rm S, KMTC}$               & $    0.047 \pm 0.005 $ & $    0.042 \pm 0.004 $ & $    0.044 \pm 0.007 $ & $    0.046 \pm 0.007 $ & $f_{\rm S, KMTC}$               & $    0.024 \pm 0.003 $ & $f_{\rm S, KMTC}$               & $    0.245 \pm 0.080 $ \\
$f_{\rm B, KMTC}$               & $    0.148 \pm 0.005 $ & $    0.154 \pm 0.004 $ & $    0.151 \pm 0.007 $ & $    0.150 \pm 0.007 $ & $f_{\rm B, KMTC}$               & $    0.171 \pm 0.003 $ & $f_{\rm B, KMTC}$               & $   -0.049 \pm 0.080 $ \\
% -----------------------------------------------------------------------------------
\enddata
\tablecomments{
${\rm HJD' \equiv HJD - 2450000.0}$. 
We note that the inequality sign of the $\rho_{\ast}$ parameters indicates 
the upper limit on $\rho_{\ast}$ (i.e., $3\sigma$).
}
\tabletypesize{\small}
\end{deluxetable}
\end{longrotatetable}
%\end{rotate}
% -----------------------------------------------------------------------------------

\subsection{\seventeenfiftyone} % Planet Candidate : KMT-2021-BLG-1751(*)
\label{sec:KB211751}

As shown in Figure \ref{fig:lc_1751}, the observed light curve of \seventeenfiftyone\ exhibits a half-covered bump-shaped anomaly at ${\rm HJD}^{\prime} \sim 9392.0$. We also present the model light curves of 2L1S, 1L2S, and 1L1S models with their residuals for comparison. We find that the anomaly can be explained by either 2L1S or 1L2S interpretations. We also find that the 2L1S interpretation can describe the anomaly by both small- and large-$\rho_{\ast}$ cases because the bump was not fully covered. We present their model parameters in Table \ref{table:model_1751}. Indeed, the heuristic analysis indicates that $s_{+}^{\dagger} = 1.757$ from $(\tau_{\rm anom}, u_{\rm anom}, t_{\rm anom}, t_{0}, t_{\rm E}, u_{0}) = (-1.173, 1.188, 9391.8, 9406.7, 12.7, 0.186)$, which is consistent with $s_{+} = 1.754$ of the large $\rho_{\ast}$ case.  

The small $\rho_{\ast}$ case indicates that the lens system could have a planet (i.e., $q\sim 0.027 < 0.03$). In contrast, the large $\rho_{\ast}$ case is unlikely to be a planetary event (i.e., $q\sim 0.044 > 0.03$). These cases (i.e., the planet/binary degeneracy) cannot be resolved using $\chi^{2}$ (i.e., $\Delta\chi^{2} = 2.90$).  

Moreover, the best-fit model is the 1L2S case, which cannot be distinguished from the 2L1S cases. Quantitatively, $\Delta\chi^{2}$ between them is less than $11$, which does not satisfy our criterion for resolving the 2L1S/1L2S degeneracy (i.e., $\Delta\chi^{2} > 15.0$). Thus, we conclude that this event should be treated as a planet candidate.

\subsection{\nineteenseven} % Planet Candidate : KMT-2021-BLG-1907(*)
\label{sec:KB211907}

\begin{figure}[t]
\epsscale{1.00}
\plotone{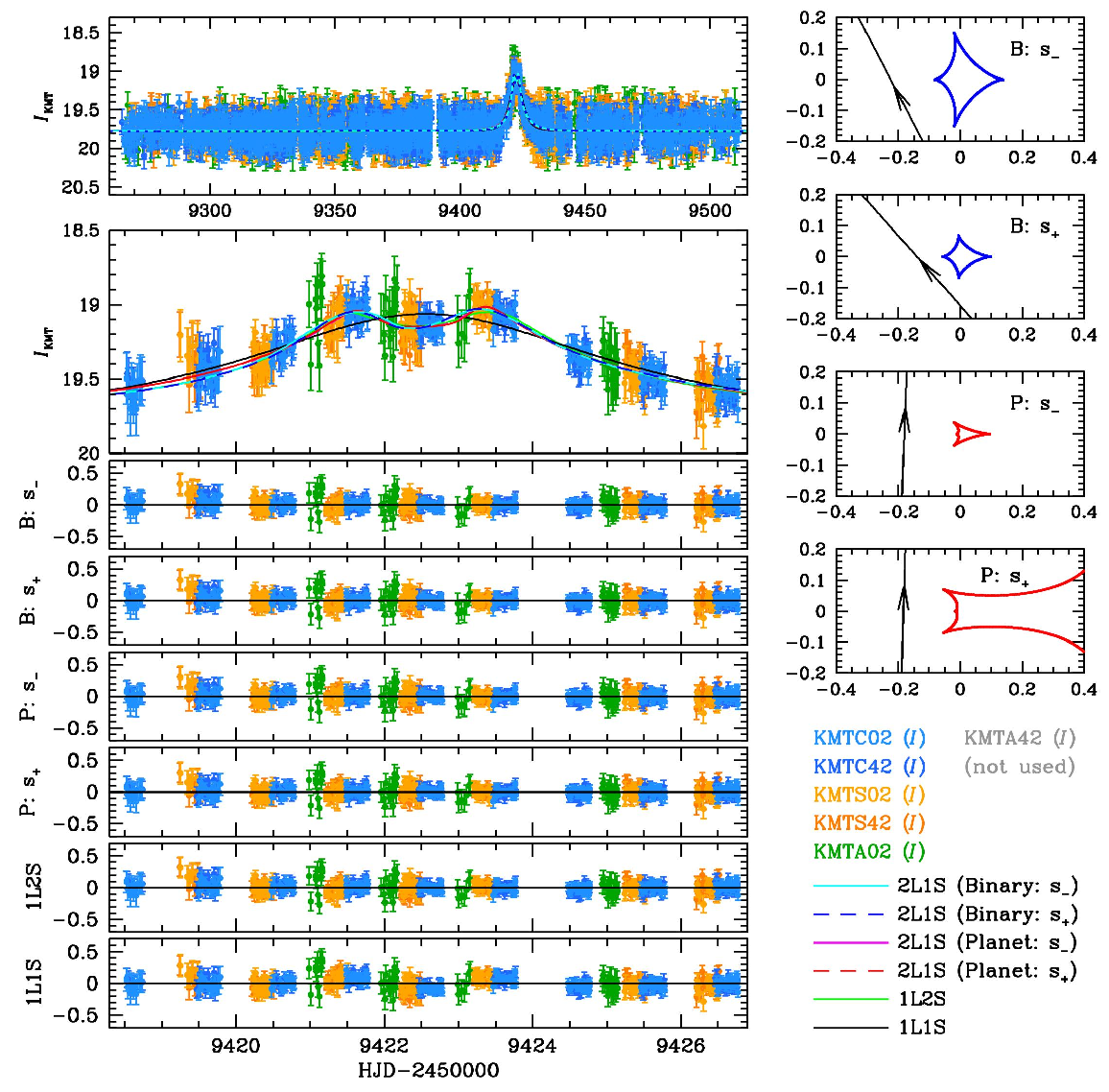}
\caption{Light curve of \nineteenseven\ \HL{(planet candidate)} with 2L1S, 1L2S, and 1L1S models.  
\label{fig:lc_1907}}
\end{figure}

In Figure \ref{fig:lc_1907}, we present the observed light curve of \nineteenseven\ with several 2L1S models, including 1L2S and 1L1S models for comparison. The light curve shows two shallow bump-shaped anomalies around the peak, which yield $\Delta\chi^{2} = 286$ relative to the 1L1S model that exhibit noticeable residuals (see 1L1S residuals in Figure \ref{fig:lc_1907}). We find that the anomalies can be well explained by both 2L1S and 1L2S interpretations. In Table \ref{table:model_1907}, we present their model parameters, together with the 1L1S case for comparison. Indeed, for the 2L1S planetary model pair, the heuristic analysis indicates that $s_{-}^{\dagger} = 0.914$ and $s_{+}^{\dagger} = 1.094$ from $(\tau_{\rm anom}, u_{\rm anom}, t_{\rm anom}, t_{0}, t_{\rm E}, u_{0}) = (-0.014, 0.181, 9422.4, 9422.5, 7.1, 0.18)$. The $s_{-}^{\dagger}$ prediction is similar to $s^{\dagger} = \sqrt{s_{-}s_{+}} = 0.922$. While, for the 2L1S binary model pair, the heuristic analysis indicates that $s_{-}^{\dagger} = 0.749$ and $s_{+}^{\dagger} = 1.335$ from $(\tau_{\rm anom}, u_{\rm anom}, t_{\rm anom}, t_{0}, t_{\rm E}, u_{0}) = (-0.551, 0.586, 9419.1, 9422.9, 6.9, 0.20)$. The $s_{+}^{\dagger}$ prediction is consistent with $s^{\dagger} = \sqrt{s_{-}s_{+}}= 1.331$. 

% Figure 14 (KB-21-1907) : planet candidate ---------------------------------------------------------
% --------------------------------------------------------------------------------------------------

Next, we check whether the bump anomaly can be explained by a 1L2S model, but we find that the 1L2S model has a worse fit compared to the best-fit 2L1S models by $\Delta\chi^{2} = 18$. Thus, we can nominally resolve the 2L1S/1L2S degeneracy considering our criterion (i.e., $\Delta\chi^{2} > 15.0$). Even though we can rule out the 1L2S model, we also find that there exists the planet/binary degeneracy in the 2L1S interpretation. There are four possible solutions resulting from the close/wide degeneracy in both binary and planetary cases. These solutions cannot be resolved by $\Delta\chi^{2}$ values, which are less than $3$. Therefore, because the nature of the lens system is unclear, we should treat this event as a planet candidate.

\section{CMD Analysis} \label{sec:CMDs}

\begin{figure*}[t]
\epsscale{1.00}
\plotone{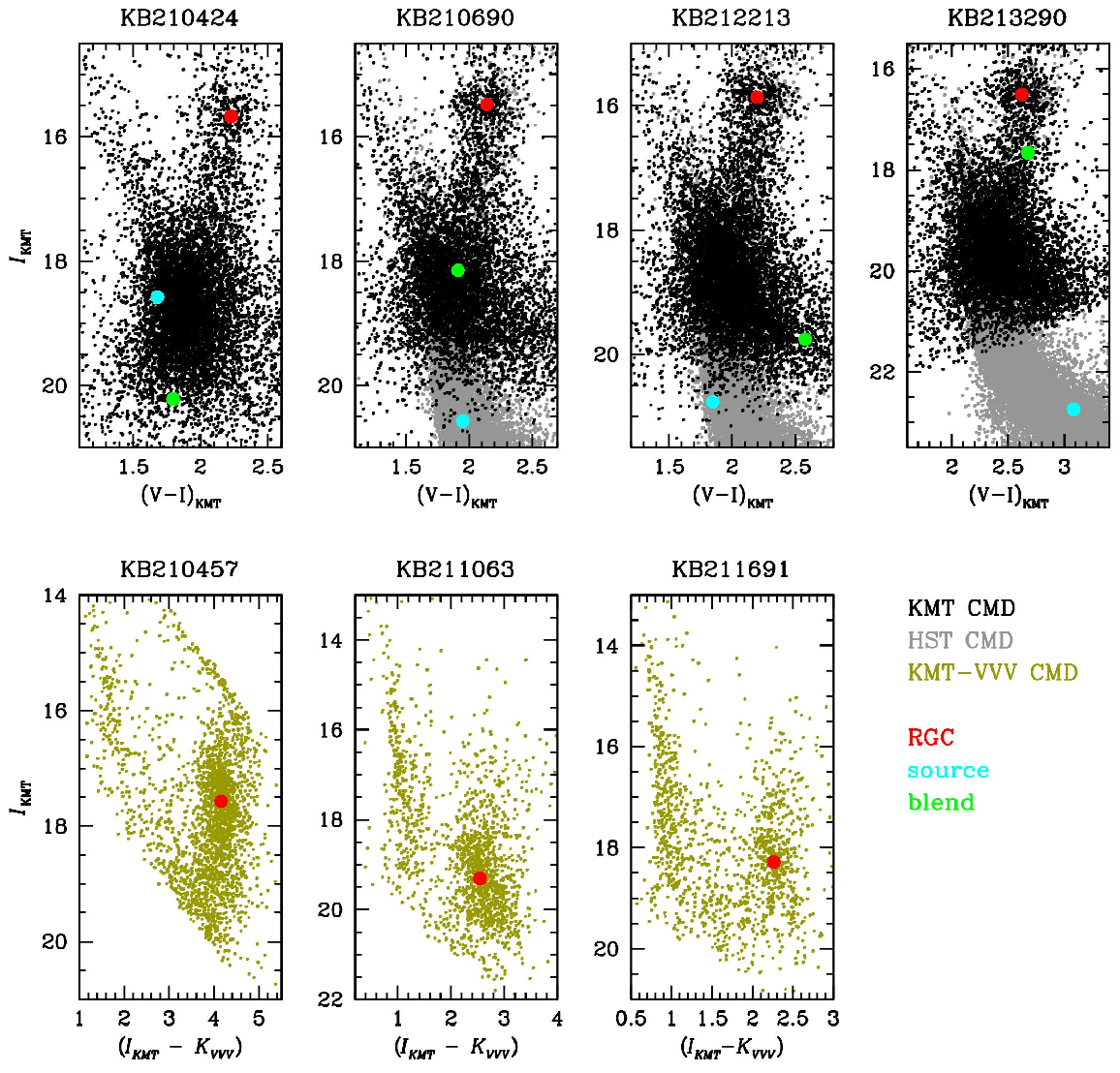}
\caption{Color-magnitude diagrams (CMDs) of planetary events. 
We use the abbreviation for event names, e.g., \fourtwentyfour\ is abbreviated as KB210424.
The HST CMD are presented for visual inspection of the faint sources and are not used for the source-color measurements.
For KB210457, KB211063, and KB211691, the CMDs are built by cross-matching with VVV and KMTNet.    
\label{fig:CMDs}}
\end{figure*}

Among the seven planetary events, we obtained reliable $V$-band observations for four events: \fourtwentyfour, \sixninety, \twentytwothirteen, and \thirtytwoninety. Thus, we can determine the angular source radius ($\theta_{\ast}$) of these based on their $V$-band light curves and KMTNet color-magnitude diagrams (CMD) using the conventional CMD analysis method described in \citet{yoo04}. In Table \ref{table:cmd}, we present the results of the CMD analyses for these four events with the Einstein angular radii ($\theta_{\rm E} \equiv \theta_{\ast} / \rho_{\ast}$) and lens-source relative proper motions ($\mu_{\rm rel}$) of each model case. If we can only measure $\rho_{\ast}$ upper limits, we present lower limits on $\theta_{\ast}$ and $\mu_{\rm rel}$. In addition, we present the KMTNet CMDs of these four events with the locations of the source, blend, and centroid of the red giant clump (RGC) in Figure \ref{fig:CMDs}.

% Figure 15 (CMDs) ----------------------------------------------------------------------------------
% --------------------------------------------------------------------------------------------------

In contrast, for \fourfiftyseven, \tensixtythree, and \sixteenninetyone, the $V$-band image qualities are too poor due to the heavy extinction. As a result, we cannot use the conventional method based on the KMTNet CMDs and the $V$-band light curves. Thus, we build cross-matched CMDs using KMTNet and VVV \citep[VISTA Variables and Via Lactea Survey;][]{minniti10} data. We determine the RGC locations on the KMTNet-VVV CMDs. In Figure \ref{fig:CMDs}, we present the KMTNet-VVV CMDs with the RGC locations. Using the $I$-band magnitudes of the RGC and the source, we estimate the mean source color on the HST CMD \citep{holtman98} using the method described in \citet{bennett08}. Then, we estimate the angular source radius using the color/surface-brightness relation \citep{kervella04}. In Table \ref{table:VVV}, we present the estimated source colors, the angular source radii, and angular Einstein ring radii (only for \sixteenninetyone; see Section \ref{sec:lens_KB211691}) for these events.

% Table 13 (CMD analysis) ----------------------------------------------------
\begin{longrotatetable}
\begin{deluxetable}{crrrrrrrr}
\tablecaption{CMD analyses of Planetary Events \label{table:cmd}}
\tablewidth{0pt}
%\tabletypesize{\scriptsize}
\tablehead{
% ---------------------------------------------------------------------------
\multicolumn{1}{c}{Event} & 
\multicolumn{1}{c}{} & 
\multicolumn{1}{c}{} & 
\multicolumn{1}{c}{} & 
\multicolumn{1}{c}{} & 
\multicolumn{1}{c}{} & 
\multicolumn{1}{c}{} &
\multicolumn{1}{c}{} &
\multicolumn{1}{c}{} \\
% ----------------------------------------------
\multicolumn{1}{c}{Case} &
\multicolumn{1}{r}{$(V-I)_{\rm RGC}$} & 
\multicolumn{1}{r}{$(V-I)_{0, \rm RGC}$} &
\multicolumn{1}{r}{$(V-I)_{\rm S}$} &
\multicolumn{1}{r}{$(V-I)_{0, \rm S}$} &
\multicolumn{1}{r}{$(V-I)_{\rm B}$} &
\multicolumn{1}{c}{$\theta_{\ast}$} &
\multicolumn{1}{c}{$\theta_{\rm E}$} &
\multicolumn{1}{c}{$\mu_{\rm rel}$} \\
% ----------------------------------------------
\multicolumn{1}{c}{} &
\multicolumn{1}{r}{$I_{\rm RGC}$} & 
\multicolumn{1}{r}{$I_{0, \rm RGC}$} &
\multicolumn{1}{r}{$I_{\rm S}$} &
\multicolumn{1}{r}{$I_{0, \rm S}$} &
\multicolumn{1}{r}{$I_{\rm B}$} &
\multicolumn{1}{c}{($\mu{\rm as}$)} &
\multicolumn{1}{c}{(mas)} &
\multicolumn{1}{c}{($\rm mas\, yr^{-1}$)}
% ------------------------------------
}
\startdata
% ------------------------------------------------------------------------------------------------------------------------------------------------------------
% KB-21-0424(*) == MB-21-111
KB210424           &            &            &                      &                      &                      &                     &                     &                     \\
APRX ($u_{0} > 0$) & $  2.225 $ & $  1.060 $ & $  1.648 \pm 0.008 $ & $  0.483 \pm 0.051 $ & $  1.718 \pm 0.092 $ & $ 0.882 \pm 0.041 $ & $ 0.702 \pm 0.075 $ & $ 4.718 \pm 0.504 $ \\
                   & $ 15.670 $ & $ 14.343 $ & $ 18.576 \pm 0.002 $ & $ 17.249 \pm 0.002 $ & $ 20.255 \pm 0.014 $ & \nodata             & \nodata             & \nodata             \\
APRX ($u_{0} < 0$) & $  2.225 $ & $  1.060 $ & $  1.648 \pm 0.008 $ & $  0.483 \pm 0.051 $ & $  1.720 \pm 0.094 $ & $ 0.886 \pm 0.041 $ & $ 0.676 \pm 0.071 $ & $ 4.611 \pm 0.486 $ \\
                   & $ 15.670 $ & $ 14.343 $ & $ 18.567 \pm 0.002 $ & $ 17.240 \pm 0.002 $ & $ 20.278 \pm 0.014 $ & \nodata             & \nodata             & \nodata             \\
\hline
% ------------------------------------------------------------------------------------------------------------------------------------------------------------
% KB-21-0690(*) == MB-21-161
KB210690                    &            &            &                      &                      &                      &                     &                     &                     \\
APRX ($u_{0} > 0$)          & $  2.142 $ & $  1.060 $ & $  1.950 \pm 0.010 $ & $  0.868 \pm 0.051 $ & $  1.915 \pm 0.010 $ & $ 0.469 \pm 0.027 $ & $ 0.036 \pm 0.003 $ & $ 0.106 \pm 0.010 $ \\
                            & $ 15.482 $ & $ 14.385 $ & $ 20.615 \pm 0.001 $ & $ 19.518 \pm 0.001 $ & $ 18.145 \pm 0.001 $ & \nodata             & \nodata             & \nodata             \\
APRX ($u_{0} < 0$)          & $  2.142 $ & $  1.060 $ & $  1.950 \pm 0.010 $ & $  0.868 \pm 0.051 $ & $  1.915 \pm 0.010 $ & $ 0.470 \pm 0.027 $ & $ 0.036 \pm 0.003 $ & $ 0.106 \pm 0.010 $ \\
                            & $ 15.482 $ & $ 14.385 $ & $ 20.612 \pm 0.001 $ & $ 19.515 \pm 0.001 $ & $ 18.146 \pm 0.001 $ & \nodata             & \nodata             & \nodata             \\
XRP                         & $  2.142 $ & $  1.060 $ & $  1.948 \pm 0.010 $ & $  0.866 \pm 0.051 $ & $  1.916 \pm 0.010 $ & $ 0.481 \pm 0.028 $ & $ > 0.370         $ & $ > 1.105         $ \\
                            & $ 15.482 $ & $ 14.385 $ & $ 20.560 \pm 0.001 $ & $ 19.462 \pm 0.001 $ & $ 18.153 \pm 0.001 $ & \nodata             & \nodata             & \nodata             \\
\hline
% ------------------------------------------------------------------------------------------------------------------------------------------------------------
% KB-21-2213(*) 
KB212213           &            &            &                      &                      &                      &                     &                     &                     \\
outer              & $  2.197 $ & $  1.060 $ & $  1.847 \pm 0.121 $ & $  0.710 \pm 0.131 $ & $  2.576 \pm 0.109 $ & $ 0.441 \pm 0.064 $ & $ 0.213 \pm 0.035 $ & $ 5.052 \pm 0.837 $ \\
                   & $ 15.867 $ & $ 14.387 $ & $ 20.760 \pm 0.020 $ & $ 19.280 \pm 0.020 $ & $ 19.757 \pm 0.009 $ & \nodata             & \nodata             & \nodata             \\ 
\hline
% ------------------------------------------------------------------------------------------------------------------------------------------------------------
% KB-21-3290(*) 
KB213290           &            &            &                      &                      &                      &                     &                     &                     \\
$s_{-}$            & $  2.624 $ & $  1.060 $ & $  3.081 \pm 0.137 $ & $  1.517 \pm 0.146 $ & $  2.677 \pm 0.003 $ & $ 0.457 \pm 0.028 $ & $ 0.439 \pm 0.054 $ & $ 6.801 \pm 0.832 $ \\
                   & $ 16.512 $ & $ 14.527 $ & $ 22.734 \pm 0.013 $ & $ 20.749 \pm 0.013 $ & $ 17.661 \pm 0.001 $ & \nodata             & \nodata             & \nodata             \\ 
$s_{+}$            & $  2.624 $ & $  1.060 $ & $  3.083 \pm 0.137 $ & $  1.519 \pm 0.146 $ & $  2.677 \pm 0.003 $ & $ 0.444 \pm 0.027 $ & $ 0.448 \pm 0.059 $ & $ 6.633 \pm 0.870 $ \\
                   & $ 16.512 $ & $ 14.527 $ & $ 22.797 \pm 0.013 $ & $ 20.812 \pm 0.013 $ & $ 17.661 \pm 0.001 $ & \nodata             & \nodata             & \nodata             \\ 
% ------------------------------------------------------------------------------------------------------------------------------------------------------------
\enddata
\tablecomments{We use the abbreviation for event names, e.g., \fourtwentyfour\ is abbreviated as KB210424.
}
%\tabletypesize{\small}
\end{deluxetable}
\end{longrotatetable}
% ------------------------------------------------------------------------------------------------------------------------------------------------------------

% Table 14 (source color) ----------------------------------------------------
% For KB-21-0457, KB-21-1063, and KB-21-1691
\begin{longrotatetable}
\begin{deluxetable}{rr|rr|rrrrrr}
\tablecaption{$\langle\theta_{\ast}\rangle$ estimations for \fourfiftyseven, \tensixtythree, and \sixteenninetyone \label{table:VVV}}
\tablewidth{0pt}
\tabletypesize{\scriptsize}
\tablehead{
% ----------------------------------------------
\multicolumn{1}{c}{Event}              & 
\multicolumn{1}{c}{\fourfiftyseven}    & 
\multicolumn{2}{|c}{\tensixtythree}    & 
\multicolumn{6}{|c}{\sixteenninetyone} \\
% ----------------------------------------------
\multicolumn{1}{c}{Case}             & 
\multicolumn{1}{c}{2L1S (outer)}     & 
\multicolumn{1}{|c}{2L1S ($s_{-}$)}  & 
\multicolumn{1}{c}{2L1S ($s_{+}$)}   & 
\multicolumn{1}{|c}{2L2S ($s_{-}$)}  & 
\multicolumn{1}{c}{2L2S ($s_{+}$)}   & 
\multicolumn{1}{c}{3L1S ($s_{-,-}$)} & 
\multicolumn{1}{c}{3L1S ($s_{-,+}$)} & 
\multicolumn{1}{c}{3L1S ($s_{+,-}$)} & 
\multicolumn{1}{c}{3L1S ($s_{+,+}$)}
% ----------------------------------------------
}
\startdata
% --------------------------------------------------------------------------------------------------------------------------------------------------------------------------------------------------------------------------------
$(I-K, I)_{\rm RGC}$                                    & (4.155, 17.567)    & (2.544, 19.312)    & (2.544, 19.312)    & (2.264, 18.281)    & (2.264, 18.281)    & (2.264, 18.281)    & (2.264, 18.281)    & (2.264, 18.281)    & (2.264, 18.281)    \\
$(V-I, I)_{0, \rm RGC}$                                 & (1.060, 14.392)    & (1.060, 14.447)    & (1.060, 14.447)    & (1.060, 14.384)    & (1.060, 14.384)    & (1.060, 14.384)    & (1.060, 14.384)    & (1.060, 14.384)    & (1.060, 14.384)    \\
$(V-I)_{\rm S1}$                                        &\nodata             &\nodata             &\nodata             &\nodata             &\nodata             &\nodata             &\nodata             &\nodata             &\nodata             \\
$I_{\rm S1}$                                            & $18.226 \pm 0.007$ & $21.614 \pm 0.012$ & $21.644 \pm 0.012$ & $24.359 \pm 0.196$ & $24.264 \pm 0.390$ & $23.823 \pm 0.023$ & $23.773 \pm 0.023$ & $23.778 \pm 0.023$ & $23.779 \pm 0.023$ \\
$(V-I)_{\rm S2}$                                        & \nodata            & \nodata            & \nodata            &\nodata             &\nodata             & \nodata            & \nodata            & \nodata            & \nodata            \\
$I_{\rm S2}$                                            & \nodata            & \nodata            & \nodata            & $24.958 \pm 0.338$ & $24.978 \pm 0.753$ & \nodata            & \nodata            & \nodata            & \nodata            \\
$\langle(V-I)_{0, \rm S1}\rangle$                       & $ 1.034 \pm 0.151$ & $ 0.972 \pm 0.095$ & $ 0.943 \pm 0.091$ & $ 1.307 \pm 0.271$ & $ 1.263 \pm 0.268$ & $ 1.024 \pm 0.145$ & $ 0.999 \pm 0.122$ & $ 1.011 \pm 0.115$ & $ 1.009 \pm 0.114$ \\
$I_{0, \rm S1}$                                         & $15.051 \pm 0.007$ & $16.749 \pm 0.012$ & $16.779 \pm 0.012$ & $20.463 \pm 0.196$ & $20.368 \pm 0.390$ & $19.926 \pm 0.023$ & $19.877 \pm 0.023$ & $19.881 \pm 0.023$ & $19.882 \pm 0.023$ \\
$\langle(V-I)_{0, \rm S2}\rangle$                       & \nodata            & \nodata            & \nodata            & $ 1.650 \pm 0.358$ & $ 1.670 \pm 0.414$ & \nodata            & \nodata            & \nodata            & \nodata            \\
$I_{0, \rm S2}$                                         & \nodata            & \nodata            & \nodata            & $21.061 \pm 0.338$ & $21.082 \pm 0.753$ & \nodata            & \nodata            & \nodata            & \nodata            \\
$\langle\theta_{\ast, \rm S1}\rangle$ $({\mu{\rm as}})$ & $ 4.468 \pm 0.749$ & $ 1.896 \pm 0.196$ & $ 1.813 \pm 0.180$ & $ 0.460 \pm 0.093$ & $ 0.468 \pm 0.123$ & $ 0.468 \pm 0.077$ & $ 0.464 \pm 0.063$ & $ 0.469 \pm 0.060$ & $ 0.467 \pm 0.059$ \\
$\langle\theta_{\ast, \rm S2}\rangle$ $({\mu{\rm as}})$ & \nodata            & \nodata            & \nodata            & $ 0.421 \pm 0.090$ & $ 0.421 \pm 0.163$ & \nodata            & \nodata            & \nodata            & \nodata            \\
$\langle\theta_{\rm E}\rangle$ $({\rm mas})$            & \nodata            & \nodata            & \nodata            & $ 0.278 \pm 0.194$ & $ 0.263 \pm 0.194$ & $ 0.224 \pm 0.114$ & $ 0.210 \pm 0.100$ & $ 0.211 \pm 0.100$ & $ 0.210 \pm 0.099$ \\
%$\langle\mu_{\rm rel}\rangle$  $({\rm mas\, yr^{-1}})$ & \nodata            & \nodata            & \nodata            &                    &                    &                    &                    &                    &                    \\
% --------------------------------------------------------------------------------------------------------------------------------------------------------------------------------------------------------------------------------
\enddata
%\tablecomments{We use the abbreviation for event names, e.g., is abbreviated as KB211063.
%}
\tabletypesize{\small}
\end{deluxetable}
\end{longrotatetable}
% ------------------------------------------------------------------------------------------------------------------------------------------------------------

\section{Planet Properties} \label{sec:lens_properties}

% Table 15 (Lens properties) -------------------------------------------------
%\begin{longrotatetable}
\begin{deluxetable}{lccrrrrr}
\tablecaption{Lens Properties of Planetary Events \label{table:lens}}
%\tabletypesize{\scriptsize}
\tablewidth{0pt}
\tablehead{
% ---------------------------------------------------------------------------
\multicolumn{1}{c}{Event}                 &
\multicolumn{1}{c}{Constraints}           &
\multicolumn{1}{c}{Case}                  &
\multicolumn{1}{c}{$M_{\rm host}$}        &
\multicolumn{1}{c}{$M_{\rm planet}$}      &
\multicolumn{1}{c}{$D_{\rm L}$}           &
\multicolumn{1}{c}{$a_{\perp}$}           &
\multicolumn{1}{c}{$\mu_{\rm rel}$}       \\
% ---------------------------------------------------------------------------
\multicolumn{1}{c}{}                      &
\multicolumn{1}{c}{}                      &
\multicolumn{1}{c}{}                      &
\multicolumn{1}{c}{($M_{\odot}$)}         &
\multicolumn{1}{c}{($M_{\rm J}$ / $M_{\rm N}$)} &
\multicolumn{1}{c}{(kpc)}                 &
\multicolumn{1}{c}{(au)}                  &
\multicolumn{1}{c}{($\rm mas\, yr^{-1}$)} 
% ------------------------------------
}
\startdata
% --------------------------------------------------------------------------------------------------------------------------------------------------------------------------------------------------------------------------------
% KB-21-0424(*) == MB-21-111
KB210424      & Direct                                & APRX ($u_{0} > 0$) & $ 0.232_{-0.050}^{+0.050} $ & $ 0.779_{-0.169}^{+0.169}\, M_{\rm N} $ & $ 2.550_{-0.394}^{+0.394} $ & $ 1.962_{-0.303}^{+0.303} $ & $ 4.718_{-0.504}^{+0.504} $ \\ 
              & $t_{\rm E} + \theta_{\rm E} + \pivec$ & APRX ($u_{0} > 0$) & $ 0.244_{-0.039}^{+0.048} $ & $ 0.822_{-0.131}^{+0.176}\, M_{\rm N} $ & $ 2.620_{-0.372}^{+0.524} $ & $ 2.071_{-0.306}^{+0.257} $ & $ 4.699_{-0.338}^{+0.815} $ \\
              & Direct                                & APRX ($u_{0} < 0$) & $ 0.157_{-0.029}^{+0.029} $ & $ 0.556_{-0.103}^{+0.103}\, M_{\rm N} $ & $ 2.044_{-0.291}^{+0.291} $ & $ 1.516_{-0.216}^{+0.216} $ & $ 4.611_{-0.486}^{+0.486} $ \\
              & $t_{\rm E} + \theta_{\rm E} + \pivec$ & APRX ($u_{0} < 0$) & $ 0.231_{-0.063}^{+0.073} $ & $ 0.822_{-0.224}^{+0.265}\, M_{\rm N} $ & $ 2.780_{-0.644}^{+0.389} $ & $ 2.038_{-0.385}^{+0.160} $ & $ 4.536_{-0.720}^{+0.686} $ \\
% --------------------------------------------------------------------------------------------------------------------------------------------------------------------------------------------------------------------------------
\hline
% --------------------------------------------------------------------------------------------------------------------------------------------------------------------------------------------------------------------------------
% KB-21-0457(*)
KB210457      & $t_{\rm E} + \rho_{\ast}$             & outer     & $ 0.227_{-0.134}^{+0.297} $ & $ 1.121_{-0.694}^{+1.473}\, M_{\rm J} $ & $ 6.870_{-1.308}^{+1.035} $ & $ 1.918_{-0.565}^{+0.628} $ & $ 8.682_{-2.256}^{+2.345} $ \\ 
% --------------------------------------------------------------------------------------------------------------------------------------------------------------------------------------------------------------------------------
\hline
% --------------------------------------------------------------------------------------------------------------------------------------------------------------------------------------------------------------------------------
% KB-21-0690(*) == MB-21-161
KB210690      & $t_{\rm E} + \theta_{\rm E} + \pivec$ & APRX ($u_{0} < 0$)          & $  0.476_{-0.280}^{+0.280} $ & $ 11.644_{-6.845}^{+6.835}\, M_{\rm J} $ & $  8.839_{-1.420}^{+1.420} $ & $  0.155_{-0.015}^{+0.015} $ & $ 0.106_{-0.013}^{+0.013} $ \\
              & $t_{\rm E} + \theta_{\rm E} + \pivec$ & APRX ($u_{0} > 0$)          & $  0.463_{-0.267}^{+0.267} $ & $ 11.167_{-6.456}^{+6.449}\, M_{\rm J} $ & $  8.703_{-1.504}^{+1.504} $ & $  0.154_{-0.016}^{+0.016} $ & $ 0.107_{-0.013}^{+0.013} $ \\  
              & $t_{\rm E} + \rho_{\ast}$             & XRP                         & $  0.757_{-0.371}^{+0.522} $ & $ 10.413_{-5.108}^{+7.182}\, M_{\rm J} $ & $  3.791_{-1.881}^{+2.499} $ & $  1.772_{-0.555}^{+0.625} $ & $ 2.736_{-1.295}^{+2.168} $ \\
% --------------------------------------------------------------------------------------------------------------------------------------------------------------------------------------------------------------------------------
\hline
% --------------------------------------------------------------------------------------------------------------------------------------------------------------------------------------------------------------------------------
% KB-21-1063(*) 
KB211063      & $t_{\rm E} + \rho_{\ast}$             & $s_{-}$     & $ 0.416_{-0.260}^{+0.332} $ & $ 0.292_{-0.189}^{+0.284}\, M_{\rm J} $ & $ 6.707_{-1.550}^{+1.110} $ & $ 1.736_{-0.625}^{+0.673} $ & $ 6.166_{-2.229}^{+2.812} $ \\
              &                                       & $s_{+}$     & $ 0.435_{-0.269}^{+0.328} $ & $ 0.701_{-0.457}^{+0.600}\, M_{\rm J} $ & $ 6.649_{-1.588}^{+1.126} $ & $ 2.924_{-1.039}^{+1.098} $ & $ 6.282_{-2.228}^{+2.833} $ \\
% --------------------------------------------------------------------------------------------------------------------------------------------------------------------------------------------------------------------------------
\hline
% --------------------------------------------------------------------------------------------------------------------------------------------------------------------------------------------------------------------------------
% KB-21-2213(*) 
KB212213      & $t_{\rm E} + \theta_{\rm E}$ &             & $ 0.228_{-0.122}^{+0.284} $ & $ 1.089_{-0.583}^{+1.357}\, M_{\rm J} $ & $ 6.964_{-1.151}^{+0.980} $ & $ 1.233_{-0.262}^{+0.266} $ & $ 5.192_{-0.808}^{+0.822} $ \\ 
% --------------------------------------------------------------------------------------------------------------------------------------------------------------------------------------------------------------------------------
\hline
% --------------------------------------------------------------------------------------------------------------------------------------------------------------------------------------------------------------------------------
% KB-21-3290(*) 
KB213290      & $t_{\rm E} + \theta_{\rm E}$ & $s_{-}$     & $ 0.542_{-0.276}^{+0.264} $ & $ 1.860_{-0.959}^{+0.916}\, M_{\rm J} $ & $ 6.488_{-1.592}^{+0.966} $ & $ 2.357_{-0.576}^{+0.424} $ & $ 6.718_{-0.910}^{+0.994} $ \\
              &                              & $s_{+}$     & $ 0.550_{-0.282}^{+0.265} $ & $ 1.885_{-0.977}^{+0.921}\, M_{\rm J} $ & $ 6.464_{-1.610}^{+0.973} $ & $ 3.191_{-0.793}^{+0.588} $ & $ 6.571_{-0.975}^{+1.084} $ \\
% --------------------------------------------------------------------------------------------------------------------------------------------------------------------------------------------------------------------------------
\enddata
\tablecomments{For the planet mass, we present the values in Jupiter ($M_{\rm J}$) or Neptune ($M_{\rm N}$) masses as appropriate. 
}
%\tabletypesize{\small}
\end{deluxetable}
%\end{longrotatetable}
% --------------------------------------------------------------------------- 

% Table 16 (Lens properties of KB-21-1691) -------------------------------------------------
%\begin{longrotatetable}
\begin{deluxetable}{lrr|rrrr}
\tablecaption{Lens Properties of \sixteenninetyone \label{table:lens_1691}}
%\tabletypesize{\scriptsize}
\tablewidth{0pt}
\tablehead{
% ---------------------------------------------------------------------------
\multicolumn{1}{c}{Case}                  &
\multicolumn{1}{c}{2L2S ($s_{-}$)}        &
\multicolumn{1}{c}{2L2S ($s_{+}$)}        &
\multicolumn{1}{|c}{3L1S ($s_{-},s_{-}$)} &
\multicolumn{1}{c}{3L1S ($s_{-},s_{+}$)}  &
\multicolumn{1}{c}{3L1S ($s_{+},s_{-}$)}  &
\multicolumn{1}{c}{3L1S ($s_{+},s_{+}$)}  
% ---------------------------------------------------------------------------
}
\startdata
% --------------------------------------------------------------------------------------------------------------------------------------------------------------------------------------------------------------------------------
$M_{\rm host}$ ($M_{odot}$)             & $ 0.469_{-0.282}^{+0.315} $ & $ 0.455_{-0.276}^{+0.317} $ & $ 0.353_{-0.215}^{+0.324} $ & $ 0.332_{-0.201}^{+0.322} $ & $ 0.336_{-0.203}^{+0.321} $ & $ 0.333_{-0.201}^{+0.321} $ \\
$M_{\rm planet,1}$ ($M_{\rm J}$)        & $ 4.922_{-3.605}^{+3.896} $ & $ 5.827_{-3.868}^{+4.357} $ & $ 0.522_{-0.352}^{+0.612} $ & $ 0.511_{-0.358}^{+0.577} $ & $ 0.929_{-0.573}^{+1.020} $ & $ 0.901_{-0.571}^{+0.941} $ \\
$M_{\rm planet,2}$ ($M_{\rm J}$)        & \nodata                     & \nodata                     & $ 0.777_{-0.478}^{+0.933} $ & $ 0.869_{-0.545}^{+0.939} $ & $ 0.644_{-0.495}^{+0.640} $ & $ 0.621_{-0.474}^{+0.636} $ \\
$D_{\rm L}$ (kpc)                       & $ 6.590_{-1.472}^{+1.018} $ & $ 6.620_{-1.443}^{+1.016} $ & $ 6.881_{-1.266}^{+0.956} $ & $ 6.955_{-1.241}^{+0.946} $ & $ 6.956_{-1.243}^{+0.944} $ & $ 6.957_{-1.241}^{+0.945} $ \\
$a_{\perp,1}$ (au)                      & $ 1.446_{-0.499}^{+0.495} $ & $ 3.758_{-1.311}^{+1.314} $ & $ 1.522_{-0.487}^{+0.517} $ & $ 1.470_{-0.457}^{+0.489} $ & $ 2.167_{-0.668}^{+0.711} $ & $ 2.153_{-0.663}^{+0.708} $ \\
$a_{\perp,2}$ (au)                      & \nodata                     & \nodata                     & $ 1.648_{-0.531}^{+0.553} $ & $ 2.138_{-0.659}^{+0.709} $ & $ 1.454_{-0.447}^{+0.492} $ & $ 2.325_{-0.732}^{+0.762} $ \\
$\mu_{\rm rel}$ (${\rm mas\, yr^{-1}}$) & $ 3.503_{-1.157}^{+1.400} $ & $ 3.657_{-1.235}^{+1.542} $ & $ 3.304_{-1.104}^{+1.590} $ & $ 2.906_{-0.892}^{+1.114} $ & $ 2.814_{-0.836}^{+0.979} $ & $ 2.844_{-0.858}^{+1.034} $ \\
% --------------------------------------------------------------------------------------------------------------------------------------------------------------------------------------------------------------------------------
\enddata
\tablecomments{For all cases, the constraints of the Bayesian analyses are $t_{\rm E}$ and $\langle\theta_{\rm E}\rangle$.
}
%\tabletypesize{\small}
\end{deluxetable}
%\end{longrotatetable}
% --------------------------------------------------------------------------- 

The lens properties can be directly determined when two observables are simultaneously measured. These observables are the angular Einstein ring radius ($\theta_{\rm E}$) and the microlensing parallax ($\pivec$). By combining the observables and model parameters, we can determine the lens properties such as the lens mass ($M_{\rm L})$, distance to the lens ($D_{\rm L}$), projected separation between lenses ($a_{\perp}$), and lens-source relative proper motion ($\mu_{\rm rel}$). These quantities are determined as  
\begin{equation}
\begin{split}
M_{L} = \frac{\theta_{\rm E}}{\kappa |\pivec|} ~~;~~ D_{L} = \frac{\rm au}{|\pivec|\theta_{\rm E} + \pi_{S}}~~;~~ \\ 
a_{\perp} = s \theta_{\rm E} D_{\rm L}~~;~~\mu_{\rm rel} = \frac{\theta_{\rm E}}{t_{\rm E}},
\label{eqn:direct}
\end{split}
\end{equation} 
where $\kappa = 8.144\, {\rm mas}\, M_{\odot}^{-1}$, $\pi_{S}$ is the parallax of the source ($\pi_{S} \equiv {\rm au}/D_{S}$), and $D_{S}$ is distance to the source.

However, The two observables are not always measured simultaneously. Indeed, in this work, we cannot measure $\theta_{\ast}$ and $\pivec$ simultaneously for six of the seven planetary events, with the exception being \fourtwentyfour. If we are unable to measure either one or both of the observables, we estimate the lens properties using Bayesian analysis based on Galactic model priors. The formalism and procedures of Bayesian analysis adopted in this work are described in \citet{shin23a, shin23b}. Because each event has different constraints for the Bayesian analysis, we denote the constraints that we used following the notation and methodology that are described in \citet{shin23a}. In summary, the event timescale ($t_{\rm E}$) serves as a fundamental constraint for Bayesian analysis, constructed using Gaussian distributions of the $t_{\rm E}$ model parameter. We refer to this $t_{\rm E}$ constraint using the notation ``$t_{\rm E}$". The angular Einstein ring radius ($\theta_{\rm E}$) serves as an additional constraint based on the value of $\theta_{\rm E}$ ($\equiv \theta_{\ast} / \rho_{\ast}$), which can be determined when both $\rho_{\ast}$ and $\theta_{\ast}$ are measured reliably. We denote this $\theta_{\rm E}$ constraint with the notation of ``$\theta_{\rm E}$". If we have an uncertain $\rho_{\ast}$ measurement, we construct a weight function based on $\Delta\chi^{2}$ distributions of the $\rho_{\ast}$ parameter. This weight function is then applied to the priors as a constraint. We refer to this weight function derived from the $\rho_{\ast}$ distribution with the notation of ``$\rho_{\ast}$''. If we have a meaningful $\pivec$ distribution, we use the $\chi^{2}$ ellipse of the $\pivec$ distribution as an additional constraint. We refer to this $\pivec$ distribution constraint with the notation of ``$\pivec$". Lastly, if we have the source proper motion information for the source published in the {\it GAIA} DATA release 3 \citep[{\it GAIA} DR3;][]{GAIADR3}, we incorporate this proper motion into the Bayesian analysis. We find the sources in the {\it GAIA} DR3 for two planetary events (\fourtwentyfour\ and \fourfiftyseven). However, for \fourtwentyfour, there is no proper motion information. Perhaps this information will be included in the {\it GAIA} final release. For \fourfiftyseven, the proper motion is $(\mu_{\alpha^{\ast}}, \mu_{\delta})_{\rm S} = (-7.201 \pm 0.421, -5.905 \pm 0.249)\, \masyr$ ($\rm RUWE = 1.008$).

Among the seven planetary events, we can directly determine the lens properties of \fourtwentyfour. For the remaining six events, we estimate their lens properties using Bayesian analyses. In Tables \ref{table:lens} and \ref{table:lens_1691}, we present their lens properties. In the following sections, we describe the characteristics of each planetary system.

\subsection{\fourtwentyfour}    %  KMT-2021-BLG-0424(*)     MOA-2021-BLG-111
\label{sec:lens_KB210424}
As described in Section \ref{sec:KB210424}, the two essential observables, $\theta_{\rm E}$ and $\pivec$, are measured for this event. Thus, we can directly determine the lens properties. However, from the light curve analysis, we cannot resolve the ecliptic degeneracy in the APRX models. The $\Delta\chi^{2}$ between two models is only $\sim 2$.

We find that, for the APRX ($u_{0} > 0$) model, the lens system consists of a sub-Neptune-mass planet ($M_{\rm planet} = 0.779 \pm 0.169\, M_{\rm N}$) and an M-dwarf host star ($M_{\rm host} = 0.232 \pm 0.050\, M_{\odot}$). The planet orbits the host with a projected separation of $1.962 \pm 0.303$ au, which is beyond the snow line \citep[$\sim 0.58$ au;][]{kennedy08}. This system is located at $2.550 \pm 0.394$ kpc from us toward the Galactic bulge. For the APRX ($u_{0} < 0$) case, the planetary lens system shows a similar analog, i.e., a sub-Neptune-mass planet orbiting an M-dwarf host. While, because of the larger $|\pivec|$ value, the masses are smaller than those of the APRX ($u_{0} > 0$) case, i.e., $M_{\rm planet} = 0.556 \pm 0.103\, M_{\rm N}$ and $M_{\rm host} = 0.157 \pm 0.029\, M_{\odot}$ with $a_{\perp} = 1.516 \pm 0.216$. In addition, the lens system is closer to us than the APRX ($u_{0} > 0$) case, i.e., $D_{\rm L} = 2.044 \pm 0.291$ kpc. 

Although we cannot resolve the ecliptic degeneracy in the APRX models using $\Delta\chi^{2}$ at this time, it would be resolved by future adaptive-optics (AO) observations because the APRX models predict different directions of the heliocentric lens-source relative proper motions, which would be distinguished by the AO observations. The heliocentric proper motion vector ($\muvechel$) is given by \citet{dong09},
\begin{equation}
\muvechel = \muvecgeo + \vEarth \frac{\pi_{\rm rel}}{\rm au};~~ \pi_{\rm rel} = \theta_{\rm E}|\pivec|,
\end{equation}
where $\muvechel$ consists of north ($\mu_{\rm rel,\odot,\it{N}}$) and east ($\mu_{\rm rel,\odot,\it{E}}$) components and $\muvecgeo$ is the geocentric lens-source relative proper motion vector, which also consists of north ($\mu_{\rm rel, \oplus, \it{N}}$) and east ($\mu_{\rm rel, \oplus, \it{E}}$) components. We note that the magnitude of $\muvecgeo$ is denoted as $\mu_{\rm rel}$ in this paper. In addition, $\vEarth$ is the projected velocity of Earth relative to the Sun at the time of peak magnification $t_{0}$. For this event, $\vEarth = (v_{\oplus,N}, v_{\oplus,E}) = (0.795, 28.125)\, {\rm km\,s^{-1}}$. And finally, $\pi_{\rm rel}$ is the lens-source relative parallax, which is inferred from $\theta_{\rm E}$ and $\pivec$. The direction of $\muvechel$ measured (north through east) as  
\begin{equation}
\phi = \tan^{-1}\left(\frac{\mu_{\rm rel, \odot, \it{E}}}{\mu_{\rm rel, \odot, \it{N}}}\right).  
\end{equation}
For the APRX ($u_{0} > 0$) and ($u_{0} < 0$) models, $\muvechel=(-3.63 \pm 0.39, -1.45 \pm 0.59)$ and $(-4.06 \pm 0.34, 0.03 \pm 0.54)\, \masyr$, which yield $\phi = (201 \pm 9)^{\circ}$ and $(178 \pm 8)^{\circ}$, respectively. Considering the predicted values $\mu_{\rm rel} \sim 4.7$ or $\sim 4.6\, \masyr$, by $2032$, the lens and source will be separated enough (more than $50\, {\rm mas}$) for the Keck AO to measure the direction, which could help resolve the ecliptic degeneracy of this event.

\subsection{\fourfiftyseven}    %  KMT-2021-BLG-0457(*)     ---
\label{sec:lens_KB210457}
For this event, we can find a unique solution by resolving the inner/outer degeneracy as described in Section \ref{sec:KB210457}. However, we can measure neither $\theta_{\rm E}$ (i.e., $\rho_{\ast}$ is not measured) nor $\pivec$ for this event due to the non-caustic-crossing feature and short timescale. Thus, we estimate the lens properties based on the Bayesian analysis applying $t_{\rm E}$ and $\rho_{\ast}$ constraints. In addition, we incorporate the {\it GAIA} proper motion information for this source into the analysis. We note that the $\rho_{\ast}$ constraint has a minor effect because the $\rho_{\ast}$ distribution is broad (it is possible only to measure an upper limit), combined with the large uncertainty of $\langle\theta_{\ast}\rangle$ that is caused by the source location on the HST CMD, which is $\Delta I = 0.66$ below RGC, where the lower red giant branch shows a somewhat broad dispersion. As a result, the Bayesian results are almost identical to those applying the $t_{\rm E}$ constraint only. The Bayesian results indicate that the lens system consists of a Jupiter-like planet ($M_{\rm planet} \sim 1.12\, M_{\rm J}$) and an M-dwarf host ($M_{\rm host} \sim 0.23\, M_{\odot}$) with a projected separation of $\sim 1.92$ au, which is beyond the snow line ($\sim 0.61$ au). This system is located at a distance of $\sim 6.87$ kpc. These lens properties are typical of microlensing planets.

\subsection{\sixninety}         %  KMT-2021-BLG-0690(*)     MOA-2021-BLG-161
\label{sec:lens_KB210690}
As described in Section \ref{sec:KB210690}, we found two types of solutions for this event, which are the APRX and XRP models. For the APRX cases, we can measure the $\rho_{\ast}$ values and the well-constrained $\pivec$ distributions. Although we have both $\rho_{\ast}$ and $|\pivec|$ values (i.e., the best-fit values), we cannot directly determine the lens properties because the $\chi^{2}$ improvements are minor and $|\pivec|$ values within $1\sigma$ can yield the lens mass varying by an order of magnitude (i.e., the lens mass ($M_{\rm L}$) determined $|\pivec|$ within $2\sigma$ can be in $0.06 \lesssim M_{\rm L}/M_{\odot} \lesssim 1.39$ range). Therefore, we estimate the lens properties for the APRX cases using Bayesian analysis by applying $t_{\rm E}$ and $\theta_{\rm E}$ constraints. The Bayesian results based on the APRX models indicate that the lens system consists of a super-Jupiter-mass planet and an M-dwarf host star, as shown in Table \ref{table:lens}. 

However, we find that the lens mass ($M_{\rm L} \simeq 0.5\, M_{\odot}$) yields that $|\pivec| \sim 0.009$ (see Equation \ref{eqn:direct}). In addition, from Equation \ref{eqn:direct}$, \pi_{\rm rel} = {\rm au}\,\left( D_{\rm L}^{-1} - D_{\rm S}^{-1}\right)$ and $\theta_{\rm E}^{2} = \kappa M_{\rm L} \pi_{\rm rel}$, we can obtain $\pi_{\rm rel} = \kappa M_{\rm L} |\pivec|^{2} \sim 0.33\, \mu{\rm as}$, which implies that $D_{\rm LS} \equiv D_{\rm S} - D_{\rm L} = {\rm au}\,\pi_{\rm rel}*D_{\rm L}*D_{\rm S} \simeq 0.33\,\mu{\rm as} \times (8\, {\rm kpc})^{2}/{\rm Mpc} \simeq 21$ pc. Indeed, we find $D_{\rm S} \sim 8.871$ and $8.873$ kpc for the APRX ($u_{0} < 0$) and ($u_{0} > 0$) cases, respectively, using the Bayesian analyses. That is, $D_{\rm LS} \simeq 32$ and $31$ pc for the APRX ($u_{0} < 0$) and ($u_{0} > 0$) cases, respectively. This small amount of phase space is a priori unlikely, a fact that adds to the much larger improbability of very low proper motion as well as the low probability of such small orbital motion, which were discussed in Section \ref{sec:KB210690_low_prob}. Although the lens system derived from the APRX cases seems unlikely, we do not have any conclusive evidence to rule out these models at this time. Hence, we present the lens properties of the APRX solutions. 

In contrast, for the XRP case, we cannot measure the $\rho_{\ast}$ value. Thus, we estimate the lens properties using Bayesian analysis by employing the constraints of the $t_{\rm E}$ and $\rho_{\ast}$ distributions. The Bayesian results indicate that the lens system consists of a super-Jupiter-mass planet ($M_{\rm planet} \simeq 10\, M_{\rm Jupiter}$) and a K-dwarf host star ($M_{\rm host} \simeq 0.76\, M_{\odot}$). This host is $1.6$ times more massive than the host in the APRX cases. Moreover, the lens system is located at a distance of $\sim 3.8$ kpc toward the Galactic bulge, which is probably in the disk of our galaxy. This system is much more likely than those of the APRX cases. If the XRP solution is correct, the binary sources could be confirmed by AO observations within $15$ years, considering the $\mu_{\rm rel} \simeq 2.8\, {\rm mas\, yr^{-1}}$. 

However, at this moment, we cannot resolve the APRX/XRP degeneracy. Despite the existence of degenerate solutions for this event, we find that the lens system has a super-Jupiter-mass planet ($M_{\rm planet} = 10 - 12\, M_{\rm Jupiter}$) orbiting an M or K dwarf host star within its snow line (i.e., $a_{\rm snow} \simeq 1.3$ and $2.0$ au for the APRX and XRP cases, respectively).

Lastly, we note that the STD cases cannot be a solution. The Bayesian analysis of the STD cases with $\theta_{\rm E}$ constraint leads to the unreliably small lens mass (i.e., $M_{\rm L} \sim 0.015\, M_{\odot} \sim 16\, M_{\rm Jupiter}$) due to the $\theta_{\rm E} \sim 0.036\, {\rm mas}$. However, the lens mass cannot be smaller than the minimum lens mass of $0.04\, M_{\odot}$, considering the $3\sigma$ range of $|\pivec|$ distributions, which are reliably constrained. Therefore, we ruled out the STD cases.

\subsection{\tensixtythree}     %  KMT-2021-BLG-1063(*)     ---
\label{sec:lens_KB211063}
Recall that this event is subject to a relatively rare degeneracy between minor-image and major-image perturbations, which we have labeled $s_{-}$ and $s_{+}$, respectively.  In contrast to the more generic ``inner/outer" and ``close/wide" degeneracies, both cases of which share the same image perturbation and therefore usually give rise to very similar $q$ values, this rare degeneracy generally has a much larger $q$ for the minor-image perturbation. And from Table \ref{table:model_1063}, this is indeed the case of \tensixtythree.

For both $s_{\pm}$ solutions of this event, we cannot robustly measure either $\theta_{\rm E}$ or $\pivec$. Thus, we estimate the lens properties based on Bayesian analyses by applying the constraints of the $t_{\rm E}$ and $\rho_{\ast}$ distributions. The methodology of building the constraint of $\rho_{\ast}$ distributions is described in \citet{shin23a}. As shown in Table \ref{table:lens}, the Bayesian results indicate that the lens system consists of a sub-Jupiter-mass planet ($M_{\rm planet} \sim 0.3$ or $0.7\, M_{\rm Jupiter}$) and an M-dwarf host star ($M_{\rm host} \sim 0.4\, M_{\odot}$). The projected separation between the planet and the host is $\sim 1.7$ au or $\sim 2.9$ au for the $s_{-}$ and $s_{+}$ solutions, respectively, which are both beyond the snow line (i.e., $a_{\rm snow} \sim 1.1$ au). This planetary system is located at a distance of $\sim 6.5$ kpc toward the Galactic bulge, which is a typical microlensing planetary system.

\subsection{\sixteenninetyone}  %  KMT-2021-BLG-1691(*)     ---
\label{sec:lens_KB211691}
For this event, we found six degenerate solutions based on the 2L2S and 3L1S interpretations, which cannot be resolved. In all cases, we can measure the values of $\rho_{\ast}$ and $\langle\theta_{\ast}\rangle$, but we cannot measure the values of $\pivec$. Thus, we estimate the lens properties for all cases based on Bayesian analysis with constraints of $t_{\rm E}$ and $\langle\theta_{\rm E}\rangle \equiv \langle\theta_{\ast}\rangle / \rho_{\ast}$. In Table \ref{table:lens_1691}, we present the lens properties of each case. 

We note that, for the 2L2S models, $\langle\theta_{\rm E}\rangle$ could, in principle, be determined using either $\rho_{\ast, {\rm S1}}$ or $\rho_{\ast, {\rm S2}}$ (i.e., $\langle\theta_{\rm E}\rangle = \langle\theta_{\ast\, {\rm S1}}\rangle/\rho_{\ast, {\rm S1}}$ or $\langle\theta_{\rm E}\rangle = \langle\theta_{\ast\, {\rm S2}}\rangle/\rho_{\ast, {\rm S2}}$). However, for our 2L2S cases, we can only measure $\rho_{\ast,{\rm S2}}$. Thus, we determine the $\langle\theta_{\rm E}\rangle$ based on $\rho_{\ast, {\rm S2}}$. For the 3L1S case, the $\rho_{\ast}$ values are well measured for all cases. Thus, we determine $\langle\theta_{\rm E}\rangle$ using the $\rho_{\ast}$ values. In Table \ref{table:VVV}, we present $\langle\theta_{\rm E}\rangle$ values for all 2L2S and 3L1S cases.

For the 2L2S cases, the Bayesian results indicate that a super-Jupiter-mass planet ($M_{\rm planet} \sim 4.9$ or $5.8\, M_{\rm Jupiter}$ for the $s_{-}$ and $s_{+}$ cases, respectively) orbiting an M-dwarf host star ($M_{\rm host} \sim 0.47$ or $0.46\, M_{\odot}$ for the $s_{-}$ and $s_{+}$ cases, respectively) with a projected separation of $a_{\perp} \sim 1.4$ or $3.8$ au for the $s_{-}$ and $s_{+}$ cases, respectively. The planetary lens system is located at a distance of $\sim 6.6$ kpc toward the bulge from us.

Regarding the 3L1S models, although the lens properties are slightly different for the four cases, the Bayesian results indicate that the planetary lens system consists of sub-Jupiter-mass ($M_{\rm planet} \sim 0.5-0.6\, M_{\rm Jupiter}$ and Jupiter-class ($M_{\rm planet} \sim 0.8-0.9\, M_{\rm Jupiter}$) planets and an M-dwarf host star ($M_{\rm host} \sim 0.3-0.4\, M_{\odot}$). For all cases, the planets are located beyond the snow line (i.e., $a_{\rm snow} \sim 0.9-1.0$ au). The planetary system is also located at a distance of $\sim 7$ kpc from us.

\subsection{\twentytwothirteen} %  KMT-2021-BLG-2213(*)     ---
\label{sec:lens_KB212213}
For this event, we can measure $\theta_{\rm E}$. However, we cannot either measure $|\pivec|$ or place meaningful constraints on $\pivec$. Hence, we estimate the lens properties using Bayesian analysis with the $t_{\rm E}$ and $\theta_{\rm E}$ constraints. The Bayesian results indicate that a Jupiter-mass planet ($M_{\rm planet} \sim 1.09 \, M_{\rm Jupiter}$) orbits an M-dwarf host star ($M_{\rm host} \sim 0.23\, M_{\odot}$) with a projected separation of $\sim 1.23$ au. The planetary system is located at a distance of $\sim 6.96$ kpc. This planet is also a typical microlensing planet beyond the snow line (i.e., $a_{\rm snow} \sim 0.62$ au).

\subsection{\thirtytwoninety}   %  KMT-2021-BLG-3290(*)     ---
\label{sec:lens_KB213290}
Similar to \twentytwothirteen, we cannot either measure $|\pivec|$ or place meaningful constraints on $\pivec$ for this event. Thus, we estimate the lens properties of this event based on the Bayesian analysis by applying $t_{\rm E}$ and $\theta_{\rm E}$ constraints. As shown in Table \ref{table:lens}, the Bayesian results indicate that the lens system is a typical microlensing planetary system, which consists of a super-Jupiter-mass planet ($M_{\rm planet} \sim 1.9\, M_{\rm Jupiter}$) and an M-dwarf host star ($M_{\rm host} \sim 0.55\, M_{\odot}$) with a projected separation ($a_{\perp} \sim 2.4$ or $\sim 3.2$ au), beyond the snow line (i.e., $a_{\rm snow} \sim 1.5$ au). This system is located at $D_{\rm L} \sim 6.5$ kpc from us.

\section{Summary and Discussion} \label{sec:discussion}

% Table 17 (Summary of 2021 AF Prime Planets) -------------------------------------------------
%\begin{longrotatetable}
\begin{deluxetable}{lccccc}
\tablecaption{Planetary Events Discovered on KMTNet Prime Fields Observed in 2021 \label{table:prime_planets}}
%\tabletypesize{\scriptsize}
\tablewidth{0pt}
\tablehead{
% ---------------------------------------------------------------------------
\multicolumn{1}{c}{Event}                                 &
\multicolumn{1}{c}{$\log_{10}(q)$}                        &
\multicolumn{1}{c}{$s$}                                   &
\multicolumn{1}{c}{Method}                                &
\multicolumn{1}{c}{Degeneracy}                            &
\multicolumn{1}{c}{Reference}                             
% ---------------------------------------------------------------------------
}
\startdata
% -----------------------------------------------------------------------------------------------------------------------
KMT-2021-BLG-0171 & -4.320 & 0.798 & eye & c/w, i/o, ecliptic   & \citet{yang22}  \\
KMT-2021-BLG-0712 & -3.248 & 1.206 & eye & ecliptic             & \citet{ryu23}   \\
KMT-2021-BLG-1253 & -3.636 & 1.074 & eye & c/w, $\rho_{\ast}/q$ & \citet{ryu22}   \\
KMT-2021-BLG-1391 & -4.441 & 1.027 & eye & i/o, $\rho_{\ast}/q$ & \citet{ryu22}   \\
KMT-2021-BLG-1689$^\ddagger$ & -3.680 & 1.157 & eye & c/w, $\rho_{\ast}/q$ & \citet{yang22}  \\
KMT-2021-BLG-2478 & -2.376 & 1.056 & eye & ecliptic             & \citet{ryu23}   \\
KMT-2021-BLG-0736 & -3.971 & 1.563 & eye & \nodata              & \citet{yang25}  \\
KMT-2021-BLG-0424 & -3.762 & 1.096 & AF  & ecliptic             & This work       \\
KMT-2021-BLG-0919 & -3.604 & 1.141 & eye & ecliptic             & Shin et al. (in prep.) \\ 
KMT-2021-BLG-0320 & -3.520 & 0.771 & eye & c/w                  & \citet{han22a}  \\
KMT-2021-BLG-0240 & -3.444 & 0.958 & eye & c/w, 2L2S/3L1S       & \citet{han22b}  \\
KMT-2021-BLG-0192 & -3.428 & 0.761 & eye & c/w, ecliptic        & \citet{shin23a} \\
KMT-2021-BLG-2294 & -3.253 & 1.062 & eye & c/w                  & \citet{shin23a} \\
KMT-2021-BLG-1063 & -3.174 & 0.816 & AF  & c/w                  & This work       \\
KMT-2021-BLG-1150 & -2.939 & 1.242 & eye & i/o                  & \citet{han23b}  \\
KMT-2021-BLG-1554 & -2.850 & 0.888 & eye & c/w                  & \citet{han22a}  \\
KMT-2021-BLG-2010 & -2.558 & 0.845 & eye & c/w                  & \citet{han23a}  \\
KMT-2021-BLG-3290 & -2.485 & 0.860 & AF  & c/w                  & This work       \\
KMT-2021-BLG-2213 & -2.341 & 0.826 & AF  & \nodata              & This work       \\
KMT-2021-BLG-0457 & -2.327 & 1.289 & AF  & \nodata              & This work       \\
KMT-2021-BLG-1691 & -1.923 & 1.688 & AF  & 2L2S/3L1S            & This work       \\
KMT-2021-BLG-0690 & -1.882 & 0.555 & eye & APRX/XRP, ecliptic   & This work       \\
% -----------------------------------------------------------------------------------------------------------------------
\enddata
\tablecomments{The ``eye" and ``AF" indicate the methods to detect anomalies in the light curves by eye or AnomalyFinder, respectively. 
The acronyms in the ``Degeneracy" column represent the degeneracy types. That is, ``c/w": the close/wide degeneracy, 
``i/o": the inner/outer degeneracy, ``ecliptic": the ecliptic degeneracy in the APRX models, 
``$\rho_{\ast}/q$": small/large $\rho_{\ast}$ degeneracy related to the caustic size (i.e., large/small $q$ value), 
``2L2S/3L1S": the 2L2S/3L1S degeneracy, ``XRP/APRX": the APRX/XRP degeneracy.
$^\ddagger$We note that, for KMT-2021-BLG-1689, the anomaly cannot be identified on the KMTNet-only light curve. 
The planetary anomaly of this event was confirmed by the follow-up observations (see details 
in Section 3.3 of \citealt{yang22}).
}
%\tabletypesize{\small}
\end{deluxetable}
%\end{longrotatetable}
% --------------------------------------------------------------------------- 

% Figure 16 (Mass-ratio distributions) --------------------------------------------------------------
\begin{figure*}[t]
\epsscale{1.00}
\plotone{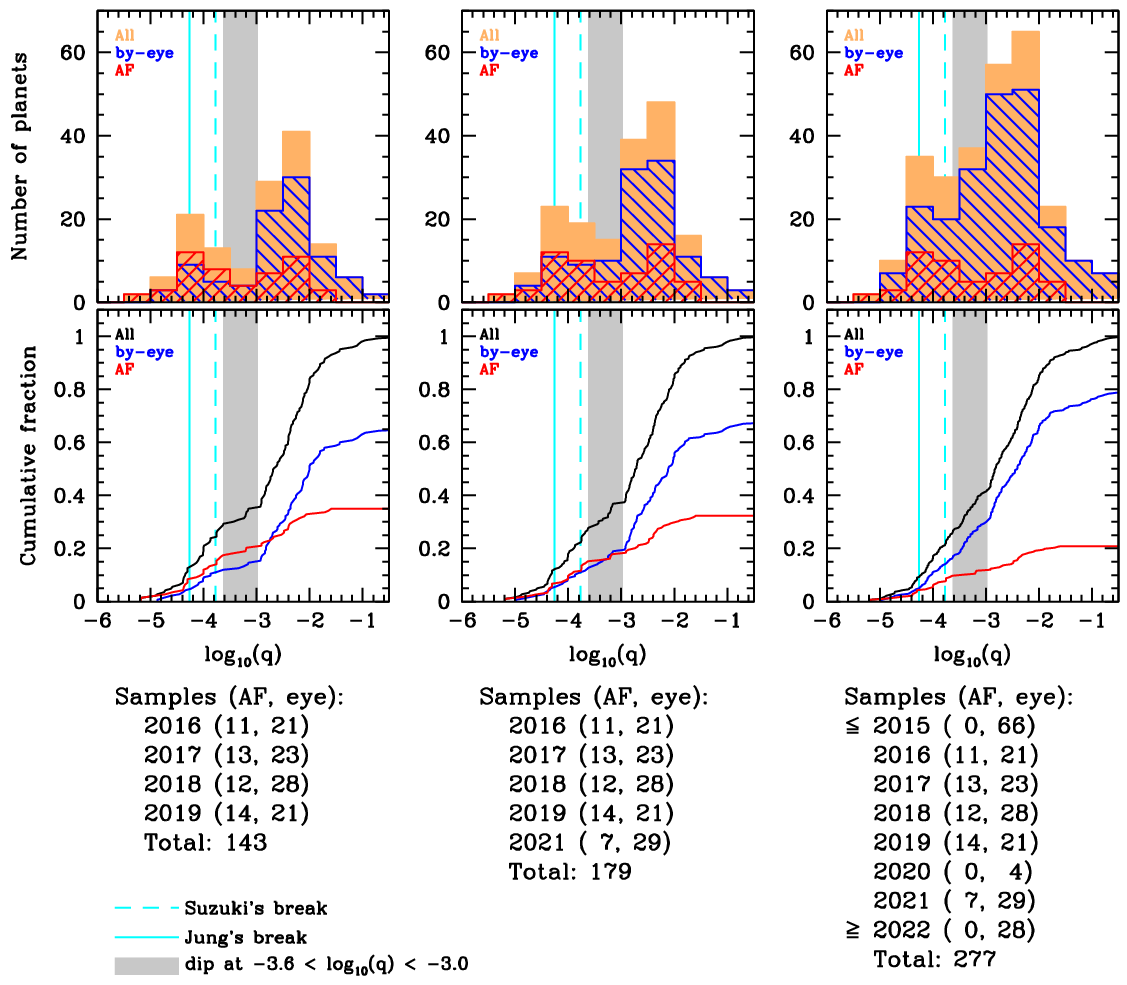}
\caption{Mass-ratio distributions based on various microlensing planet samples. 
The first sample is from planets discovered in $2016$ -- $2019$ by eye and systematic search series (left-side panels). 
The second sample includes additional planets from this work and those found by eye in $2021$ (middle panels). 
The third sample includes planets found before the KMTNet survey ($2015$) and after $2022$ (right-side panels). 
Upper panels present the distribution in histograms. Lower panels present them in cumulative fractions, 
where the fraction is normalized by the total number of planet detections of each sample. 
At the bottom, we show the number of planets detected via the AnomalyFinder and the by-eye method 
by year for each of the three samples.
\label{fig:q_dist}}
\end{figure*}
% --------------------------------------------------------------------------------------------------

In this work, we found seven planetary events and three planet candidates hidden in the $2021$ KMTNet Prime fields. Most of these new planets are typical microlensing planets that are orbiting M-dwarf hosts beyond their snow lines (except \sixninety, which has a tighter orbit). In addition, the planetary systems are mostly located near our Galactic bulge (except \fourtwentyfour, which is located in the disk). In Table \ref{table:prime_planets}, we list microlensing planets discovered on KMTNet Prime fields observed in 2021 through the by-eye and systematic planetary anomaly search methods. This work contributes about $33\%$ of the total microlensing planets in the $2021$ Prime fields. Indeed, during the first phase of the systematic planetary anomaly search series, approximately $35\%$ of the total microlensing planet discoveries from 2016 to 2019 were made by the systematic search method, which reflects a similar fraction of hidden planets discovered in this work. Hence, although the conventional by-eye search is the primary method for identifying planetary anomalies, we expect that one-third of microlensing planets may still be hidden in the data archive. This fact clearly demonstrates the importance of continuing systematic anomaly search work in building a complete microlensing planet sample.

The complete sample is crucial for statistical studies, as an incomplete or biased sample can lead to erroneous \HL{interpretations. Indeed,} several previous studies present mass-ratio distributions and their characteristics, which were conducted based on the latest samples at that time. For example, \citet{suzuki16} claims a break of the broken power law at $q \simeq 1.7\times10^{-4}$ in the mass-ratio distributions. In addition, \citet{udalski18} presented a broken power-law mass-ratio distribution including a very low-mass sample (i.e., $q <1\times10^{-4} $), which is similar to the result of \citet{suzuki16}. In contrast, \citet{jung19} claims that the break is located at $q \simeq 5.5\times10^{-4}$. \citet{gui24} and \citet{yang20} claim a dip in the distributions at the range of $-3.6 < \log_{10}(q) < -3.0$. \citet{zang25} present the latest mass-ratio distributions based on the sample built by the first phase of the systematic search series. They claim the distributions can be better described by a double Gaussian form than a broken power law by $-2\Delta\ln\mathcal{L} = 22.6$ ($\sim 3.8\sigma$).

In Figure \ref{fig:q_dist}, we present mass-ratio distributions built by three types of microlensing planet samples. The first sample is built by planets discovered by the first phase of systematic search series and by-eye searches from the $2016$ to $2019$ seasons. This sample is similar to those used for the studies of \citet{gui24} and \citet{zang25}. The second sample contains additional planets discovered by this work and by-eye searches in $2021$. The third sample is built using all microlensing planets to date. Note that, because sample selection is beyond the scope of present work, we do not apply the selection criteria used in the previous studies to these samples. The left-hand panels show an almost identical shape to those in \citet{gui24} and \citet{zang25} as expected. One noticeable feature is that neither break (i.e., $q\simeq 1.7\times10^{-4}$ or $5.5\times10^{-4}$) seems to exist. It is clearer that the breaks have disappeared in the second panel, which includes the $2021$ planet samples (see cyan lines in Figure \ref{fig:q_dist}). For both cases, the dip is clear, and the distributions can be described by a double Gaussian. However, for the third panel, which includes all planets, even those from before the KMTNet survey and systematic searches, the dip feature is not clear. Moreover, the breaks are also unclear. Perhaps the morphology changes in the distributions may be attributed to the contamination in the samples, such as the publication bias. However, although these distributions are built from uncontrolled samples, the discrepancy among them clearly shows the importance of a complete sample for statistical studies. This is a major reason to continue this series.

\hbox{}
% =================================
% Acknowledgments
% =================================
%\begin{acknowledgments}
% KMTNet
This research has made use of the KMTNet system
operated by the Korea Astronomy and Space Science Institute
(KASI) at three host sites of CTIO in Chile, SAAO in South
Africa, and SSO in Australia. Data transfer from the host site to
KASI was supported by the Korea Research Environment
Open NETwork (KREONET). This research was supported by KASI
under the R\&D program (project No. 2026-1-904-01) supervised
by the Ministry of Science and ICT.
% MOA
The MOA project is supported by JSPS KAKENHI Grant Number JP16H06287, JP22H00153, JP23KK0060, and JP25H00668.
I.G.S. and W.Z. acknowledges the support from the Harvard-Smithsonian Center for Astrophysics through the SAO and CfA Fellowships.
% Jennifer C. Yee
J.C.Y. acknowledges support from U.S. NASA Grant No. 80NSSC25K7146. 
% Weicheng & Hongjing
W.Z., H.Y., and S.M. acknowledge support by the National Natural Science Foundation of China (Grant No. 12133005). 
H.Y. acknowledges support by the China Postdoctoral Science Foundation (No. 2024M762938). 
% Han, Cheongho
Work by C.H. was supported by the grants of National Research Foundation of Korea (2019R1A2C2085965 and 2020R1A4A2002885).
% Yossi Shvartzvald
Y.S. acknowledges support from BSF Grant No. 2020740.
% Hydra
The computations in this paper were conducted on the Smithsonian High Performance Cluster (SI/HPC), Smithsonian Institution (\url{https://doi.org/10.25572/SIHPC}).
%\end{acknowledgments}
% =================================

\appendix 

% Table 18 (Appendix: Non-planetary events) -------------------------------------------------
%\begin{longrotatetable}
\begin{deluxetable}{llccrrl}
\tablecaption{Summary of Non-Plnaetary Events' Characteristics \label{table:non_planet}}
%\tabletypesize{\scriptsize}
\tablewidth{0pt}
\tablehead{
% ---------------------------------------------------------------------------
\multicolumn{1}{c}{Event}                                 &
\multicolumn{1}{c}{Best-fit model}                        &
\multicolumn{1}{c}{$s$}                                   &
\multicolumn{1}{c}{$q$}                                   &
\multicolumn{1}{c}{$t_{\rm E}$ (days)}                    &
\multicolumn{1}{c}{$\rho_{\ast}$}                         &
\multicolumn{1}{c}{Lens System (expected)}                            
% ---------------------------------------------------------------------------
}
\startdata
% -----------------------------------------------------------------------------------------------------------------------
KMT-2021-BLG-0148      & 1L2S (APRX) & \nodata & \nodata    &  63.1 & $< 0.406$, $< 0.228$  & Single star       \\
\bf{KMT-2021-BLG-0338} & 2L1S (STD)  & 2.460   & \bf{0.081} & 294.5 & 0.030                 & \bf{Binary w/ BD} \\ 
KMT-2021-BLG-0341      & 2L1S (STD)  & 0.449   & 0.292      &  14.2 & $< 0.067$             & Binary star       \\ 
KMT-2021-BLG-0567      & 2L1S (STD)  & 0.835   & 0.138      &  22.8 & 0.004                 & Binary star       \\ 
KMT-2021-BLG-0657      & 2L1S (STD)  & 0.486   & 0.201      &  54.3 & $\sim 0.014$          & Binary star       \\ 
KMT-2021-BLG-0876      & 2L1S (STD)  & 0.762   & 0.528      &  35.4 & $0..68\times10^{-3}$  & Binary star       \\ 
KMT-2021-BLG-2340      & 2L1S (STD)  & 3.221   & 0.308      &  16.7 & 0.056                 & Binary star       \\ 
\bf{KMT-2021-BLG-2358} & 2L1S (STD)  & 4.428   & \bf{0.050} &  32.7 & $< 0.011$             & \bf{Binary w/ BD} \\ 
KMT-2021-BLG-2737      & 1L2S (STD)  & \nodata & \nodata    &  34.2 & $< 0.022$, $< 0.326$  & Single star       \\ 
\bf{KMT-2021-BLG-2745} & 1L2S (STD)  & \nodata & \nodata    &   7.9 & $< 0.274$, $< 0.240$  & Single star       \\ 
                       & 2L1S (STD)  & 1.481   & 0.149      &  \bf{6.2} & $< 0.294$         & \bf{Binary w/ BD} \\ 
KMT-2021-BLG-3112      & 2L1S (STD)  & 1.088   & 0.128      &  17.4 & 0.002                 & Binary star       \\ 
KMT-2021-BLG-3140      & 2L1S (STD)  & 0.199   & 0.408      &  24.9 & $< 0.028$             & Binary star       \\ 
                       & 1L2S (STD)  & \nodata & \nodata    &  22.2 & 0.011, 0.074          & Single star       \\ 
% -----------------------------------------------------------------------------------------------------------------------
\enddata
\tablecomments{
The acronym ``BD" refers to ``Brown dwarf". KMT-2021-BLG-2745 and KMT-2021-BLG-3140 exhibit severe 2L1S/1L2S degeneracy. 
Thus, we present both 2L1S and 1L2S cases. 
In the $\rho_{\ast}$ column, we present $\rho_{\ast, S1}$ and $\rho_{\ast, S2}$, respectively, if the best-fit model is 
the 1L2S case. 
In the last column, we present an ``expected" component(s) of each lens system, which is not a conclusive lens property.
}
%\tabletypesize{\small}
\end{deluxetable}
%\end{longrotatetable}
% --------------------------------------------------------------------------- 

\section{Non-planetary events} \label{sec:appendix_binaries}
From the preliminary analysis based on the pipeline data sets, we found a total of $22$ planet-like events in the $2021$ KMTNet Prime fields. These events can be described by at least one model with $q < 0.06$, which is a loose mass-ratio criterion for selecting events caused by potential planetary lens systems, given the quality of the pipeline datasets. From the detailed analyses based on the best quality re-reduced datasets (hereafter, referred to as TLC (tender-loving-care) datasets), we found $7$ planets and $3$ planet candidates, as described in the main article. However, we find that the remaining $12$ events were caused by non-planetary lens systems, such as 1L2S or binary-lens systems (i.e., all fiducial solutions have $q > 0.03$). Although these are not planetary events, we briefly describe them for the record. This documentation can help prevent redundant efforts in searching for planetary events in the 2021 KMTNet Prime fields. \HL{In Table \ref{table:non_planet}, we summarize the main characteristics of these non-planetary events. In addition, non-planetary events highlighted in boldface in Table \ref{table:non_planet} may have interesting lens components, such as brown dwarfs. We briefly discuss them in Section \ref{sec:BD_candidates}.}

\subsection{\onefourtyeight} % KMT-2021-BLG-0148 == MOA-2021-BLG-031(*) | 1L1S event
The light curve of \onefourtyeight\ exhibits a bump-shaped anomaly at $\HJD \sim 9410.0$. We find plausible 2L1S models, which imply that this event would be caused by a planetary system (i.e., $q < 0.03$). However, no 2L1S model can adequately describe the observed light curve, even when including higher-order effects such as APRX and/or OBT. The 2L1S+OBT case is the most likely case among our tested models. However, there are still residuals in the fits. Hence, we search for a 1L2S model, which we find describes the full light curve, including the anomaly, much better. The best-fit case is a 1L2S$+$APRX model. Quantitatively, $\Delta\chi^{2}$ between 2L1S and 1L2S models is $778$, which conclusively distinguishes between them. Hence, we conclude that \onefourtyeight\ was caused by a binary source with a single lens rather than a planetary system. For the record, we present the best-fit 1L2S$+$APRX model parameters: $(t_{0,{\rm S1}}, t_{0,{\rm S2}}, u_{0,{\rm S1}}, u_{0,{\rm S2}}, t_{\rm E}, q_{\rm flux}, \rho_{\ast, {\rm S1}}, \rho_{\ast, {\rm S2}}, \pi_{{\rm E},N}, \pi_{{\rm E},E}, \\ f_{\rm S}, f_{\rm B}) = (9307.604, 9401.133, 0.375, 0.105, 63.105, 0.048, \\ < 0.406, < 0.228, 0.286, 0.217, 0.598, -0.032)$. The inequality in $\rho_{\ast}$ values indicate the upper limits ($3\sigma$) of the $\rho_{\ast}$ distributions, i.e., $\rho_{\ast}$ values are not measured. 

\subsection{\threethirtyeight} % KMT-2021-BLG-0338(*) == MOA-2021-BLG-082 | Binary event
The light curve of \threethirtyeight\ shows asymmetric features relative to the 1L1S model. From the initial analysis, we find that the asymmetric anomaly could be explained by 2L1S models having $q < 0.06$. However, from the detailed analysis based on the TLC datasets, we find that the best-fit model has $q \sim 0.081$ and there are no competing solutions having $q < 0.03$. Hence, we conclude that this event was caused by a binary system. Note that the MOA light curve produced by their pipeline exhibits an additional planet-like anomaly at $\HJD = 9347.97$. However, the re-reduced MOA data confirm that this anomaly is not a real feature. For the record, we present the best-fit 2L1S model parameters: $(t_{0}, u_{0}, t_{\rm E}, s, q, \alpha, \rho_{\ast}, f_{\rm S}, f_{\rm B}) = (9342.798, 0.029, 294.525, 2.460, 0.081, 4.754, 0.030, 0.261, \\ 8.277)$. We note that parameter errors are omitted because we stopped the analysis during the grid search stage when we could conclude that the event is a non-planetary case. We also note that the value of $\rho_{\ast}$ we presented cannot be considered conclusive for the same reason.

\subsection{\threefourtyone} % KMT-2021-BLG-0341(*) | Binary event
The light curve of \threefourtyone\ exhibits shallow double bump anomalies at the peak. From the initial analysis, we find that the anomalies can be explained by a binary-lens model (i.e., $q \sim 0.14$), however, it also can be well described by a planet-like model (i.e., $q \sim 0.026$). Thus, we investigate this event based on the TLC data sets. We find that the best-fit model indicates $q = 0.292 \pm 0.074$ and the lowest mass ratio model suggests $q = 0.073 \pm 0.014$. In both cases, the mass ratios imply that this event was caused by a binary-lens system. For the record, we present the best-fit 2L1S model parameters: $(t_{0}, u_{0}, t_{\rm E}, s, q, \alpha, \rho_{\ast}, f_{\rm S}, f_{\rm B}) = (9320.858, 0.218, 14.173, 0.449, 0.292, 2.484, < 0.067, \\ 0.142, 0.227)$. The inequality in the $\rho_{\ast}$ value indicates the upper limit ($3\sigma$) of the $\rho_{\ast}$ distribution, i.e., $\rho_{\ast}$ is not measured.

\subsection{\fivesixtyseven} % KMT-2021-BLG-0567(*) | Binary event
The light curve of \fivesixtyseven\ shows a horn-shaped anomaly (partly covered) from $\HJD \sim 9334.5$ to $9336.5$, which is a typical anomaly caused by a binary-lens system. However, from the initial analysis, we find that the 2L1S model that can explain the anomaly has $q \sim 0.057$, which satisfies our criterion to conduct a more detailed analysis. (i.e., $q < 0.06$). Thus, we conduct the analysis using TLC datasets. We find that the best-fit 2L1S STD model shows $q = 0.138 \pm 0.006$. Hence, we conclude that \fivesixtyseven\ was caused by a binary-lens system. For the record, we present the best-fit 2L1S model parameters: $(t_{0}, u_{0}, t_{\rm E}, s, q, \alpha, \rho_{\ast}, f_{\rm S}, f_{\rm B}) = (9331.739, 0.433, 22.771, 0.835, 0.138, 5.429, 0.004, 0.184, \\ 0.017)$.

\subsection{\sixfiftyseven} % KMT-2021-BLG-0657(*) | Binary event
The light curve of \sixfiftyseven\ exhibits weak bump-shaped anomalies at the peak. These anomalies can be well described by binary or planetary models. However, we find that the best-fit model is a binary case ($q \sim 0.20$). In contrast, the plausible planetary case ($q \sim 0.028$) shows worse fits by $\Delta\chi^{2} \sim 126.5$. Hence, we conclude that this event was caused by a binary-lens system. For the record, we present the best-fit 2L1S model parameters: $(t_{0}, u_{0}, t_{\rm E}, s, q, \alpha, \rho_{\ast}, f_{\rm S}, f_{\rm B}) = (9369.089, 0.189, 54.322, 0.486, 0.201, 3.773, 0.014, 0.130, \\ 0.021)$. The $\rho_{\ast}$ value is uncertain.

\subsection{\eightseventysix} % KMT-2021-BLG-0876(*) | Binary event
\eightseventysix\ exhibits a clear U-shaped anomaly feature, which is caused by caustic crossing, on its low-magnification 1L1S-like light curve. Based on the pipeline data sets, we found that the anomaly can be explained by a planetary model (i.e., $q \sim 0.029$). However, in the analysis based on the re-reduced data sets, we find that the best-fit model indicates that this event is caused by a binary-lens system (i.e., $q \sim 0.087 \pm 0.007$). Hence, we conclude that \eightseventysix\ was caused by a binary-lens system. For the record, we present the best-fit 2L1S model parameters: $(t_{0}, u_{0}, t_{\rm E}, s, q, \alpha, \rho_{\ast}, f_{\rm S}, f_{\rm B}) = (9330.225, 0.429, 35.425, 0.762, 0.528, 5.156, 0.00068, \\ 0.043, 0.329)$.

\subsection{\twentythreefourty} % KMT-2021-BLG-2340(*) == MOA-2021-BLG-367 | Binary event
The light curve of \twentythreefourty\ shows an asymmetric pattern from the peak to the decreasing side. This anomaly can be described by the 2L1S interpretation. The best-fit model indicates that a binary-lens system with a mass ratio, $q = 0.309 \pm 0.0104$, causes the asymmetric anomaly. We find that the alternative model having the lowest mass ratio, $q = 0.052 \pm 0.006$, can describe the bump-like anomaly at ${\rm HJD}^{\prime} \sim 9473.0$. However, the mass ratio does not satisfy our criterion of planet detection. Moreover, this model shows worse fits by $\Delta\chi^{2} = 21$. Thus, we can reject this solution considering our $\chi^{2}$ criterion. Indeed, we find that this bump-like anomaly at ${\rm HJD}^{\prime} \sim 9473.0$ is likely to be correlated with the rising pattern of background levels. Thus, this bump-like anomaly is highly likely to be a false positive. Hence, we conclude that this event was caused by a binary-lens system. For the record, we present the best-fit 2L1S model parameters: $(t_{0}, u_{0}, t_{\rm E}, s, q, \alpha, \rho_{\ast}, f_{\rm S}, f_{\rm B}) = (9461.216, 0.191, 16.717, 3.221, 0.308, 2.019, 0.056, 0.440, \\ -0.256)$. 

\subsection{\twentythreefiftyeight} %  KMT-2021-BLG-2358(*) == MOA-2021-BLG-380 | Binary event
For \twentythreefiftyeight\ event, the pipeline data exhibit a bump-like anomaly at ${\rm HJD}^{\prime} \sim 9468.6$ on its light curve with an asymmetric pattern at the peak. However, the TLC datasets reveal that the bump-like anomaly is a false positive. Although the bump-like anomaly is fake, the light curve of the TLC datasets still shows the asymmetric peak. We find that this light curve can be explained by 2L1S models. The degenerate models ($\Delta\chi^{2} < 1$) caused by the close/wide degeneracy show relatively low-mass ratios, $q = 0.050 \pm 0.006$ and $0.044 \pm 0.004$. However, these do not satisfy our planet detection criterion. Hence, we conclude that this event was caused by a binary-lens system, which would have a very low mass companion, such as a brown dwarf. For the record, we present the best-fit 2L1S model parameters: $(t_{0}, u_{0}, t_{\rm E}, s, q, \alpha, \rho_{\ast}, f_{\rm S}, f_{\rm B}) = (9469.028, 0.016, 32.689, 4.428, 0.050, 5.334, < 0.011, \\ 0.190, 0.031)$.

\subsection{\twentyseventhirtyseven} %  KMT-2021-BLG-2737(*) | 1L2S event
The light curve of \twentyseventhirtyseven\ exhibits a long-term bump-shaped anomaly from ${\rm HJD}^{\prime} \sim 9485.0$ to $9505.0$. We find that the 1L2S interpretation can perfectly describe the anomaly rather than any 2L1S interpretation. In fact, the most plausible 2L1S model shows a worse fit than the 1L2S best-fit model by $\Delta\chi^{2} \sim 94$. Hence, we conclude that this event was caused by a binary-source system. For the record, we present the best-fit 1L2S model parameters: $(t_{0,{\rm S1}}, t_{0,{\rm S2}}, u_{0,{\rm S1}}, u_{0,{\rm S2}}, t_{\rm E}, q_{\rm flux}, \rho_{\ast, {\rm S1}}, \rho_{\ast, {\rm S2}}, f_{\rm S}, f_{\rm B}) = \\ (9474.659, 9495.519, 0.016, 0.168, 34.173, 0.372, \\ < 0.022, < 0.326, 0.067, 0.112)$. 

\subsection{\twentysevenfourtyfive} %  KMT-2021-BLG-2745(*) | Binary or 1L2S event
The light curve of \twentysevenfourtyfive\ exhibits a bump-shaped anomaly at ${\rm HJD}^{\prime} \sim 9276.6$. We find that this anomaly can be explained by either a 2L1S or 1L2S interpretation. The best-fit case is a 1L2S model. The 2L1S model can also describe the anomaly, which shows a slightly worse fit by $\Delta\chi^{2} = 3$. Although the 2L1S/1L2S degeneracy cannot be resolved, the 2L1S model indicates that this event was caused by a binary-lens system (i.e., $q = 0.149 \pm 0.046$) rather than a planetary lens system. Hence, we conclude that this event occurred by either a binary-source or a binary-lens system. For the record, we present both 2L1S and 1L2S model parameters: $(t_{0}, u_{0}, t_{\rm E}, s, q, \alpha, \rho_{\ast}, f_{\rm S}, f_{\rm B}) = (9276.801, 0.368, 6.791, 1.481, 0.149, 6.225, < 0.294, 0.235, \\ -0.038)$ and $(t_{0,{\rm S1}}, t_{0,{\rm S2}}, u_{0,{\rm S1}}, u_{0,{\rm S2}}, t_{\rm E}, q_{\rm flux}, \rho_{\ast, {\rm S1}}, \rho_{\ast, {\rm S2}}, \\ f_{\rm S}, f_{\rm B}) = (9276.622, 9282.963, 0.142, 0.268, 7.944, 8.088, \\ < 0.274, < 0.240, 0.134, 0.062)$. We note that, considering the short $t_{\rm E}$ of the 2L1S case, the lens system could have a very low-mass companion, such as a brown dwarf.

\subsection{\thirtyonetwelve} %  KMT-2021-BLG-3112(*) | Binary event
The light curve of \thirtyonetwelve\ shows a long-term caustic-crossing anomaly from ${\rm HJD}^{\prime} \sim 9483.5$ to $9490.6$. We find that the best-fit 2L1S model with $(s,q) = (1.088 \pm 0.022, 0.128 \pm 0.015)$ can perfectly describe the anomaly, which indicates that the lens is a binary system. There exists a competing binary lens model with $(s,q) = (0.999 \pm 0.015, 0.126 \pm 0.016)$ that is disfavored by $\Delta\chi^{2} = 5.06$.  We also find a plausible planetary model with $(s, q) = (1.060 \pm 0.014,  0.010 \pm 0.002)$. However, this model shows a worse fit than the best-fit model by $\Delta\chi^{2} = 13$. We reject this model considering our $\chi^{2}$ criterion. Hence, we conclude that this event was caused by a binary-lens system. For the record, we present the best-fit 2L1S model parameters: $(t_{0}, u_{0}, t_{\rm E}, s, q, \alpha, \rho_{\ast}, f_{\rm S}, f_{\rm B}) = (9489.886, 0.041, 17.386, 1.088, 0.128, 3.926, 0.002, 0.016, \\ 0.415)$.

\subsection{\thirtyonefourty} %  KMT-2021-BLG-3140(*) | Binary event
The light curve of \thirtyonefourty\ shows an anomaly at the peak, which is poorly covered. We find that both 2L1S and 1L2S models can describe the anomaly. The $\chi^{2}$ difference between the 2L1S and 1L2S models is $11.8$. Thus, the 2L1S/1L2S degeneracy cannot be resolved because our criterion to resolve this degeneracy is $15$. The best-fit model is a 2L1S case with $(s,q) = (0.199 \pm 0.020, 0.408 \pm 0.186)$, which indicates the lens is a binary system. We also find a competing planetary model with $(s,q) = (1.078_{-0.020}^{+0.745}, \left(15.870_{-0.389}^{+36.149}\right)\times10^{-4})$, which has poorer fits by $\Delta\chi^{2} = 11$. Considering our $\chi^{2}$ criterion for resolving the planet/binary degeneracy (i.e., $\Delta\chi^{2} > 10$), this planetary model can be nominally rejected. Indeed, although the $\Delta\chi^{2}$ value is at the border line of the criterion, the $\Delta\chi^{2}$ mostly comes from the fits of the peak region of the anomaly (i.e., from ${\rm HJD}^{\prime} = 9488.0$ to $9492.0$), which is a crucial part to determine the models. The $\Delta\chi^{2}$ considering only the peak part is $14$. Hence, we conclude that this event was caused by either a binary lens or a binary-source system rather than a planetary lens system. For the record, we present both 2L1S and 1L2S model parameters: $(t_{0}, u_{0}, t_{\rm E}, s, q, \alpha, \rho_{\ast}, f_{\rm S}, f_{\rm B}) = (9490.162, 0.032, 24.901, 0.199, 0.408, 1.153, < 0.028, \\ 0.039, 0.158)$ and $(t_{0,{\rm S1}}, t_{0,{\rm S2}}, u_{0,{\rm S1}}, u_{0,{\rm S2}}, t_{\rm E}, q_{\rm flux}, \rho_{\ast, {\rm S1}}, \\ \rho_{\ast, {\rm S2}}, f_{\rm S}, f_{\rm B}) = (9490.408, 9490.059, -0.001, 0.058, \\ 22.246, 24.946, 0.011, 0.074, 0.047, 0.151)$.

\subsection{Potentially Interesting Low-mass, Non-planetary Companions} % BD candidates
\label{sec:BD_candidates}
Summarizing, three of these $12$ events have companions that are plausibly brown dwarfs and therefore might be the focus of future studies. Two of these have mass ratios $q<0.1$, namely \threethirtyeight\, ($q = 0.08$) and \twentythreefiftyeight\, ($q = 0.05$). 

For the third, \twentysevenfourtyfive, the 2L1S solution has a substantially larger mass ratio ($q = 0.15$).  However, it has a very short timescale ($t_{\rm E} = 6.8$ days).  Hence, it is likely that the host is itself either a brown dwarf or a very low-mass star, in which case the companion would likely be an extremely low-mass brown dwarf.  For a similar case, KMT-2019-BLG-0371, see \citet{kim21}. For a list of short events with low-$q$ companions, see \citet{ryu21}. To confirm the 2L1S hypothesis, one would have to rule out the 1L2S interpretation, which might require, for example, spectroscopy using next-generation extremely large telescopes.  We do not delve into these issues here but simply point out that this event is of potential interest.

% References ----------------------------------------------------------------   


\begin{thebibliography}{999}

\bibitem[Albrow et al.(2009)]{albrow09} Albrow, M.~D., Horne, K., Bramich, D.~M., et al.\ 2009, \mnras, 397, 2099. doi:10.1111/j.1365-2966.2009.15098.x
\bibitem[Alard \& Lupton(1998)]{alard98} Alard, C. \& Lupton, R.~H.\ 1998, \apj, 503, 325. doi:10.1086/305984
\bibitem[Bennett et al.(2008)]{bennett08} Bennett, D.~P., Bond, I.~A., Udalski, A., et al.\ 2008, \apj, 684, 663. doi:10.1086/589940
\bibitem[Bond et al.(2001)]{bond01} Bond, I.~A., Abe, F., Dodd, R.~J., et al.\ 2001, \mnras, 327, 868. doi:10.1046/j.1365-8711.2001.04776.x
\bibitem[Dominik(1998)]{dominik98} Dominik, M.\ 1998, \aap, 329, 361. doi:10.48550/arXiv.astro-ph/9702039
\bibitem[Dong et al.(2009)]{dong09} Dong, S., Gould, A., Udalski, A., et al.\ 2009, \apj, 695, 2, 970. doi:10.1088/0004-637X/695/2/970
\bibitem[Gaia Collaboration et al.(2023)]{GAIADR3} Gaia Collaboration, Vallenari, A., Brown, A.~G.~A., et al.\ 2023, \aap, 674, A1. doi:10.1051/0004-6361/202243940
\bibitem[Gaudi(1998)]{gaudi98} Gaudi, B.~S.\ 1998, \apj, 506, 533. doi:10.1086/306256
\bibitem[Gould(1992)]{gould92} Gould, A.\ 1992, \apj, 392, 442. doi:10.1086/171443
\bibitem[Gould(1997)]{gould97} Gould, A.\ 1997, \apj, 480, 1, 188. doi:10.1086/303942
\bibitem[Gould et al.(2014)]{gould14} Gould, A., Udalski, A., Shin, I.-G., et al.\ 2014, Science, 345, 6192, 46. doi:10.1126/science.1251527
\bibitem[Gould et al.(2022)]{gould22} Gould, A., Han, C., Zang, W., et al.\ 2022, \aap, 664, A13. doi:10.1051/0004-6361/202243744 % AF Paper 05
\bibitem[Gould(2022)]{MASADA} Gould, A.\ 2022, arXiv:2209.12501. doi:10.48550/arXiv.2209.12501
\bibitem[Gonzalez et al.(2012)]{gonzalez12} Gonzalez, O.~A., Rejkuba, M., Zoccali, M., et al.\ 2012, \aap, 543, A13. doi:10.1051/0004-6361/201219222
\bibitem[Griest \& Hu(1992)]{griest92} Griest, K. \& Hu, W.\ 1992, \apj, 397, 362. doi:10.1086/171793
\bibitem[Gui et al.(2024)]{gui24} Gui, Y., Zang, W., Zhai, R., et al.\ 2024, \aj, 168, 2, 49. doi:10.3847/1538-3881/ad4ce5 % AF Paper 12
\bibitem[Han \& Gould(1997)]{han97} Han, C. \& Gould, A.\ 1997, \apj, 480, 1, 196. doi:10.1086/303944
\bibitem[Han et al.(2022a)]{han22a} Han, C., Kim, D., Gould, A., et al.\ 2022a, \aap, 664, A33. doi:10.1051/0004-6361/202243484
\bibitem[Han et al.(2022b)]{han22b} Han, C., Kim, D., Yang, H., et al.\ 2022b, \aap, 664, A114. doi:10.1051/0004-6361/202243161
\bibitem[Han et al.(2023a)]{han23a} Han, C., Lee, C.-U., Zang, W., et al.\ 2023a, \aap, 674, A90. doi:10.1051/0004-6361/202346298
\bibitem[Han et al.(2023b)]{han23b} Han, C., Jung, Y.~K., Bond, I.~A., et al.\ 2023b, \aap, 675, A36. doi:10.1051/0004-6361/202346596
\bibitem[Holtzman et al.(1998)]{holtman98} Holtzman, J.~A., Watson, A.~M., Baum, W.~A., et al.\ 1998, \aj, 115, 1946. doi:10.1086/300336
\bibitem[Hwang et al.(2022)]{hwang22} Hwang, K.-H., Zang, W., Gould, A., et al.\ 2022, \aj, 163, 43. doi:10.3847/1538-3881/ac38ad % AF Paper 02
\bibitem[Jung et al.(2019)]{jung19} Jung, Y.~K., Gould, A., Zang, W., et al.\ 2019, \aj, 157, 2, 72. doi:10.3847/1538-3881/aaf87f
\bibitem[Jung et al.(2022)]{jung22} Jung, Y.~K., Zang, W., Han, C., et al.\ 2022, \aj, 164, 262. doi:10.3847/1538-3881/ac9c5c % AF Paper 06.
\bibitem[Jung et al.(2023)]{jung23} Jung, Y.~K., Zang, W., Wang, H., et al.\ 2023, \aj, 165, 226. doi:10.3847/1538-3881/accb8f % AF Paper 08.
\bibitem[Kennedy \& Kenyon(2008)]{kennedy08} Kennedy, G.~M. \& Kenyon, S.~J.\ 2008, \apj, 673, 502. doi:10.1086/524130 
\bibitem[Kervella et al.(2004)]{kervella04} Kervella, P., Th{\'e}venin, F., Di Folco, E., et al.\ 2004, \aap, 426, 297. doi:10.1051/0004-6361:20035930
\bibitem[Kim et al.(2016)]{kim16} Kim, S.-L., Lee, C.-U., Park, B.-G., et al.\ 2016, Journal of Korean Astronomical Society, 49, 37. doi:10.5303/JKAS.2016.49.1.37 
\bibitem[Kim et al.(2018)]{kim18} Kim, D.-J., Kim, H.-W., Hwang, K.-H., et al.\ 2018, \aj, 155, 76. doi:10.3847/1538-3881/aaa47b
\bibitem[Kim et al.(2021)]{kim21} Kim, Y.~H., Chung, S.-J., Yee, J.~C., et al.\ 2021, \aj, 162, 1, 17. doi:10.3847/1538-3881/abf930
\bibitem[Kuang et al.(2022)]{kuang22} Kuang, R., Zang, W., Jung, Y.~K., et al.\ 2022, \mnras, 516, 2, 1704. doi:10.1093/mnras/stac2315
\bibitem[Minniti et al.(2010)]{minniti10} Minniti, D., Lucas, P.~W., Emerson, J.~P., et al.\ 2010, \na, 15, 5, 433. doi:10.1016/j.newast.2009.12.002 % VVV
\bibitem[Nataf et al.(2013)]{nataf13} Nataf, D.~M., Gould, A., Fouqu{\'e}, P., et al.\ 2013, \apj, 769, 2, 88. doi:10.1088/0004-637X/769/2/88
\bibitem[Paczynski(1997)]{paczynski97} Paczynski, B.\ 1997, , astro-ph/9711007. doi:10.48550/arXiv.astro-ph/9711007
\bibitem[Pecaut \& Mamajek(2013)]{dwarftable} Pecaut, M.~J. \& Mamajek, E.~E.\ 2013, \apjs, 208, 1, 9. doi:10.1088/0067-0049/208/1/9
\bibitem[Poindexter et al.(2005)]{poindexter05} Poindexter, S., Afonso, C., Bennett, D.~P., et al.\ 2005, \apj, 633, 914. doi:10.1086/468182
\bibitem[Refsdal(1966)]{refsdal66} Refsdal, S.\ 1966, \mnras, 134, 3, 315. doi:10.1093/mnras/134.3.315
\bibitem[Ryu et al.(2021)]{ryu21} Ryu, Y.-H., Hwang, K.-H., Gould, A., et al.\ 2021, \aj, 162, 96. doi:10.3847/1538-3881/ac062a      
\bibitem[Ryu et al.(2022)]{ryu22} Ryu, Y.-H., Kil Jung, Y., Yang, H., et al.\ 2022, \aj, 164, 5, 180. doi:10.3847/1538-3881/ac8d6c
\bibitem[Ryu et al.(2023)]{ryu23} Ryu, Y.-H., Shin, I.-G., Yang, H., et al.\ 2023, \aj, 165, 3, 83. doi:10.3847/1538-3881/acab6b
\bibitem[Ryu et al.(2024)]{ryu24} Ryu, Y.-H., Udalski, A., Yee, J.~C., et al.\ 2024, \aj, 167, 3, 88. doi:10.3847/1538-3881/ad1888 % AF Paper 10
\bibitem[Shin et al.(2012)]{shin12} Shin, I.-G., Han, C., Choi, J.-Y., et al.\ 2012, \apj, 755, 2, 91. doi:10.1088/0004-637X/755/2/91
\bibitem[Shin et al.(2019)]{shin19} Shin, I.-G., Yee, J.~C., Gould, A., et al.\ 2019, \aj, 158, 5, 199. doi:10.3847/1538-3881/ab46a5
\bibitem[Shin et al.(2023a)]{shin23a} Shin, I.-G., Yee, J.~C., Gould, A., et al.\ 2023, \aj, 165, 8. doi:10.3847/1538-3881/ac9d93
\bibitem[Shin et al.(2023b)]{shin23b} Shin, I.-G., Yee, J.~C., Zang, W., et al.\ 2023, \aj, 166, 104. doi:10.3847/1538-3881/ace96d % AF Paper 09
\bibitem[Shin et al.(2024)]{shin24} Shin, I.-G., Yee, J.~C., Zang, W., et al.\ 2024, \aj, 167, 6, 269. doi:10.3847/1538-3881/ad3ba3 % AF Parper 11
\bibitem[Smith et al.(2003)]{smith03} Smith, M.~C., Mao, S., \& Paczy{\'n}ski, B.\ 2003, \mnras, 339, 925. doi:10.1046/j.1365-8711.2003.06183.x
\bibitem[Sumi et al.(2003)]{sumi03} Sumi, T., Abe, F., Bond, I.~A., et al.\ 2003, \apj, 591, 204. doi:10.1086/375212
\bibitem[Suzuki et al.(2016)]{suzuki16} Suzuki, D., Bennett, D.~P., Sumi, T., et al.\ 2016, \apj, 833, 2, 145. doi:10.3847/1538-4357/833/2/145
\bibitem[Szyma{\'n}ski et al.(2011)]{OIIIcat} Szyma{\'n}ski, M.~K., Udalski, A., Soszy{\'n}ski, I., et al.\ 2011, \actaa, 61, 2, 83. doi:10.48550/arXiv.1107.4008
\bibitem[Tomaney \& Crotts(1996)]{tomaney96} Tomaney, A.~B. \& Crotts, A.~P.~S.\ 1996, \aj, 112, 2872. doi:10.1086/118228
\bibitem[Udalski et al.(2015)]{udalski15} Udalski, A., Szyma{\'n}ski, M.~K., \& Szyma{\'n}ski, G.\ 2015, \actaa, 65, 1
\bibitem[Udalski et al.(2018)]{udalski18} Udalski, A., Ryu, Y.-H., Sajadian, S., et al.\ 2018, \actaa, 68, 1, 1. doi:10.32023/0001-5237/68.1.1
\bibitem[Wang et al.(2022)]{wang22} Wang, H., Zang, W., Zhu, W., et al.\ 2022, \mnras, 510, 1778. doi:10.1093/mnras/stab3581 % AF Paper 03
\bibitem[Yang et al.(2020)]{yang20} Yang, H., Zhang, X., Hwang, K.-H., et al.\ 2020, \aj, 159, 98. doi:10.3847/1538-3881/ab660e
\bibitem[Yang et al.(2022)]{yang22} Yang, H., Zang, W., Gould, A., et al.\ 2022, \mnras, 516, 2, 1894. doi:10.1093/mnras/stac2023
\bibitem[Yang et al.(2023)]{yang23} Yang, H., Yee, J.~C., Hwang, K.-H., et al.\ 2023, arXiv:2311.04876
\bibitem[Yang et al.(2025)]{yang25} Yang, H., Yee, J.~C., Zhang, J., et al.\ 2025, \aj, 169, 6, 295. doi:10.3847/1538-3881/adc73e
\bibitem[Yee et al.(2015)]{yee15} Yee, J.~C., Gould, A., Beichman, C., et al.\ 2015, \apj, 810, 2, 155. doi:10.1088/0004-637X/810/2/155
\bibitem[Yoo et al.(2004)]{yoo04} Yoo, J., DePoy, D.~L., Gal-Yam, A., et al.\ 2004, \apj, 603, 139. doi:10.1086/381241
\bibitem[Zang et al.(2021)]{zang21} Zang, W., Hwang, K.-H., Udalski, A., et al.\ 2021, \aj, 162, 163. doi:10.3847/1538-3881/ac12d4 % AF Paper 01
\bibitem[Zang et al.(2022a)]{zang22a} Zang, W., Shvartzvald, Y., Udalski, A., et al.\ 2022a, \mnras, 514, 4, 5952. doi:10.1093/mnras/stac1631
\bibitem[Zang et al.(2022b)]{zang22b} Zang, W., Yang, H., Han, C., et al.\ 2022b, \mnras, 515, 928. doi:10.1093/mnras/stac1883 % AF Paper 04.
\bibitem[Zang et al.(2023)]{zang23} Zang, W., Jung, Y.~K., Yang, H., et al.\ 2023, \aj, 165, 103. doi:10.3847/1538-3881/acb34b % AF Paper 07.
\bibitem[Zang et al.(2025)]{zang25} Zang, W., Jung, Y.~K., Yee, J.~C., et al.\ 2025, Science, 388, 6745, 400. doi:10.1126/science.adn6088
\bibitem[Zhang et al.(2022)]{zhang22} Zhang, K., Gaudi, B.~S., \& Bloom, J.~S.\ 2022, Nature Astronomy, 6, 782. doi:10.1038/s41550-022-01671-6

\end{thebibliography}
\end{document}